\documentclass[preprints,article,accept,moreauthors,pdfTeX]{Definitions/mdpi}
\firstpage{1} 
\pubvolume{12}
\issuenum{1}
\articlenumber{132}
\pubyear{2026}
\copyrightyear{2026}
\datereceived{3 April 2026} 
\daterevised{26 April 2026 } 
\dateaccepted{27 April 2026} 
\datepublished{3 May 2026} 
\hreflink{https://doi.org/10.3390/universe12050132} 

\usepackage{bm, bbm, amssymb, latexsym, tensor}
\usepackage{amsthm, thmtools, natbib}
\DeclareMathOperator{\sech}{sech}
\usepackage{dcolumn,multicol}

\usepackage{slantsc}
\usepackage{slashed}
\usepackage{cancel}

\newcommand{\bd}{{\bf d}}

\newcommand{\bpartial}{\boldsymbol{\partial}}
\newcommand{\rd}{{\mathrm d}}

\newcommand{\CR}{\mathcal R}
\newcommand{\CK}{\mathcal K}
\newcommand{\CF}{\mathcal F}

\font\bigastfont=cmr10 scaled \magstep 1
\def\bdot{\hbox{\bigastfont .}}

\newcommand{\mnu}{{\mu\nu}}

\newcommand{\p}{\partial}
\newcommand{\nn}{\nonumber}
\newcommand{\I}{{\rm I}}
\newcommand{\II}{{\rm II}}
\newcommand{\III}{{\rm III}}

\newcommand{\diffb}{{\mathbf d}}

\newcommand{\ddt}[1]{\frac{{\mathrm d} #1}{{\mathrm d} t}}

\definecolor{myblue}{rgb}{0.2,0.3,0.7}
\definecolor{darkgreen}{rgb}{0,0.3,0}
\definecolor{mygreen}{rgb}{0,0.5,0}
\definecolor{grey}{rgb}{0.5,0.5,0.5}
\definecolor{orange}{rgb}{1,0.5,0}
\definecolor{mybloodyred}{rgb}{0.7,0,0}
\definecolor{chromeyellow}{rgb}{1.0, 0.65, 0.0}
\definecolor{yellow(ncs)}{rgb}{1.0, 0.83, 0.0}
\definecolor{saffron}{rgb}{0.96, 0.77, 0.19}
\definecolor{laserlemon}{rgb}{1.0, 1.0, 0.13}
\Title{Novel Realizations of Warp Drive Spacetimes\\ as Solutions of General Relativity}
\TitleCitation{Warp drive spacetimes}

\Author{Thomas Buchert *\orcidA{} and Antony Frackowiak \orcidB{}}

\AuthorNames{Thomas Buchert, Antony Frackowiak}

\AuthorCitation{Buchert, T.; Frackowiak, A.}

\address{Universit\'e Lyon 1, ENS de Lyon, CNRS, CRAL, UMR 5574, Lyon, France;
antonyfrackowiak@gmail.com}

\corres{Correspondence: TB --- buchert@ens--lyon.fr} 

\abstract{
We first take a closer look at the original warp drive proposal by Alcubierre, examine its kinematics in the context of a covariant 3+1 setting, and explain some drawbacks of this construction. In this model, changes of the velocity profile are suppressed, apart from an externally given amplitude. We then discuss Einstein's equations for currently employed spacetime restrictions, and provide the governing equations for the Nat\'ario class of metrics with one-component coordinate velocity in a subcase. Following Synge's G-method we determine the constraints on realizations for two examples:  assuming the form of the solution \textit{a priori} as in Alcubierre's model,  and determining the solution through an assumption imposed along geodesics. We analyze in detail the role of coordinate acceleration and coordinate vorticity, providing illustrations for both example solutions. For the second we find an expected generic instability of the warp field.
We then propose a framework that allows for spatial curvature and the description of warp field dynamics within a relativistic Lagrangian perturbation approach, also including exact solutions of the Szekeres class II.
These generalizations allow us to link studies on warp fields to relativistic cosmology.
A direct correspondence between solutions of Newtonian gravity and general relativity is exploited. 
We conclude by discussing possible future paths towards physical warp drives within tilted fluid flows.
}

\begin{document}
\setcounter{tocdepth}{2}
\tableofcontents
%
\bigskip
\section{The warp drive context}
\label{I}

In 1994, Miguel Alcubierre \cite{alcubierre} introduced the concept of a warp drive metric. It is claimed that this metric form
would allow to travel at an apparent superluminal speed. After this publication, many others have followed with the aim of improving or studying the properties of this metric in more detail and, in particular, to respond to the problems of negative energy such as the work carried out by Nat\'{a}rio \cite{natario} and many others. For a comprehensive introduction and references we refer the reader to the recent investigation of restricted motion in general relativity \cite{BBV}, providing a review on warp drive proposals in the literature, as well as investigations of inherent problems, asymptotic flatness, global hyperbolicity, energy conditions and misconceptions in the literature, all of which are not touched upon in this work; see also \cite{Santiago} for another critical review. 

The Alcubierre model faces substantial physical and structural challenges.
Notably, the shape and dynamics of the warp field are imposed in an \textit{ad hoc} manner, rather than being derived from self-consistent solutions of the Einstein field equations. The velocity profile, typically encoded via a shift vector, is externally prescribed, and the warp field's spatial structure does not exhibit dynamical changes. Subjected to Einstein's equations, it requires negative energy densities that violate the classical energy conditions \cite{Santiago}. These limitations raise fundamental concerns about the physical realism and predictive power of the model. In contrast, the present work aims to provide the framework to derive warp field configurations by imposing assumptions on degrees of freedom left by the metric ansatz, allowing to determine all aspects of the warp field, its shape, evolution, and energy-momentum content, that satisfy Einstein's equations.

In a recent \textit{communication} \cite{BBL}, we discuss the kinematical architecture of the most studied warp drive models. By inspection of the proposed metric form, we classified the imposed restrictions into ${\rm\bf R1}$ (flow-orthogonality, i.e. the $4$-velocity and the normal to the hypersurfaces are aligned); ${\rm\bf R2}$ (the lapse function is assumed constant, i.e. the spacetime foliation is geodesic, while the shift vector is non-vanishing and equals minus the coordinate velocity in view of ${\rm\bf R1}$); ${\rm\bf R3}$ (spatial hypersurfaces are flat). 
Although these constraints effectively suppress covariant motion, covariant vorticity, and spatial curvature, which we consider essential for realistic descriptions of motion through curved spacetime, the present work retains but fully exploits these restrictions, and we extend these lines of inquiry by contrasting two approaches: a metric-based approach that, however, specifies the complete velocity model \textit{a priori}, and a metric-based approach that analyses an alternative assumption on the kinematical properties, both as opposed to a matter-based strategy where one specifies the energy-momentum content \textit{a priori}. In the metric-based approach, commonly known as Synge's G-method \cite{Synge,EllisGarfinkle}, one begins with a chosen spacetime geometry, defined through the lapse, shift, and spatial metric, and subsequently derives the implied energy-momentum tensor. While both metric- and matter-based methods are legitimate within general relativity, the former approach is particularly well-suited to the exploration of exact solutions and constraints on physically admissible configurations, especially in cases involving spacetime engineering such as warp drives, while the matter-based strategy allows to make physical assumptions on the matter content and contact with common strategies to evolve initial data according to the Cauchy problem. 

Here, we consider a class of 3+1 decomposed spacetimes with constant lapse and a shift vector describing motion. This setting encompasses the Alcubierre model \cite{alcubierre} as a subcase of one-component velocity models, while also allowing for more general configurations based on the so-called Nat\'ario class of metrics \cite{natario} that admits a general three-component velocity model (not to be confused with Nat\'ario's zero-expansion warp drive, for details and refinements see \cite{RodalNatario} and \cite{BBV}).
Our objective is to characterize admissible warp fields that respect the full structure of the Einstein equations in explicit form by employing global inertial coordinates and by looking at the warp drive from the perspective of Eulerian observers that follow the normal congruence of the foliation. 

We shall furthermore suggest an approach to study the dynamics and morphology of warp fields (not warp drives) in a flow-orthogonal foliation. This more direct description aligns the time-vector with the normal of the foliation: the coordinate velocity $\bm V$, and hence the shift vector $\bm N$, vanishes, thus altering  ${\rm\bf R2}$. This suggestion allows to follow a matter-based approach, however, keeping a constant lapse function only encompasses  the matter model of irrotational dust, i.e. vanishing stress tensor sources. Such models are listed in \cite{BBL} as Lagrangian ${\rm\bf R1}$-warp.
In this approach it is necessary but also advantageous to relax ${\rm\bf R3}$ in order to obtain physically nontrivial  models within a flow-orthogonal foliation. Within such a setting we can exploit a direct correspondence between solutions of Newtonian gravity and general relativity, summarized in \cite{Bcorrespondence}. 
We are then in the position to describe a warp field with intrinsic curvature by transforming known Newtonian solutions and approximation schemes. A further introduction of a background spacetime will allow us to make contact to well-studied models in relativistic cosmology, also including special classes of exact solutions. 
The drawback will be that we have to confine ourselves to irrotational warp fields, considered e.g. in \cite{Lenz1,Lenz2,Lavinia, Rodal}. However, as we already discussed in \cite{BBL}, this setting forms an intermediate step towards warp drives in a tilted foliation allowing for nonvanishing covariant velocity, acceleration and vorticity. While  within ${\rm\bf R1}$-models we can study dynamics, morphology and stability of warp fields under realistic conditions, we think that we necessarily have to move to the tilted setting to come closer to a realization of physical warp drives.

This paper is organized as follows.
In Section~\ref{II}, Alcubierre's proposal is reviewed and its kinematical properties are studied in detail, followed by a discussion of its drawbacks. Section~\ref{III} then derives the Einstein equations for three-component coordinate velocity fields in the framework of commonly used warp metrics, and we also specify to one-component velocity fields thereafter. This section complements the presentation in the recent critical review \cite{BBV} by only employing active, fluid-based variables rather than variables of the geometry that are commonly used, and we employ vectorial notation where possible.
We focus the presentation on the coordinate acceleration field to make contact with a correspondence between Newtonian gravity and general relativity, and we include the cosmological constant. We also evaluate the one-component case in components, where most of this material is detailed in appendices.
In Section~\ref{IV} we discuss different methods for deriving solutions, both in the spirit of Synge's G-method, where we first give the dynamical properties and the sources of Alcubierre's model as an example solution of general relativity, the second one proposing an alternative assumption to model dynamical warp fields for Alcubierre initial data. Section~\ref{V} proposes a novel approach to study the warp field's dynamics and morphology through a systematic construction of solutions of Einstein's equations with intrinsic curvature from gravitational vector theories. Section~\ref{VI} is then dedicated to perspectives on generalizations that are possible in a flow-orthogonal and in a tilted setting.

\vspace{5pt}

\noindent
NOTATIONS: Bold notation will be used for vectors and forms. The vector product is denoted by $\times$, the 
tensor product by $\otimes$, with components $({\bm a} \otimes {\bm b})_{ij} = a_i b_j$. We define symmetrization and antisymmetrization of indices by $v_{(i,j)} = \tfrac{1}{2}(v_{i,j}+v_{j,i})$, $v_{[i,j]} = \tfrac{1}{2}(v_{i,j}-v_{j,i})$, respectively; four-dimensional coordinate indices are denoted by greek letters; three-dimensional coordinate indices are denoted  $i,j,k \ldots  = 1,2,3$, while indices starting with $a,b,c   = 1 \ldots, 3 $ denote counter indices. A semicolon $;$ stands for the 4-covariant derivative, a comma  for the partial derivative with respect to $x^j$, $\partial / \partial x^j$, sometimes also denoted by $\p_i$ (for globally rectangular (inertial, nonrotating) coordinates $\bm x$), a vertical slash $|$ for the derivative with respect to $X^k$, $\partial / \partial X^k$ (in local Lagrangian coordinates $\bm X$); a double vertical slash $||$ stands for the covariant spatial derivative; an overdot will be used to represent the covariant time derivative of a vector field $\bm F$ along the fluid flow, $u^\mu \bm F_{; \mu}$, and $\frac{\rm d}{{\rm d}t}F^i= \partial_t F^i + V^k F^i_{\ || k}$ the Lagrangian (or total) coordinate time derivative convected with the coordinate velocity $\bm V$, the latter two coincide for scalar fields; summation over repeated indices is understood. We will set $c = 1$.\\

\section{Alcubierre's proposal of a warp drive model}
\label{II}

We give a brief recap on the assumptions made in \cite{alcubierre} and study the kinematical properties of the warp field, also beyond the expansion rate commonly presented. We conclude this section by a list of drawbacks of this construction
on top of the restricting assumptions made on the spacetime. 

\subsection{The Alcubierre model}
\label{model}

Alcubierre works within a 3+1 foliation of spacetime, foliated into spatial leaves $\Sigma_t$ with the unit normal $\bm n$.
The components of $\bm n$ and its metric dual $1$-form,\footnote{Underlined variables denote the metric dual $1$-form of a given vector field tangent to the given manifold.}
are given by 
\begin{equation}\label{eq:n_vec}
	\bm{n} = \frac{1}{N} \left( 1, - \bm{N} \right)
\,,
\quad
	\underline{\bm n} = - N \, ( 1, \bm 0 ) \, ,
\end{equation}
where $N$ is the lapse function and $\bm N = N^i \bm{\partial}_i$ is the shift vector.
The general line-element reads: 
\begin{equation}
\label{Alcubierremetric}
    \rd s^2 = g_{\mu\nu}\rd x^\mu\rd x^\nu 
    = - (N^2-N_i N^i)\rd t^2+2 N_i\rd x^i\rd t+ h_{ij}\rd x^i\rd x^j\ .
\end{equation}
The induced metric components in space are $h_{ij}$, which are diffeomorphic to $\delta_{ij}$ for flat spatial sections.
Alcubierre considers the local coordinates $x^i =(x,y,z)$ as global inertial (nonrotating) coordinates, hence the metric coefficients assume their natural form, $h_{ij} = \delta_{ij}$. He further simplifies the model to a one-component coordinate velocity and specifies lapse and shift in the above metric as follows:
\begin{equation}
\label{Alcubierremetricvalues}
    N = 1 \ , 
\quad
    N^x = - V^x \ ,
\quad
    N^y = N^z = 0 \ ,
\end{equation}
where the coordinate velocity $V^x ({\bm x},t) =: V_S$ for short, is separated into an externally given velocity of the spaceship $v_S (t)$, tangent to the trajectory defined $x_S (t)= f_S ({\bm X}_S,t)$ for the spaceship $S$ at any constant position ${\bm X}_S$ in a Lagrangian coordinate system that is attached to the center of the warp field:
\begin{equation}
\label{velocity_and_distance}
    v_S (t) = \frac{\p}{\p t}\Big\vert_{\bm X_S} f_S ({\bm X}_S, t)\ ,
\end{equation}
and an explicit form of the warp field by writing 
the radial distance from the trajectory in Eulerian space in terms of an Euclidean spherical region probing the radius $r_S$ of the warp field around the moving spaceship,
\begin{equation}
    r_S ({\bm x},t) = [(x-x_S(t))^2+y^2+z^2)]^{1/2} \ , 
\end{equation}
and he assumes the separation: 
\begin{equation}
\label{VSAlc}
V_S({\bm x},t) = \bm v_S(t) W(r_S({\bm x},t))\ .
\end{equation}
The function $W(r_S(t,{\bm x}))$ determines the shape of the warp field, hence the coordinate velocity profile: 
\begin{equation}
\label{warpshape}
    W(r_S({t,\bm x})) = 
    \frac{\tanh{(\sigma(r_S + R))}-\tanh{(\sigma(r_S-R))}}{2 \tanh{(\sigma R)}}\ ,
\end{equation}
with $R$ a fixed Eulerian radius, and $\sigma$ a constant that determines the inverse thickness of the wall of the `warp bubble'.

\begin{figure}[H]
     \centering
        \includegraphics[width=0.55\textwidth]{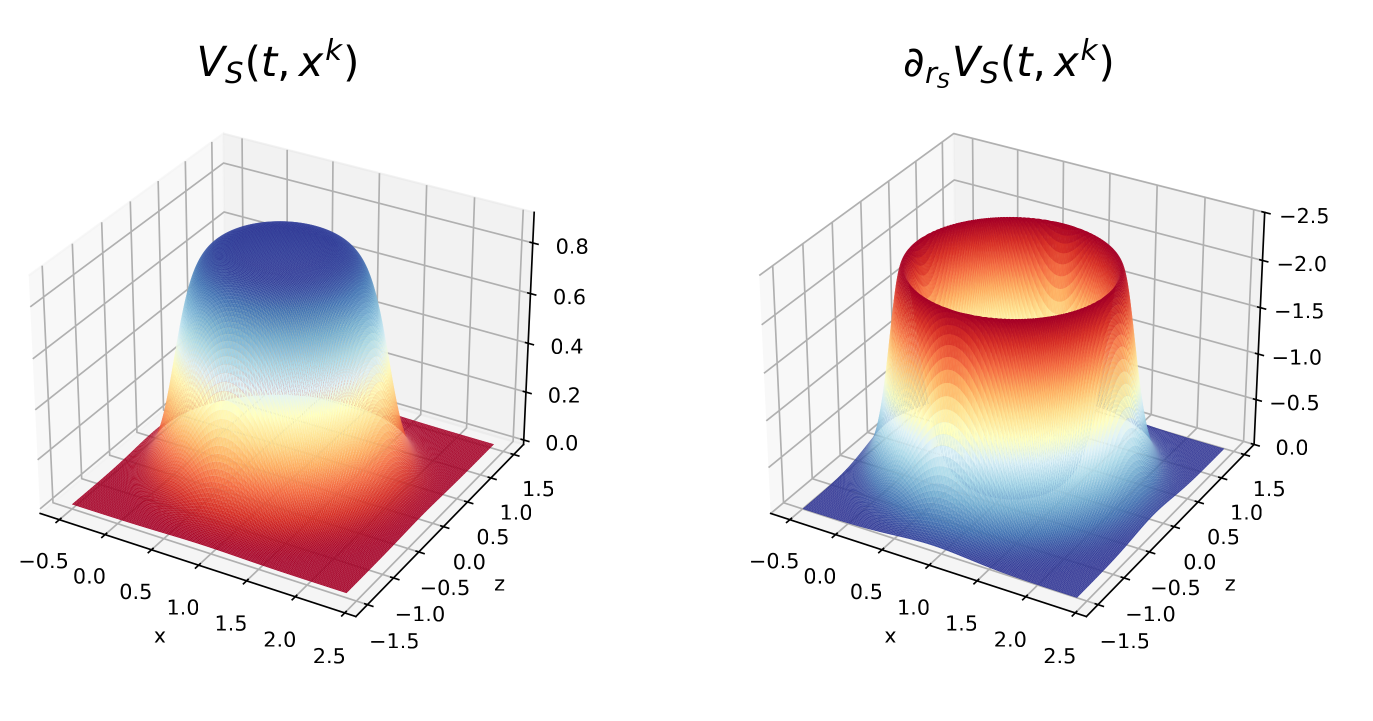}               
        \hfill
         \includegraphics[width=0.44\textwidth]{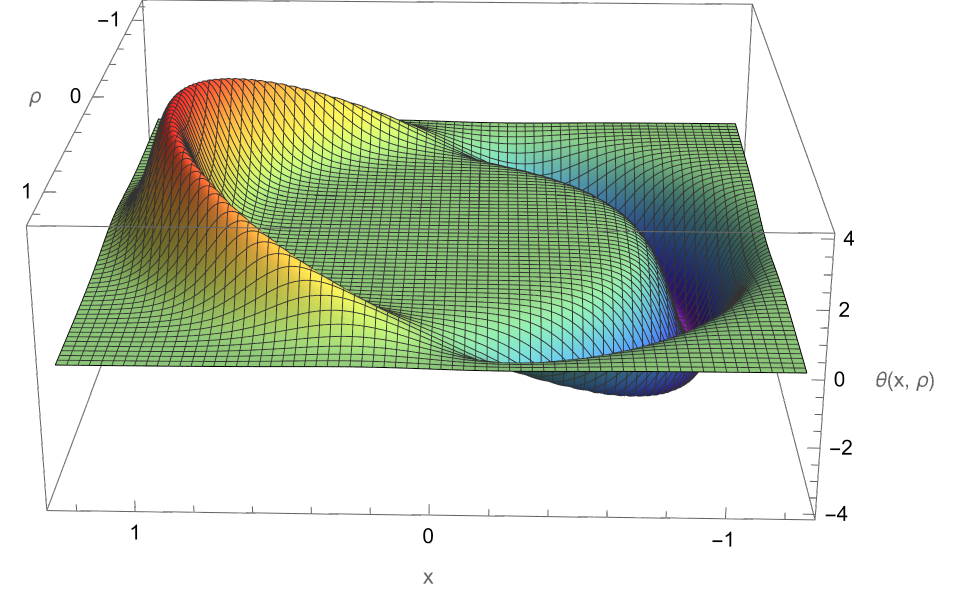}      
     \caption{Representation of the window function $W$ in 3D (left), its derivative with respect to $r_S$ (middle),  
     and the expansion of the normal volume elements (right), with $\rho = \sqrt{y^2+z^2}$, $\sigma = 8$, $R = 1$ and $x_s = 0$, using \textit{Python 3.11.11} and \textit{Mathematica 14.0}.}
     \label{fig1}  
 \end{figure}

In figure \ref{fig1} we can see the characteristic form of this function, together with its derivative with respect to $r_S$ that we will need later:
\begin{equation}
\label{warpshapederivative}
    \frac{\p W(r_S({t,\bm x}))}{\p r_S} = 
    \frac{1}{2} \coth(R \sigma) \left[ -\sigma\, \text{sech}^2\left[\sigma ( r_S - R)\right] + \sigma \, \text{sech}^2\left[\sigma (r_S + R)\right] \right] . 
\end{equation}
We know the extrinsic curvature tensor or the expansion tensor coefficients, respectively, $K_{ij} = -\Theta_{ij}$,
and we illustrate the expansion rate $\Theta$ as seen by an Eulerian observer.
\begin{equation}
    \Theta_{ij} = V_{(i,j)}\quad , \quad  
\label{theta}
    \Theta = \mathrm{Tr}(\Theta_{ij}) = v_S(t) \frac{\partial W(r_S(t,{\bm x}))}{\partial r_S(t,{\bm x})} \Big( \frac{x-x_S(t)}{r_S(t,{\bm x})} \Big)\ . 
\end{equation}
We will now further analyze the kinematics of Alcubierre's model.

\subsection{Kinematical description of the warp field}
\label{kinematics}
Recall that the Alcubierre metric is realized in rectangular coordinates with a given coordinate velocity field. 
We employ the kinematical decomposition of the coordinate velocity gradient:
\begin{equation}
   \label{Kinematic_decomposition}
    V_{i,j} = V_{(i,j)}+V_{[i,j]}
    =\Theta_{ij} + \Omega_{ij} 
    =\frac{1}{3}\Theta\delta_{ij}+\Sigma_{ij}+\Omega_{ij}\ ,
\end{equation}
where $\Theta_{ij}$ and $\Theta$ are the expansion tensor and the rate of expansion, respectively, $\Sigma_{ij}$ are the coefficients of the shear tensor, and $\Omega_{ij}$ those of the coordinate vorticity tensor. 
We also define the rate of shear scalar $\Sigma$ and the rate of vorticity scalar $\Omega$: 
\begin{equation}
   \label{vorticityscalar}
   \Sigma^2 := \frac{1}{2} \Sigma_{ij} \Sigma^{ij}\ , 
\quad
   \Omega^2 :=\frac{1}{2}\Omega_{ij} \Omega^{ij}\ . 
\end{equation}
In the case of the Alcubierre model in a rectangular coordinate system, we assume a motion in $x$ direction for the one-component velocity $\bm V^i(t,{\bm x}) = V_S(t,{\bm x})\delta_x^i$. 

\newpage

\begin{figure}[H]
      \centering
          \includegraphics[trim={12cm 0 12cm 0},clip, width=0.23\textwidth]
          {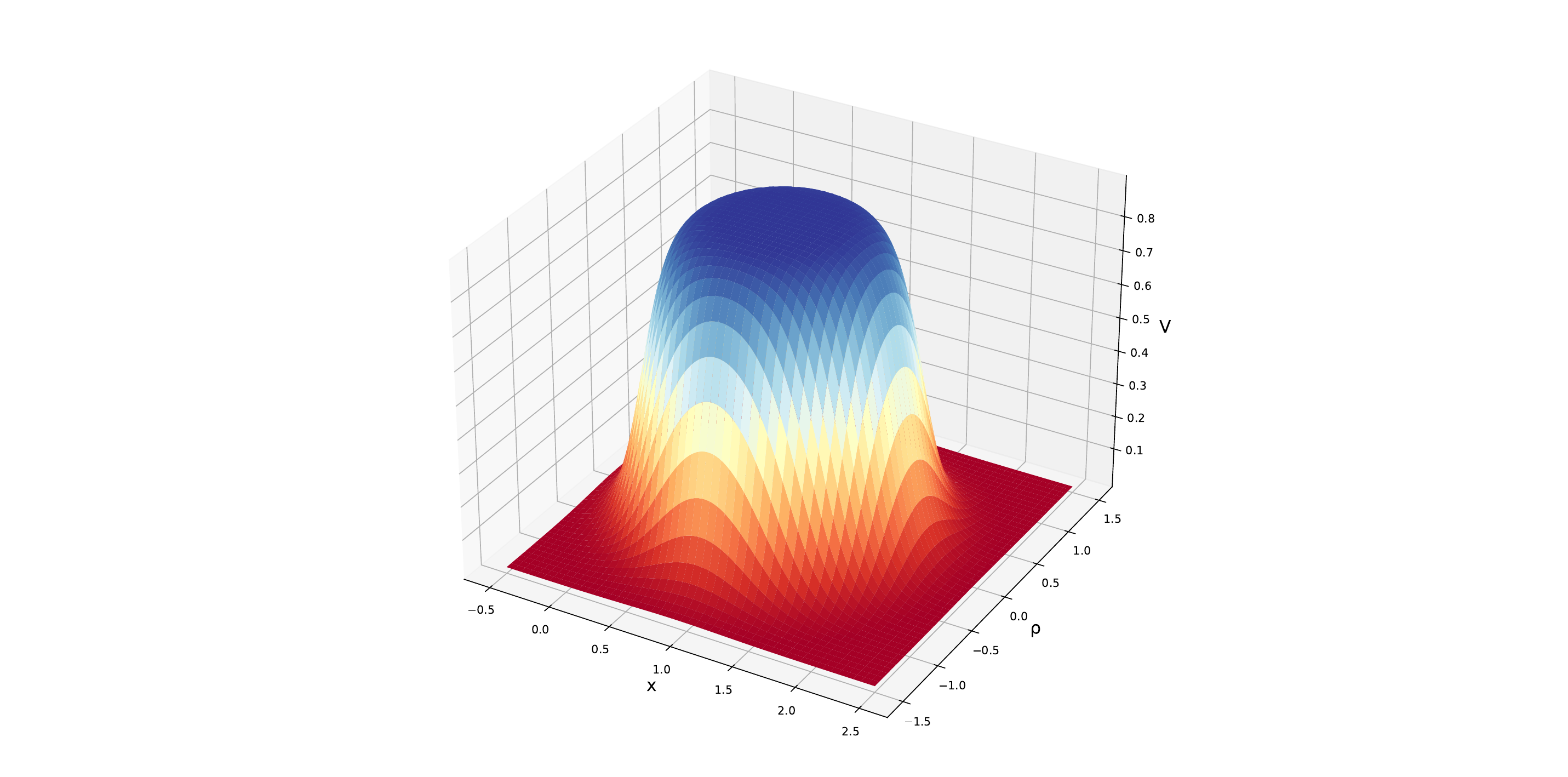}
      \hfill
          \includegraphics[trim={12cm 0 12cm 0},clip, width=0.23\textwidth]{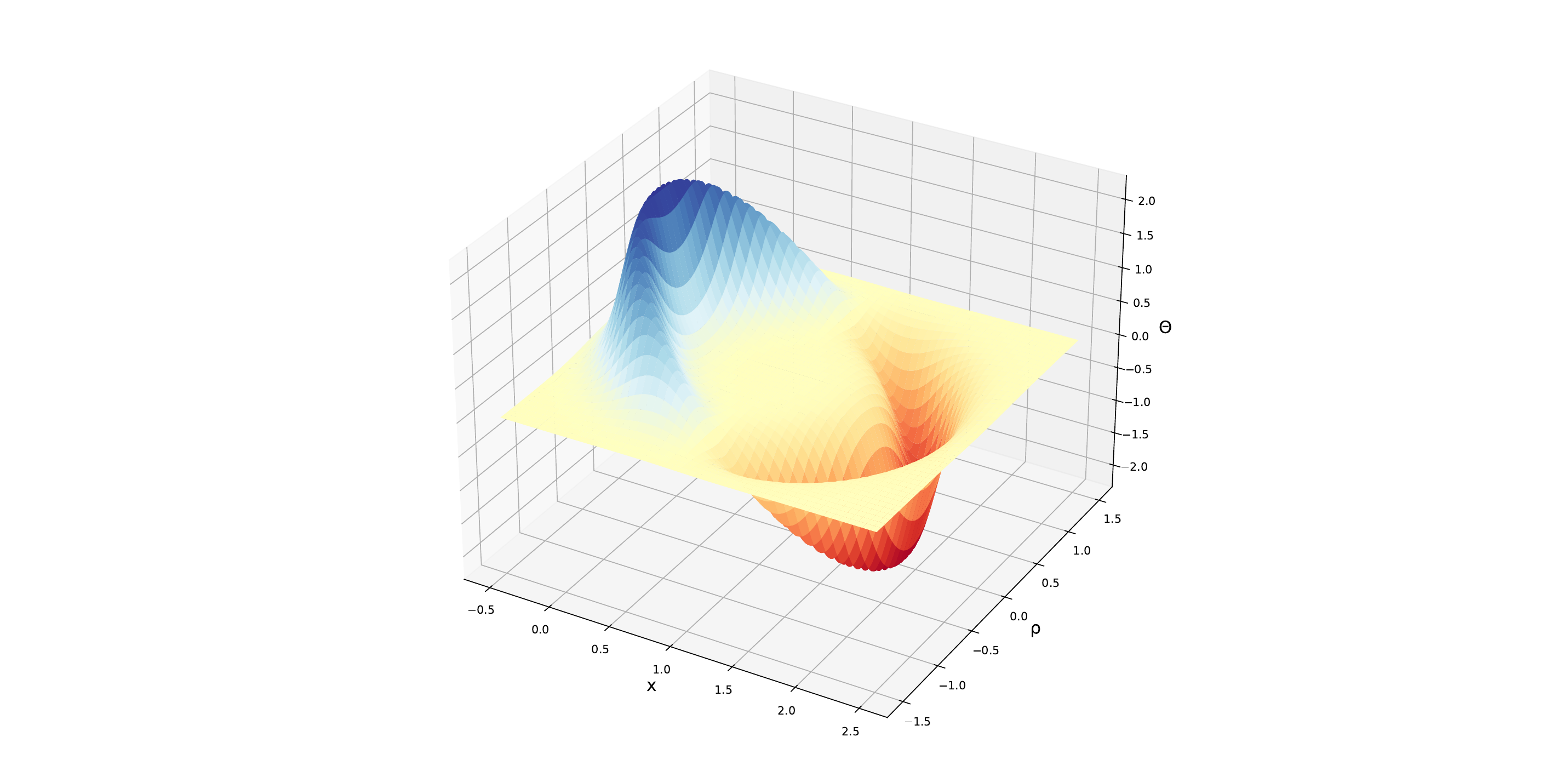}  
      \hfill      
          \includegraphics[trim={12cm 0 12cm 0},clip, width=0.23\textwidth]{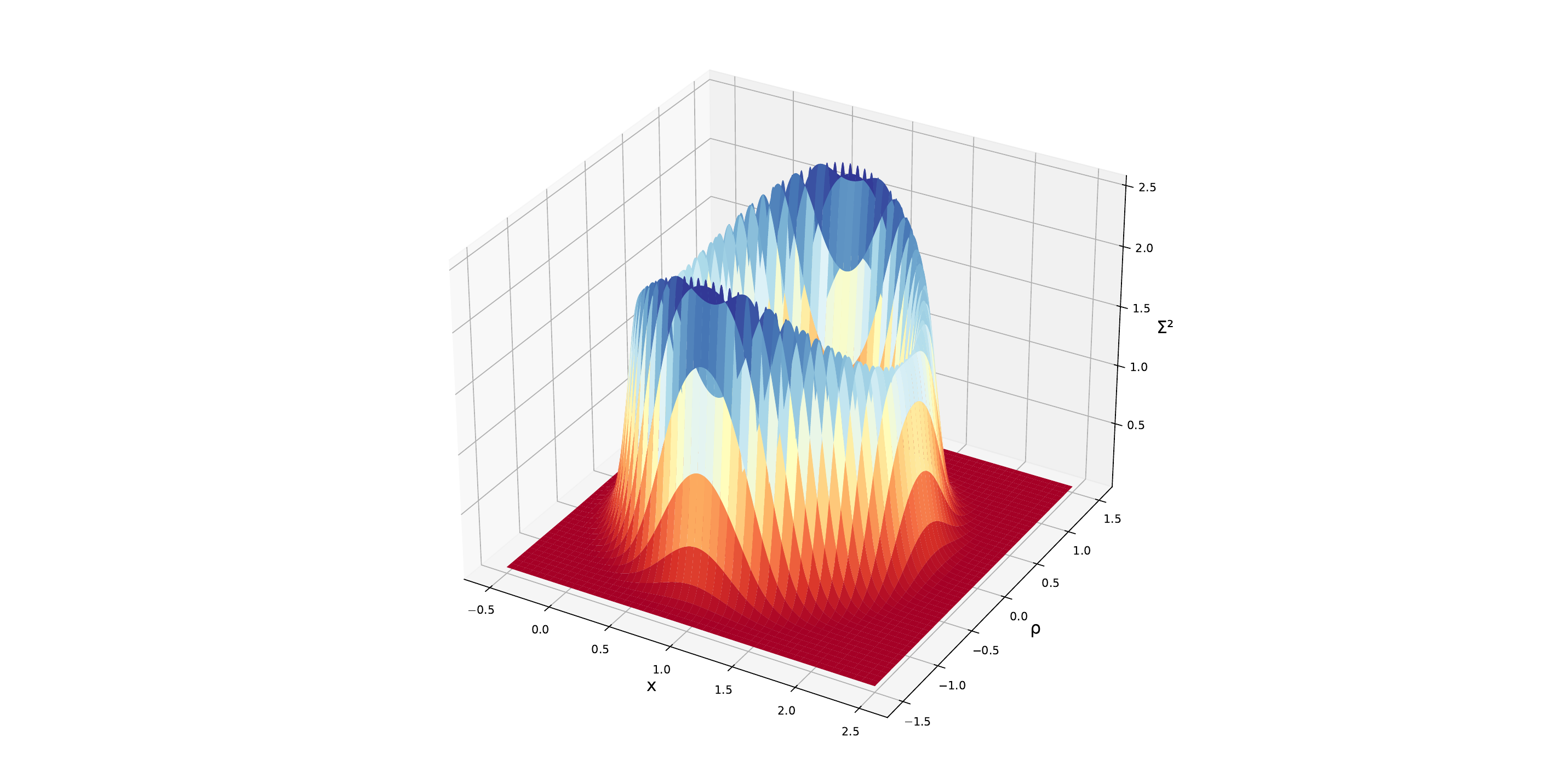}
       \hfill
         \includegraphics[trim={12cm 0 12cm 0},clip, width=0.23\textwidth]{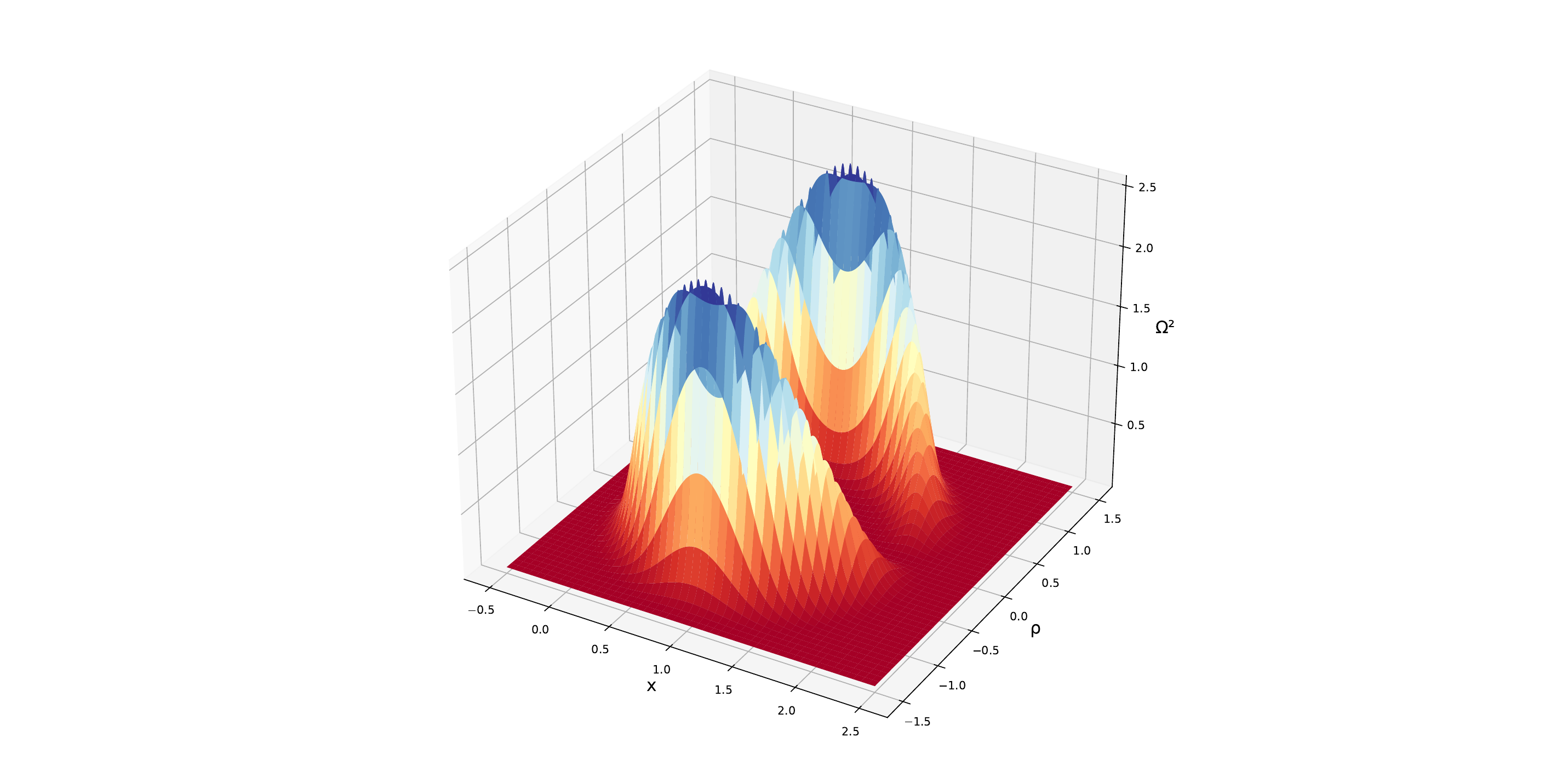}
      \caption{3D representation of kinematical quantities. From left to right: the velocity profile $V_S$, the rate of expansion $\Theta$, the shear scalar $\Sigma^2$, and the vorticity scalar $\Omega^2$, using \textit{Python 3.11.11}.}
      \label{t1_kinematics_values}
  \end{figure}

Figures \ref{t1_kinematics_values} result from numerical computation for a choice of arbitrary parameters $\sigma = 5$, $R = 1$ and for a chosen time $t=1$. We plot in three dimensions the projection of the kinematical properties in relation to the plane $(x, \rho)$, where $\rho = \sqrt{y^2 + z^2}$. 

When time evolves, we have a displacement in $x$-direction proportional to the externally given velocity $v_S (t)$ with no change in `warp bubble' shape as e.g. represented by the velocity profile. Only the amplitudes are increasing in the course of evolution if a time-dependence on $v_S (t)$ is imposed. We also see from figure \ref{t1_kinematics_values} that the interior of the warp field guarantees vanishing expansion, shear and vorticity amplitudes by construction. An interesting modification of the warp field, while keeping these properties, has been recently given \cite{White25}.

In Appendix~\ref{app:alcubierrekinematics} we give the explicit expressions for the shear and vorticity tensors as well as for the rates of expansion, shear and vorticity,
both for general one-component velocity fields and for Alcubierre's model.

\subsection{Reflections on Alcubierre's construction}
\label{drawbacks}

We observe a number of features of the Alcubierre warp drive model that do not allow for a dynamical interpretation of an evolving warp field. Notwithstanding, Alcubierre's model furnishes a solution of general relativity via Synge's G-method, but for rather exotic sources that we will determine in Section~\ref{III}.

We first recall the restrictions imposed on the spacetime, listed in the introduction: 
flow-orthogonality, i.e. the alignment of the spaceship's $4$-velocity $\bm u$ with the normal $\bm n$ to the spatial hypersurfaces, named ${\rm\bf R1}$, does not \textit{per se} impose strong dynamical restrictions apart from the fact that any covariant motion within the hypersurfaces is suppressed, i.e. the covariant velocity, ${\bm v} = 1/N ({\bm N} + {\bm V})$ (\textit{c.f.} \cite{BBL}), and the covariant acceleration both vanish. The spaceship will follow the normal congruence of the foliation, identified say by the tangent space of an Eulerian observer. A tilt of the $4$-velocity with respect to the normal is needed to allow the spaceship to have a covariant spatial velocity. Intrinsic curvature then allows the spaceship to leave its initial tangent space.
However, the morphology of a warp field can change in the course of evolution. Nontrivial density, Ricci and Weyl curvature fields up to the evolution of gravitational waves can be described within the class of ${\rm\bf R1}$-restricted metrics, as we will investigate in detail in Section~\ref{IV}. 

Setting the lapse function $N=1$, first part of ${\rm\bf R2}$, assumes a geodesic slicing that, even for vanishing shift, still leaves nontrivial source terms such as an irrotational dust matter model with or without a cosmological constant. For vanishing shift, a flow-orthogonal setting can be easily generalized to include pressure gradients by allowing for nonconstant lapse function. However, in warp drive models, motion is identified with a shift vector that  equals the negative of the  coordinate velocity, second part of ${\rm\bf R2}$. Since covariantly the spaceship is in free fall along the normal congruence, the introduction of a coordinate velocity has to be compensated by the shift, ${\bm N} = - {\bm V}$. 
Finally, to assume flat spatial hypersurfaces, ${\rm\bf R3}$, turns out to be a too strong restriction for zero shift vector, \textit{c.f.} Section~\ref{IV}. 

Second, considering a given space- and time-dependent velocity model \textit{a priori} is resulting in a solution \textit{via} Synge's G-method. Leaving e.g. a one-component coordinate velocity as a free function in the metric, Einstein's equations provide relations to the stress-energy sources that also leave one free function. If only one assumption is made, the Einstein equations form a closed system that fully determines the coordinate velocity field. We face the situation that either we know the solution (Alcubierre's model) and require corresponding sources to exist, or we determine the admissible velocity fields by some more physical assumption on the motion or the sources. In this latter approach we can always use an \textit{a priori} velocity model as initial data. 

Third, beyond giving a velocity model \textit{a priori}, Alcubierre's separation ansatz within a global inertial coordinate system results in no dynamical change of the velocity profile and the kinematics of the warp field, apart from a change in amplitude in the case of a time-dependent $v_S(t)$. 
A velocity profile, by its very definition,  generically induces different velocities at different places. Assuming it to remain constant is a forcing condition. Kinematical properties, given at some time $t_0$ would generically change in the course of evolution. 
We can also look at some average properties of the warp field, e.g. 
the spatial average of $\langle\Theta\rangle_{\mathbb{S}}$ up to the radius $R$ of the Alcubierre warp field $\mathbb{S}$:
\begin{align*}
&\text{we put}\quad
\left\{
\begin{array}{ll}
       q_x = x-x_S (t) = r_S \sin \theta \cos \phi
       \\
       q_y= y = r_S \sin \theta \sin \phi
       \\
       q_z= z = r_S \cos \theta
\end{array}  
\right.\nn\\
    &\Rightarrow \Theta = v_S(t) \frac{\p W(r_S)}{\p r_S} \frac{x-x_S}{r_S} 
    = v_S(t) \frac{\p W(r_S)}{\p r_S} \sin \theta \cos \phi 
\,,
\\[5pt]
    &\langle\Theta\rangle_{\mathbb{S}} \textrm{Vol}(\mathbb{S}) = v_S(t) \int_0^R \int_0^{2\pi} \int_0^{\pi} \frac{\p W(r_S)}{\p r_S} \sin \theta \cos \phi \,dV 
\\[5pt]
    &= v_S(t) \int_0^R \frac{\p W(r_S)}{\p r_S} r_S^2 \,\rd r_S \int_0^{2\pi} \cos \phi \,\rd \phi \int_0^\pi (\sin \theta)^2 \rd\theta = 0
\quad\text{with}\quad  \int_0^{2\pi} \cos \phi \,\rd\phi = 0
\,.
\end{align*}
We also deduce the conservation of the comoving volume $\frac{\rd}{\rd t}\textrm{Vol}(\mathbb{S})=0$ from the calculation of the volume of the warp field:
\begin{equation}
   \textrm{Vol}(\mathbb{S}) = \int_0^R \int_0^{2\pi} \int_0^\pi r_S^2 \sin \theta\, \rd\theta \,\rd\phi \,\rd r_S= \frac{4}{3}\pi R^3 \ .
\end{equation}
Hence, the volume-averaged rate of expansion remains zero in time. Furthermore, the R-support of the warp field remains a sphere.

Fourth, addressing the expectation that this model provides a mechanism for warp drive is not backed by any physical relations, even if technological considerations are put aside. A warp drive should provide dynamical relations between motion and by the spaceship controllable variables. Sometimes, in the warp literature, the expression \textit{reverse engineering} is employed to signal Synge's G-method, but this term makes only sense after one knows that the warp drive works. \\(For further aspects and other warp drive proposals, see \cite{BBV}.)\\

If we first aim at the description of the dynamics and morphology of warp fields in general relativity, we can keep the current restricted setting for a general coordinate velocity field and derive dynamical warp configurations, while keeping Alcubierre's idea of a warp field as \textit{initial condition}. We will investigate such models as solutions or approximations to Einstein's equations in Section~\ref{IV}. As an intermediate step, we then propose to describe warp field dynamics in a flow-orthogonal setting more directly by setting the shift to zero in Section~\ref{V}. 
Finally, covariant motion can then be described by tilting the $4$-velocity with respect to the normal congruence as outlined in \cite{BBL}. We will then have the full program of describing covariant motion with covariant accelerations, covariant vorticity and intrinsic curvature.\\

\section{\textbf{R}-Motion: solutions of Einstein's equations for arbitrary coordinate velocity fields}
\label{III}

Our definition of \textbf{R}-Motion covers the classic case of a ``restricted warp drive", called \textbf{R}-Warp in \cite{BBL}. We recall again the three main restrictions.
First, named \textbf{R1}, flow-orthogonality, the 4-velocity $\bm u$ is assumed to follow the normal congruence defined by $\bm n$. So, $\bm u = \bm n$, we have no tilt. Second, named \textbf{R2}, lapse function and shift vector are fixed as a constant lapse function (geodesic slicing) and a shift vector identified with the spatial coordinate velocity, denoted by ${\bm N} : = -{\bm V} (t,\bm x)$. Third, we only consider a reduced class of solutions on flat spatial hypersurfaces, \textbf{R3}. In conclusion we have for the spaceship $\bm u_S = (1, \bm V_S)$ and $\underline{\bm u}_S=(-1,\bm 0)$ for \textbf{R}-Motion. 

The four-dimensional line-element for the restriction to \textbf{R}-Motion attains the form of the so-called Nat\'{a}rio class of metrics \cite{natario} with the line-element:
\begin{equation}\label{restricted_R_metric}
    \rd s^2  = - (1-V_k V^k)\rd t^2 - 2 V_i\rd x^i\rd t+ h_{ij}\rd x^i\rd x^j
    \, ,
\end{equation}
where $h_{ij}$ are the coefficients of a flat metric, diffeomorphic to $\delta_{ij}$.
\subsection{Stress-energy tensor and conservation laws}

Contracting the stress-energy tensor by the 4-velocity and using its standard decomposition 
through the projection onto the local rest frames of the fluid orthogonal to $\bm u$, 
$b_{\mu\nu} : = g_{\mu\nu} + u_{\mu} u_{\nu}$, with 
\begin{equation}
\label{b_definition}
    b_{\alpha\mu} u^{\alpha} = 0 
\ , \
    \tensor{b}{^\mu_\alpha} \tensor{b}{^\alpha_\nu} = \tensor{b}{^\mu_\nu} 
\ , \
    b^{\alpha\beta} b_{\alpha\beta} = 3 
\  ,
\end{equation}
we obtain: 
\begin{equation}\label{EMT}
    T_{\mu\nu} = \epsilon u_\mu u_\nu + 2 q_{(\mu} u_{\nu)}+ p b_{\mu\nu}+\pi_{\mu\nu}\ ,
\end{equation}
with $\epsilon := u^\alpha u^\beta T_{\alpha\beta}\ $, $\ \ q_\mu := -\tensor{b}{^\alpha_\mu} u^\beta T_{\alpha\beta}\ $, $\ \ p_{\mu\nu} := p b_{\mu\nu}+\pi_{\mu\nu} := \tensor{b}{^\alpha_\mu} \tensor{b}{^\beta_\nu} T_{\alpha\beta}\ $, $\ \ b^{\mu\nu}\pi_{\mu\nu} = 0$, 
and where $\epsilon$ denotes the energy density of the fluid in its rest frames, $q_\mu$ the spatial momentum flux vector, $p$ the isotropic pressure, and $\pi_{\mu\nu}$ the spatial and traceless anisotropic stress.

We introduce the decomposition of the 4-covariant velocity $u$ into the 4-acceleration and the 
kinematical fields \cite{BMR}:\footnote{Compared with the decomposition \eqref{Kinematic_decomposition}, we here denote covariant shear and vorticity components in small letters. Since the expansion tensor (here always denoted with capital letters) is covariant, we will use for its symmetric trace-free part $\sigma_{ij} = \Sigma_{ij}$ interchangeably. This distinction matters for the vorticity components, since the covariant vorticity vanishes, $\omega_{ij}=0$, while the coordinate vorticity is non-vanishing, $\Omega_{ij} \ne 0$.}
\begin{equation}
    u_{\nu ;\mu} = - u_\mu a_\nu + \frac{1}{3} \Theta b_{\mu\nu} + \sigma_{\mu\nu} + \omega_{\mu\nu} \ ,
\end{equation}
with $\Theta = u^{\mu}_{\ ; \mu}$, $\ a_{\mu} := u^\mu u_{\mu ; \nu}\ $, $\ \ \sigma_{\mu\nu} = \tensor{b}{^\alpha_\mu} \tensor{b}{^\beta_\nu} u_{(\beta ; \alpha)}\ $, 
$\ \ \omega_{\mu\nu} =  \tensor{b}{^\alpha_\mu} \tensor{b}{^\beta_\nu} u_{[\beta ; \alpha]}$, 
and where $\Theta$ is the expansion rate, $\bm a$ is the acceleration of the fluid, $\bm\sigma$ is the shear tensor and $\bm\omega$ the vorticity tensor.

From the property $\tensor{T}{^{\mu\nu}_{; \nu}} = 0$ we deduce the energy conservation law,
\begin{equation}\label{energy_conservation_law}
    u_{\mu} \tensor{T}{^{\mu\nu}_{;\nu}} = 0 
    \quad
    \Leftrightarrow \quad
    \dot\epsilon + \Theta (\epsilon + p) = - a_\mu q^\mu - q^\mu_{\ ; \mu} - \pi^{\mu\nu}\sigma_{\mu\nu}\ ,
\end{equation}
and the momentum conservation law,
\begin{equation}\label{momentum_conservation_law}
    b_{\alpha\mu} \tensor{T}{^{\mu\nu}_{;\nu}} = 0 
\quad
\Leftrightarrow\quad
  - (\epsilon + p)  a_\mu =  \tensor{b}{^\alpha_\mu} p_{;\alpha} + b_{\mu\alpha}\pi^{\alpha\beta}_{\ \ ;\beta}+ b_{\mu\alpha} \dot q^{\alpha} + \frac{4}{3} \Theta q_{\mu} + q^{\alpha} (\sigma_{\alpha\mu} + \omega_{\alpha\mu})\ .
\end{equation}

In the flow-orthogonal case, \textbf{R1}, the induced metric due to projection onto the hypersurfaces orthogonal to $\bm n$, $\bm h := h_{\mu\nu} \bm \rd x^{\mu} \otimes \bm \rd x^{\nu}$ is identical to the projection onto the rest frames of the fluid, $\bm b := b_{\mu\nu} \bm \rd x^{\mu} \otimes \bm \rd x^{\nu}$.
Note that $\bm b$ and $\bm h$ usually differ because of the tilt between $\bm u$ and $\bm n$.

\subsection{3+1-Einstein dynamics for the Nat\'{a}rio class of metrics}
\label{Einstein3D}

In this subsection, we briefly introduce the 3+1 form of the Einstein equations (for more detailed representations see, e.g., \cite{Eric}, \cite{BMR}, \cite{BBV}. The evolution equations in a flow-orthogonal setting, \textbf{R1}, where the extrinsic curvature is related to the expansion tensor by $\CK_{ij} = - \Theta_{ij}$, are:
\begin{align}
	\partial_t  h_{ij} 
    &= 2 N \Theta_{ij} + N_{i || j} + N_{j || i} \ , \label{Einsteinevolution1} \\
	\partial_t \tensor{\Theta}{^i_j} 
    &= - N \Big( \CR^i_{\ j} + \Theta \tensor{\Theta}{^i_j} + 4 \pi G \big[ 
            \left( 3p - \epsilon \right) \tensor{\delta}{^i_j} - 2 \tensor{p}{^i_j} \big]
		- \Lambda \tensor{\delta}{^i_j}\Big) \nn \\
    & + \tensor{N}{^{|| i}_{|| j}} + N^k \tensor{\Theta}{^i_{j || k}} +
    \tensor{\Theta}{^i_k} \tensor{N}{^k_{|| j}} - \tensor{N}{^i_{ || k}}\tensor{\Theta}{^k_j}\ , \label{Einsteinevolution2}
\end{align}
subjected to the energy and momentum constraints:
\begin{align}
    \CR + \Theta^2 - \Theta^{ij} \Theta_{ij} &= 16 \pi G \epsilon + 2 \Lambda \ . \label{energyconstraint} \\
    \tensor{\Theta}{^j_{i || j}} - \Theta_{|| i} &= - 8 \pi G q_i \ . \label{momentumconstraints} 
\end{align}
Here, the notation ${}_{||}$ stands for the $3$-covariant derivative associated with $\mathbf{h}$, $\CR_{ij}$ are the components of the $3$-Ricci tensor and $\Theta_{ij} =  -\tensor{h}{^\alpha_i} \tensor{h}{^\beta_j}  n_{\alpha;\beta}$
the components of the expansion tensor. 

For the Nat\'ario class of metrics we restrict these equations by setting $N=1$ and $\bm N = - \bm V$, \textbf{R2},
and by assuming flat space sections, \textbf{R3}.
\subsection{\textbf{R}-motion in global inertial coordinates}

Since space sections are flat, we are entitled to choose the coordinate representation in terms of globally inertial, nonrotating coordinates, as in most of the literature on warp drive spacetimes. 

With the aim to make contact with a correspondence strategy between Newton's and Einstein's gravitational theories \cite{Bcorrespondence} that will be exploited throughout this work, we focus the presentation of the basic equations on the \textit{coordinate acceleration vector} and its gradient:
\begin{equation}\label{coordinate_acceleration}
    \bm A = A^i \bm\p_{x^i} \quad , \quad A^i(t, x^j) := \frac{\rd}{\rd t} V^i \quad , \quad \bm \p \bm A := A_{i,j} \diffb x^i \otimes \diffb x^j \ , 
\end{equation}
where 
\begin{equation}\label{coordinate_velocity}
    \frac{\rd}{\rd t} := \p_t + V^j \p_j
\quad , \quad A^i(t, x^j) = \p_t V^i + V^j \p_j V^i
\quad , \quad A_{i,j} = \frac{\rd}{\rd t} V_{i,j} + V_{i,k} \tensor{V}{^k_{,j}}
\ .
\end{equation}
We decompose the \textit{coordinate velocity gradient} $\bm \p \bm V := V_{i,j} \diffb x^i \otimes \diffb x^j$ into the covariant expansion tensor ${\bm \Theta} = \Theta_{i j} \diffb x^i \otimes \diffb x^j$ and the coordinate vorticity tensor $\boldsymbol{\Omega}$, $V_{i,j} = V_{(i,j)} +V_{[i,j]} = \Theta_{ij} + \Omega_{ij}$.   
We define the vorticity vector in terms of the \textit{angular velocity vector}:
\begin{equation}
\label{angularvelocity}
    \bm\Omega : = \frac{1}{2}\bm\p\bm\times\bm V \ ;
\end{equation}
its components $\Omega^i$ and the coordinate vorticity scalar $\Omega^2$ are related as follows:
\begin{equation}\label{vorticity_def}
    \Omega^i = - \frac{1}{2} \epsilon^{ijk} \Omega_{jk} 
\,,\quad \Omega_{ij} = - \epsilon_{ijk} \Omega^k
\,,\quad 
    {\bm\Omega}^2 = \Omega^i\Omega_i = \frac{1}{2} \Omega_{ij}\Omega^{ij}=:\Omega^2 
\, .
\end{equation}
The principal scalar invariants of the velocity gradient can be related to 
the principal scalar invariants of the expansion tensor as follows
\cite{ehlersbuchert}:\footnote{\textit{Erratum}: The last line in \cite[Eq.~(8c)]{ehlersbuchert} should read: 
$\III (\bm {\p V}) = \frac{1}{27} \Theta^{3} + \frac{1}{3} \Theta
    \left(\omega^2 - \sigma^{2}\right)+ \frac{1}{3}\sigma^i_{\ j} \sigma^j_{\ k} \sigma^k_{\ i} - 
    \omega^i \omega^j \sigma_{ij}$.}
\begin{align}
\label{relationtoOmega1}
\I (\bm{\p V}) &= V^k_{\ ,k} = \I ({\bm\Theta}) 
\ ,
\\
\label{relationtoOmega2}
2\II (\bm {\p V}) &= (V^k_{\ ,k})^2 - V^k_{\ ,\ell}V^\ell_{\ ,k} = 2\II ({\bm\Theta}) + 2\Omega^2 \ 
, 
\\
\label{relationtoOmega3}
    3\III (\bm {\p V}) &= \frac{1}{2}\left( (V^k_{\ ,k})^3 + 2 V^i_{\ ,j}V^j_{\ ,k}V^k_{\ ,i} 
    - 3 V^k_{\ ,k} V^i_{\ ,j}V^j_{\ ,i} \right)
    = 3\III ({\bm\Theta}) + \Theta \Omega^2 - 3 \Omega^i \Omega^j \Sigma_{ij} 
\ ,
\end{align}
where the expansion tensor alone is related to the rate of expansion $\Theta$ and the rate of shear $\Sigma^2:=(1/2) \Sigma_{ij}\Sigma^{ij} = (1/2) \sigma_{ij}\sigma^{ij}$ as follows \cite{ehlersbuchert}:
\begin{equation}
    {\rm I}(\bm\Theta) = \Theta \quad , \quad
    {\rm II}(\bm\Theta) = \frac{1}{3} \Theta^2 - \Sigma^2  \quad , \quad 
    {\rm III}(\bm\Theta) = \frac{1}{27} \Theta^3 + \frac{1}{3} \Sigma^i_{\ j} \Sigma^j_{\ k} \Sigma^k_{\ i} - \frac{1}{3}\Theta\Sigma^2 \ .
\end{equation}
They obey the following hierarchy of evolution equations \cite{GBC}:
\begin{eqnarray}
\label{Iev}
    &&\frac{\rd}{\rd t} \I ({\bm \Theta}) 
    = - \I^2 ({\bm \Theta}) + 2 \II({\bm \Theta}) - 4 \pi G  \left( \epsilon + 3p \right) + \Lambda\ ;
\\
\label{IIev}
    &&\frac{\rd}{\rd t} \II ({\bm \Theta}) 
    = -8 \pi G (\epsilon +p) \I({\bm \Theta}) -8 \pi G \pi^i_{\ j}\sigma^j_{\ i}\ ;
\\
\label{IIIev}
    &&\frac{\rd}{\rd t} \III({\bm \Theta}) 
    = - 3 \I ({\bm \Theta})\cdot \III({\bm \Theta}) +  \II({\bm \Theta}) \left[ 4 \pi G \left( \epsilon - p \right) + \Lambda \right] \nonumber \\
   &&\qquad\qquad\quad\ \
 - 8 \pi G \tensor{\pi}{^i_j} \left[ \tensor{\sigma}{^j_k} \tensor{\sigma}{^k_i}
 - \frac{1}{3} \I ({\bm \Theta})  \tensor{\sigma}{^j_i} \right] \ .
\end{eqnarray}
The invariants of the full velocity gradient are divergences of vector fields \cite{buchert94,ehlersbuchert}:
\begin{align}
    \label{divI}
    {\rm I}(\bm\p {\bm V}) & = \bm\p \cdot \bm V\ ,\\    
    \label{divII}
    2{\rm II}(\bm\p {\bm V})& = \bm\p \cdot [\, \bm V (\bm\p \cdot \bm V)
    - (\bm V \cdot \bm\p) \bm V \,] \ ,\\
    \label{divIII}
    3{\rm III}(\bm\p {\bm V})& =\bm\p \cdot [\,
    \frac{1}{2}  \bm\p \cdot [\bm V (\bm\p \cdot \bm V) - (\bm V \cdot \bm\p) \bm V] \bm V 
    - [\bm V (\bm\p \cdot \bm V) - (\bm V \cdot \bm\p) \bm V)]\cdot \bm\p \bm V \,]\ .
\end{align}
From here we can replace the expression for the velocity gradient in \eqref{coordinate_velocity}:
\begin{equation}
    A^i_{\ ,j} = \frac{\rd}{\rd t} \Theta^i_{\ j} + \frac{\rd}{\rd t} \Omega^i_{\ j} + (\Theta^i_{\ k}+\Omega^i_{\ k})(\tensor{\Theta}{^k_j}+\tensor{\Omega}{^k_j})
\,.
\end{equation}
We shall later refer to the trace of $A^i_{\ ,j} $ and its antisymmetric part, resulting in kinematical identities:
\begin{align}
\label{partsofgradA}
    \tensor{A}{^k_{,k}} &= 
    \frac{\rd}{\rd t} \Theta + (\tensor{V}{^k_{,\ell}})(\tensor{V}{^\ell_{,k}}) 
= \frac{\rd}{\rd t} {\rm I}(\bm\Theta) + {\rm I}(\bm\Theta)^2 - 2 {\rm II}(\bm\Theta) - 2\Omega^2
\\[10pt]
    &= \frac{\rd}{\rd t} {\rm I}(\bm\p \bm V) + {\rm I} (\bm\p \bm V)^2 - 2 {\rm II}(\bm\p \bm V)
    \nonumber
\,,\\
    A_{[i,j]} &=
    \frac{\rd}{\rd t} \Omega_{ij} + 2  \Theta_{k[i} \Omega^{k}_{\, j]}
\,.
\end{align}
The last equation is equivalent to the Kelvin-Helmholtz transport equation for the coordinate vorticity vector or equivalently the angular velocity vector \eqref{angularvelocity}: 
\begin{equation}
\label{conservation_equation_vorticity}
    \frac{1}{2}\bm\p \times \bm A = \frac{\rd}{\rd t} \bm\Omega + \bm\Omega(\bm\p\cdot \bm V)-(\bm\Omega\cdot\bm\p) \bm V \ .
\end{equation}

We now provide the constraining relations between the coordinate vorticity and coordinate acceleration that follow from the 3+1 Einstein equations for \textbf{R}-Motion.\\

\noindent
\textbf{Lemma 1. (Constraints on coordinate acceleration from Einstein's equations)}

\textit{The coordinate vorticity and coordinate acceleration are related, for \textbf{R}-Motion in global inertial coordinates, \textit{via} the stress-energy sources by the following field equations:
\begin{equation}
\label{divA}
A^k_{\ , k}  = - 2\Omega^2 + \Lambda - 4\pi G (\epsilon + 3p) \ ,
\end{equation}
\begin{equation}
\label{rotA}
    A_{[i,j]} = 
    \frac{\mathrm{d}}{\mathrm{d}t} \Omega_{ij} 
    +\frac{2}{3} \Theta \Omega_{ij} + 2\sigma_{k[i} \tensor{\Omega}{^k_{j]}}  
\ .
\end{equation}
The trace-free symmetric part links its components to the trace-free anisotropic stresses:
\begin{align}
\label{tidalA}
A_{(i,j)} - \frac{1}{3} A^k_{\ k}\delta_{ij} &=
\tensor{\Theta}{_i_k} \tensor{\Theta}{^k_j} - \Theta \tensor{\Theta}{_i_j} - \frac{1}{3}
    \left( \Theta_{\ell k}\Theta^{k\ell} - \Theta^2 \right)\delta_{ij} \nn \\ 
    &+ \tensor{\Omega}{_i_k}\tensor{\Theta}{^k_j} - \tensor{\Theta}{_i_k} \tensor{\Omega}{^k_j} 
    + \tensor{\Omega}{_i_k}\Omega{^k}{_j} + \frac{2}{3}\Omega^2 \delta_{ij} + 8\pi G \pi_{ij} \ .
\end{align}
} 
The proof of \textbf{Lemma 1} can be found in Appendix \ref{ProofofLemma1}.

We are now in the position to give the full system of Einstein equations for \textbf{R}-Motion in global inertial coordinates.\\

\noindent
\textbf{Theorem 1. (Einstein equations for R-Motion)}

\textit{For \textbf{R}-Motion in global inertial coordinates $\bm x$, the Einstein equations admit vectorial coordinate velocity and acceleration fields,
$\bm V (t,\bm x)$ and $\bm A (t,\bm x)= \rd V / \rd t$, where the coordinate acceleration obeys the field equations,
\begin{equation}
\label{divrotE}
\bm\p \cdot \bm A = - 2\Omega^2 +\Lambda - 4\pi G (\epsilon + 3p) \quad , \quad
\bm\p \times \bm A = 2\left( \frac{\rd}{\rd t} \bm\Omega + \bm\Omega(\bm\p\cdot \bm V)-(\bm\Omega\cdot\bm\p) \bm V \right)\ ,
\end{equation}
\begin{equation}
\label{trace-freeE}
A_{(i,j)} - \frac{1}{3}A^k_{\; ,k}\delta_{ij} = V_{i,k}V^k_{\ ,j} - V^k_{\; ,k}V_{i,j} + \frac{2}{3}{\rm II}(\bm\p \bm V)\delta_{ij} - 2 V_{i,k}\Omega^k_{\ j} + V^k_{\; ,k}\Omega_{ij}
+ 2 \Omega_{ik}\Omega^k_{\ j} + 8\pi G \pi_{ij} \ ,
\end{equation}
subjected to the energy and momentum constraints,
\begin{equation}
\label{constraintsE}
    8\pi G \epsilon + \Lambda = \II ({\bm \p \bm V}) - \Omega^2 \quad , \quad 
    8\pi G {\bm q} = \bm\p \times \bm\Omega \ ,
\end{equation}
and obeying the energy and momentum conservation laws: 
\begin{equation}
\label{conservationR}
\frac{\rd\epsilon}{\rd t} + \I ({\bm \Theta}) (\epsilon + p) = - \pi^i_{\ j}\sigma^{j}_{\ i}  \quad , \quad
\frac{\rd q^i}{\rd t} + \frac{4}{3}\Theta q^i + \tensor{\sigma}{^i_j} q^j - \tensor{\Omega}{^i_j}q^j + \p_j \tensor{p}{^j^i} = 0 \ .
\end{equation}
Counting the equations and variables, we can choose the set of $9$ dependent variables $\{ \bm V,\bm A, \bm \Omega \}$, and the $10$ source functions $\{\epsilon, p, \pi^i_{\ j}, q^i \}$,
which make together $19$ functions that are determined through $9$ defining equations, $\bm A = \rd \bm V / \rd t$, $\bm \Omega = (1/2)\bm\p \times \bm V$, $\bm q = (1/8\pi G)\bm\p \times \bm\Omega$ (where the momentum constraints are here listed as the last defining equation), and the remaining $7$ independent equations (trace and trace-free symmetric part of the gradient of $\bm A$, and the energy constraint), which makes overall $16$ equations for $19$ functions.\footnote{
Notice that the Einstein evolution equations are symmetric; the antisymmetric part of the gradient of $\bm A$ is an identity and therefore not an independent equation.} 
(together with a free choice of $\Lambda$ that we do not count here). 
Therefore, three functions have to be given to close the system that corresponds to the three undetermined functions $\bm V$ in the metric. The conservation laws hold by construction and do not further constrain the system. 
}

\newpage

\noindent
\textbf{Remark 1} (Warp drives with positive energy density)

\textit{The energy constraint for \textbf{R}-Motion in \eqref{constraintsE} defines warp drives with positive energy density in terms of restrictions on the kinematics and $\Lambda$, seen by Eulerian observers (see, however, \cite{Santiago}, \cite{BBV}):
\begin{equation}
\label{Remark1}
0 \, \le \, \epsilon = \frac{1}{8\pi G} ({\rm II}(\bm\Theta) - \Lambda) \ , \ \text{i.e.} \ \ 
\frac{1}{3}\Theta^2 - \Sigma^2 - \Lambda \, \ge \, 0 \ .
\end{equation}
}

\noindent
\textbf{Proof of Theorem 1}

The field equations \eqref{divrotE} for $\bm A$ follow from Lemma 1, the vector expression for the antisymmetric part of $\bm\p\bm A$ via the identity $\Omega_{ij} = -\epsilon_{ijk}\Omega^k$, \eqref{conservation_equation_vorticity}. $\Box$

To show the trace-free symmetric field equation, we start from \eqref{trace-freeA} of the proof of Lemma 1, 
and insert the decomposition $V^i_{\ ,j} = \Theta^i_{\ j} + \Omega^i_{\ j}$, using also \eqref{relationtoOmega1} and \eqref{relationtoOmega2}, to obtain \eqref{trace-freeE}. $\Box$

Assuming $\CR = 0$ and inserting $\Theta^i_{\ j} = V^i_{\ ,j} - \Omega^i_{\ j}$ into the energy constraint \eqref{energyconstraint} and using \eqref{relationtoOmega1}, \eqref{relationtoOmega2}, we obtain the first equation of \eqref{constraintsE}. $\Box$

The second equation of \eqref{constraintsE} is obtained by inserting 
$\Theta^i_{\ j} = V^i_{\ ,j} - \Omega^i_{ \ j}$ into the momentum constraints \eqref{momentumconstraints},
and noting that the covariant spatial derivative reduces to the partial derivative, to obtain:
$$
    \tensor{V}{^k_{\ ,i k}} -\tensor{\Omega}{^k_{i , k}} - V^k_{\ ,k i} 
    = -\tensor{\Omega}{^k_{i , k}} = - 8 \pi G q_i \quad, \ 8\pi G \bm q = \bm\p \times \bm \Omega \quad, \ \bm\p \cdot \bm q = 0 \ ,
$$
with the identities $\Omega^{ki} = -\epsilon^{ki}_{\ \; \ell}\Omega^\ell$, hence
$\Omega^{ki}_{\ \; ,k} = - \epsilon^{ki}_{\ \; \ell}\Omega^\ell_{\ ,k} =  \epsilon^{ik}_{ \ \; \ell}\Omega^\ell_{\ ,k}= (\bm\p \times \bm\Omega)^i$. $\Box$\\

For the conservation laws we note the following: in view of the vanishing of the covariant acceleration, $a_\mu = 0$, the covariant vorticity, $\omega_\mnu = 0$, the divergence-free nature of the momentum flux density, $q^\mu_{\ ;\mu}=0$,
and the fact that $b_{\mu\nu} = h_{\mu\nu}$, the term $b_{\mu \alpha} p^{\alpha\beta}_{\ \;;\beta}$ leaves only spatial components and the covariant spatial derivative reduces to a partial derivative in global inertial coordinates, $\tensor{p}{^j_{i||j}} = \p_j \tensor{p}{^j_i}$. 

We further note that the covariant time derivative of $\epsilon$ appearing in \eqref{energy_conservation_law} is identical to the Lagrangian time derivative, $\dot\epsilon = \frac{\rd\epsilon}{\rd t}$, while the covariant time derivative for vectors appearing in \eqref{momentum_conservation_law} can be expressed in terms of the Lagrangian time derivative of $\bm q$ as follows (with $u^\alpha = (1, V^i)^T$, $q^0 = 0$, and $^4\Gamma^i_{\ 0j} = -V^i V^k \Theta_{jk} - \Omega^i_{\ j}$, $^4\Gamma^i_{\ jk} = V^i \Theta_{jk}$):\footnote{We here use the vector form unlike the equation in \cite{BBV} that is written for the co-vector (as in Equation (6.21) of \cite{Eric}). The second equation of \eqref{conservationR} can be inferred from York's Equation (41) of his seminal work \cite{York}.}
\begin{eqnarray}
\label{covq}
{\dot q}^i = u^\alpha q^i_{\ ;\alpha} = u^\alpha (q^i_{\ ,\alpha} + ^4\Gamma^i_{\ \beta\alpha}q^\beta) = 
u^\alpha q^i_{\ ,\alpha} + ^4\Gamma^i_{\ \alpha\beta}u^\alpha q^\beta = \frac{\partial}{\partial t} q^i + V^k q^i_{\ , k} + ^4\Gamma^i_{\ 0j}q^j + ^4\Gamma^i_{\ jk}V^j q^k \nonumber\\
=\frac{\rd q^i}{\rd t} - \Omega^i_{\ j} q^j = \frac{\rd q^i}{\rd t} - ({\bm\Omega} \times{\bm q})^i \ , \ 
\Omega^i_{\ j} = - \varepsilon^i_{\ jk}\Omega^k \ , \
({\bm q}\times\bm\Omega)^i = - \varepsilon^i_{\ jk}\Omega^j q^k \, .\qquad 
\end{eqnarray}
\\
Therefore, the conservation laws
\eqref{energy_conservation_law} and \eqref{momentum_conservation_law} 
attain the form of \eqref{conservationR}. $\Box$ \\

\noindent
\textbf{Remark 2} (Covariant time-derivative and non-inertial accelerations)

\textit{
In line with the calculation of the covariant time-derivative of $\bm q$, \eqref{covq}, we have the following relation for any 3-vector $\bm F$:
\begin{equation}
\dot{\bm F} = \frac{\rd}{\rd t}{\bm F} - {\bm\Omega}\times{\bm F} \ .
\end{equation}
In particular, we have $\dot {\bm x} = \frac{\rd}{\rd t} {\bm x} - {\bm\Omega}\times{\bf x}$, i.e. the overdot can here be interpreted as the total time-derivative in a non-inertial rotating frame $\mathcal R$, $\dot{\bm x} = :{\bm V}_{\mathcal R}$, so that ${\bm V} = {\bm V}_{\mathcal R} + {\bm\Omega} \times {\bm x}$ is the usual transformation to the coordinate velocity in a rotating frame in which the acceleration includes, besides the translational acceleration $\bm A$, Coriolis, centrifugal and Euler accelerations: ${\bm A}_{\mathcal R} = \ddot{\bm x}={\bm A} - 2{\bm\Omega}\times {\bm V}_{\mathcal R} - {\bm\Omega}\times({\bm\Omega}\times{\bm x}) - \dot{\bm\Omega}\times{\bm x}$, with $\dot{\bm\Omega} = \frac{\rd}{\rd t}{\bm\Omega} + {\bm\Omega}\times{\bm\Omega} = \frac{\rd}{\rd t}{\bm\Omega}$.
\label{covA}}\\

\newpage

\noindent
\textbf{Corollary 1} (Irrotational \textbf{R}-motion)

\textit{Setting $\bm\Omega = \bm 0$, a restriction sometimes considered in the literature \cite{Lenz1,Lenz2,Lavinia,Rodal}, the field equations of \textbf{Theorem 1} reduce to the set: 
\begin{equation}
\begin{aligned}
&\bm\p \cdot \bm A = \Lambda - 4\pi G (\epsilon + 3p) \quad , \quad
\bm\p \times \bm A = \bm 0 \ ,\\
&A_{(i,j)} - \frac{1}{3} A^k_{\; ,k}\delta_{ij} = V_{(i,k}V^k_{\ ,j)} - V^k_{\; ,k}V_{(i,j)} 
+ \frac{2}{3}{\rm II}(\bm\p \bm V)\delta_{ij} + 8\pi G \pi_{ij} \ ,
\end{aligned}
\end{equation}
subjected to the energy and momentum constraints,
\begin{equation}
\label{Cconstraints}
    8\pi G \epsilon + \Lambda = \II ({\bm \Theta}) \quad , \quad {\bm q} = \bm 0  \ ,
\end{equation}
and obeying the energy and momentum conservation laws,
\begin{equation}
\frac{\rd\epsilon}{\rd t} + \I ({\bm \Theta}) (\epsilon + p) = - \pi^i_{\ j}\sigma^{j}_{\ i}  \quad , \quad
\p_j \tensor{p}{^j^i} = 0 \ .
\end{equation}
A subcase is furnished by the assumption of a perfect fluid source, already defined by $\bm q = \bm 0$, but also $\pi^i_{\ j} = 0$.\footnote{\label{potentialomega}The physical class corresponding to perfect fluid sources is not larger, since ${\bf q}={\bf 0}$ implies that the coordinate vorticity is a gradient field, ${\bm\Omega} = \bm\nabla Z$, and since $\bm\nabla \cdot {\bm\Omega} = 0$, we also have that $Z$ is a harmonic, $\Delta Z = 0$, which can be set to vanish for suitable boundary conditions.}
}

\subsection{One-component coordinate velocity}

With this background established, we are going to focus on a subcase that we shall deal with in detail. We consider a single-component velocity field, $\bm V(t,x,y,z) =: V(t,x,y,z){\bm e}_x$.
We here choose to provide explicit component expressions to provide a comprehensive treatment of \textbf{R}-motion for one-component coordinate velocity fields. We shall discuss the consequences and solutions.

For one-component coordinate velocity fields we have $\bm V = :(V,0,0)$ and $\bm A = : (A,0,0)$.
If $\bm V(t,x)$ only depends on one spatial coordinate we do not have coordinate vorticity (the case of Corrolary 1), on the contrary if $\bm V(t,x,y,z)$ we can have it with ${\bm V}\cdot{\bm\Omega} = 0$.
Component expressions for kinematical quantities can be found in Appendix \ref{app:alcubierrekinematics}.\\

\noindent
\textbf{Remark 3} (One-component coordinate velocity warp drives with positive energy density)

\textit{The principal scalar invariants of the coordinate velocity gradient $\bm\p \bm V$ are divergences of vector fields; these vector fields vanish for $\rm II (\bm\p\bm V)$ and $\rm III (\bm\p \bm V)$ for 
the restriction $V^i = V \delta^i_{\; x}$ 
as a result of a straightforward calculation of \eqref{relationtoOmega2} and \eqref{relationtoOmega3}. 
In view of \textbf{Remark 1}, \eqref{Remark1}, positivity of the energy density is then only possible for a sufficiently large and negative cosmological constant: 
\begin{equation}
\label{Remark3}
8\pi G \epsilon = {\rm II}(\bm \Theta) - \Lambda = {\rm II}(\bm\p \bm V) - \Omega^2 - \Lambda = - (\Omega^2 + \Lambda)\, \ge \, 0 \ ; \quad
{\rm II} (\bm\p\bm V) = 0 \ .
\end{equation}
}

\noindent
\textbf{Remark 4} (Evolution of principal scalar invariants for one-component coordinate velocity)

\textit{For one-component velocity fields, the hierarchy of evolution equations \eqref{Iev}, \eqref{IIev} and \eqref{IIIev} is truncated at the first equation, since $\II (\bm\partial \bm V) = 0$, $\III (\bm\partial \bm V) = 0$ (following easiest from \eqref{divII} and \eqref{divIII}), and we are left with the truncated Raychaudhuri equation \eqref{Iev}:
\begin{equation}
\label{RAY1D}
\frac{\rd}{\rd t} \I ({\bm \Theta}) 
    = - \I^2 ({\bm \Theta})  - 4 \pi G  \left( \epsilon + 3p \right) + \Lambda\quad,\quad \I (\bm\Theta) = \I (\bm\partial\bm V) \ .
\end{equation}
}

We now consider the restricted case of a one-component coordinate velocity field.

\newpage

\noindent
\textbf{Corollary 2} (\textbf{R}-motion for one-component coordinate velocities)

\textit{
Any solution to the Einstein equation in a 3+1 formalism of general relativity, with lapse $N=1$, shift vector $\bm N=(-V(t,x,y,z), 0, 0)$, where the single coordinate velocity component $V(t,x,y,z)$ depends on three spatial coordinates and a temporal coordinate, and for an Euclidean spatial metric $h_{ij}=diag(1,1,1)$, corresponds to a spacetime with the following properties.}

\textit{The metric can be written in global inertial (nonrotating) coordinates:
\begin{align}
\label{metricTheorem1}
    \rd s^2 &= -(1-V^2(t,x,y,z))\rd t^2 + 2 V(t,x,y,z) \rd t \rd x + \rd x^2 +\rd y^2 + \rd z^2 
\,,
\end{align}
where $V(t,x,y,z)$ obeys $\p_t V + V \p_x V = A(t,x,y,z)$, the energy density and the momentum flux density attain the following forms:
\begin{equation}\label{energy_density_and_flux}
    \epsilon = -\frac{1}{8 \pi G} \left( \Omega^2 + \Lambda \right)
\,,
\quad
    \bm q = \frac{1}{8 \pi G}(\bm\p \times \bm\Omega)
\, .
\end{equation}
Energy and momentum conservation imply:
\begin{equation}
\frac{\rd\epsilon}{\rd t} + \Theta (\epsilon + p) = - \pi^i_{\ j}\sigma^{j}_{\ i}  \quad , \quad
     \frac{\rd q^i}{\rd t} + \frac{4}{3} \Theta q^i + \tensor{\sigma}{^i_j}q^j - \tensor{\Omega}{^i_j}q^j + \p_j p^{ji} = 0 
\,,
\end{equation}
(which can be simplified according to Appendix~\ref{elementsCorollary2}).\\
The only non-trivial equations that constrain the coordinate acceleration gradient ${\bm\partial}A$ read:
\begin{equation}
\begin{aligned}
\p_x A &= \Lambda - 4\pi G (\epsilon + 3p) - 2 \Omega^2
    = \frac{3}{2}\left(\Lambda -\Omega^2 - 8\pi G p \right) 
    = 3 \Omega^2 + 12\pi G \tensor{\pi}{^x_x}\,;\\
\p_y A &= -\p_x V \p_y V +  16 \pi G \pi^x_{\ y} \ ;\\
\p_z A &= - \p_x V \p_z V +  16 \pi G \pi^x_{\ z} 
\ ,
\end{aligned}
\end{equation}
the first of which implies the equation of state relation through elimination of $\Omega^2$ in favour of $\epsilon$:
\begin{equation}
\label{EOS1D}
p + \pi^x_{\ x} = 3\epsilon + \frac{\Lambda}{2\pi G} \ .
\end{equation}
The divergence and the curl of $A$ are both sourced by the components of the trace-free symmetric stresses:
in the first line, the first equation follows from the divergence, the second by inserting the energy constraint, and the third from the trace-free symmetric part, as the derivatives in the second and third line.\\ 
Counting equations and variables, the counting of \textbf{Theorem 1} here reduces the $9$ dependent variables to $4$, 
the $9$ defining equations to $6$, and the $10$ source functions to $8$; the remaining equations for trace and trace-free symmetric part of $\bm\p A$ together with the energy constraint provide $5$ additional non-trival equations, so that overall we have $11$ equations that determine $12$ functions, leaving one function free that corresponds to the single free function $V$ in the metric.\\
}

\noindent
The explicit proof of \textbf{Corollary 2} can be found in Appendix~\ref{elementsCorollary2}., where we also list further useful coordinate expressions.

As a further Corollary we restrict the above to irrotational flows.\\

\noindent
\textbf{Corollary 3} (Irrotational \textbf{R}-motion for one-component coordinate velocity)

\textit{A one-component coordinate velocity with zero coordinate vorticity, $\bm\Omega = {\bm 0}$, only depends on one independent spatial coordinate, $V (t,x)$. \textbf{Corollary 2} reduces to the following equations:
\begin{eqnarray}\label{Omega01D}
    \epsilon &=& -\frac{\Lambda}{8 \pi G} \quad , \quad  
    \bm q = {\bm 0}\quad, \quad 
\p_x V \left(p + \pi^x_{\ x} - \frac{\Lambda}{8\pi G} \right) = 0 \ ; \nonumber\\
\p_x A &=& \Lambda - 4\pi G (\epsilon + 3p)
    = \frac{3}{2}\left(\Lambda - 8\pi G p \right) = 12\pi G \tensor{\pi}{^x_x}\ ; \ \pi^y_{\ y} = \pi^z_{\ z} = -\frac{1}{2}\pi^x_{\ x} \ ;\nonumber \\
\p_y A &=& 16 \pi G \pi^x_{\ y} = 0 \quad;\quad
\p_z A = 16 \pi G \pi^x_{\ z} = 0 \quad ; \quad  \p_y p = - \p_z p \ .
\end{eqnarray}
}
\noindent
\textbf{Proof of Corollary 3}

Putting $\bm\Omega = \bm 0$ in \textbf{Corollary 2} implies $\epsilon = -\frac{\Lambda}{8\pi G}$, $\bm q = {\bf 0}$, and $\p_y A = \p_z A = 0$, hence $\pi^x_{\ y} = \pi^x_{\ z} = 0$. The remaining equation determines $\p_x A= (3/2)\Lambda- 12\pi G p = 12\pi G \pi^x_{\ x}$, i.e. $p= \frac{\Lambda}{8\pi G}- \pi^x_{\ x}$, which can be directly obtained from 
the equation of state relation \eqref{EOS1D}. 
The stress tensor components, $p^i_{\ j} = p \delta^i_{\, j} + \pi^i_{ \ j}$, $p^k_{\ k} = 3p$, (\textit{c.f.} Appendix~\ref{elementsCorollary2}) read:
\begin{eqnarray}
p^i_{\ j} &=& diag \left(\frac{\Lambda}{8\pi G}\ ,\  - \frac{\Lambda}{16\pi G}+ \frac{3}{2}p\ ,\  - \frac{\Lambda}{16\pi G} + \frac{3}{2}p\right)  \nonumber \\ 
&=& diag \left( \frac{\Lambda}{8\pi G}\ ,\  \frac{\Lambda}{8\pi G}-  \frac{3}{2}\pi^x_{\ x}\ ,\  \frac{\Lambda}{8\pi G} 
- \frac{3}{2}\pi^x_{\ x}   \right) \ ,
\end{eqnarray}
with null divergence, $\p_j p^{ji} =  0 \ \Leftrightarrow\ (3/2)(\p_y p + \p_z p) = -(3/2)(\p_y \pi^x_{\ x}+\p_z \pi^x_{\ x}) =0$. $\Box$\\

For $\Lambda = 0$, the full energy-momentum tensor \eqref{EMT} reduces to the spatial stress tensor components
$p^i_{\ j} = \frac{3}{2} diag \left(0\ ,\  p\ ,\  p\right) = -\frac{3}{2}diag \left( 0\ ,\  \pi^x_{\ x}\ ,\  \pi^x_{\ x} \right)$. (This also shows that if we assume an irrotational perfect fluid source, $\pi^x_{\ x}=0$, then the spacetime is Minkowski.)

Notice also that an irrotational Alcubierre warp drive does not exist (or is also Minkowski \cite{BBV}), a fact that can easily be seen in the formulas of Appendix~\ref{app:alcubierrekinematics}.\\

\noindent
\textbf{Remark 5} (Solution of the hierarchy of principal scalar invariants for one-component coordinate velocity)

\textit{As noted in \textbf{Remark 4}, the hierarchy of evolution equations of principal scalar invariants is truncted at the first equation for one-component coordinate velocity fields. The remaining equation \eqref{RAY1D} can be cast into an equation for the Jacobian \eqref{jacobian} by virtue of the Jacobi identity, $\frac{\rd}{\rd t} J + J\, \I = 0$, to yield:
\begin{equation}
\frac{\rd^2}{\rd t^2} J - J \left(\Lambda - 4\pi G (\epsilon + 3p)\right) = 0 \ ,
\end{equation}
which reduces for vanishing sources to $\frac{\rd^2}{\rd t^2} J = 0$, compatible with the Jacobian for inertial motion (and for non-vanishing sources discussed in Subsection \ref{illustrationofinertial}).
}\\

We are now going to discuss possible assumptions for the solution to the equations of \textbf{Corollary 2} for 
one-component coordinate velocity fields.

\section{Examples of solutions of Einstein's equations}
\label{IV}

We have the freedom of choice of a single function to close the system of Einstein equations of \textbf{Corollary 2}. 
Two examples are considered in the following, both in the spirit of Synge's G-method. The first determines the solution for $V = V_S (\bm x ,t)$
\textit{a priori} (Eulerian assumption). For this we shall consider Alcubierre's model. 
The second is imposing an assumption on the dynamics along geodesics  for the single free function (Lagrangian assumption).
For this solution we shall consider initial data that correspond to Alcubierre's model for comparison. We also put the matter-based approach into perspective, discussing assumptions on the sources according to the Cauchy problem. 

\subsection{Alcubierre's model as a solution}

An \textit{a priori} given solution for $V_S (t,\bm x)$ is, according to Synge's G-method, a solution. The Alcubierre model is an example, although physically less interesting, since the warp field is forced to stay spherical and its kinematical properties frozen in the coarse of time, besides other problems \cite{BBV}. We are now in the position to derive the necessary sources to support the forcing conditions for the Alcubierre solution. 

In the following, we study the general representation of Alcubierre's warp field and calculate its coordinate acceleration field from the governing equations of \textbf{Corollary 2}. The externally given proper speed of the bubble $v_S(t) = \p_t x_S(t)$ and, more importantly, the window function $W(r_S)$  \eqref{warpshape} can induce coordinate acceleration.

In this general case it is possible to obtain the exact form of the only remaining stress components $\tensor{\pi}{^x_x}$, $\tensor{\pi}{^x_y}$, $\tensor{\pi}{^x_z}$ that support Alcubierre's warp field. From Einstein's equations restricted to the case of \textbf{Corollary 2} we have the relations:
\begin{equation}
\label{stresses}
\begin{aligned}
    \tensor{\pi}{^x_x} &= \frac{1}{12 \pi G} \left[\p_x A_S(t,x^k) -3\Omega^2 \right]
    = \frac{1}{12 \pi G} \left[\p_x \left(\frac{\rd V_S (t,x^k)}{\rd t}\right) -\frac{3}{4}[(\p_y V)^2 + (\p_z V)^2] \right]
\,,
\\[2pt]
    \tensor{\pi}{^x_y} &= \frac{1}{16\pi G} [\p_y A_S(t,x^k)-\p_y V \p_x V] \ \ ,\ \ 
    \tensor{\pi}{^x_z} = \frac{1}{16\pi G} [\p_z A_S(t,x^k) - \p_z V \p_x V]\,.
\end{aligned}
\end{equation}
The components of the spatial gradient of the coordinate acceleration $A_S (t,\bm x)$ enter these equations; we give their form in Appendix~\ref{app:acceleration}. 
We present in figure \ref{acceleration} the derivative with respect to $r_S$ of the coordinate acceleration $A_S$. We can observe that the shape of the warp field derives from the derivative of the window function and that the coordinate acceleration is oriented in $x$-direction, with a positive amplitude at the front against a negative one at the back of the `bubble wall'. Furthermore, we observe that this acceleration encompasses the shape of the `bubble', which is situated between two accelerating fronts that tend to offset each other. This is more visible in a 2D view of the amplitude (right of figure \ref{acceleration}). This helps us understand the dynamics of such a rigid warp field. 
\begin{figure}[H]
        \centering
        \includegraphics[trim={0cm 0cm 0cm 0cm},clip, width=0.85\textwidth]{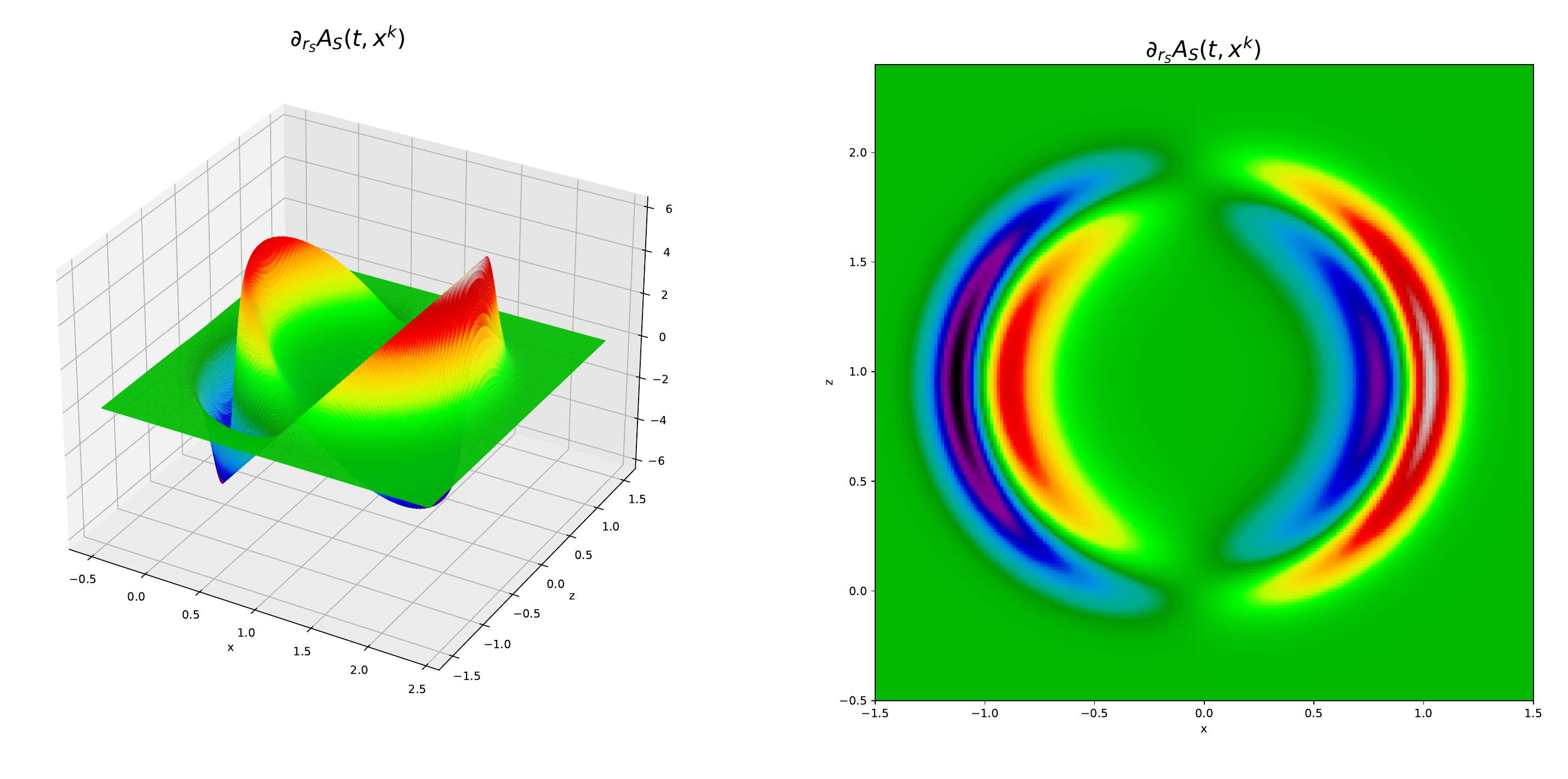}
    \caption{Recalling the representation of the Alcubierre velocity field and its $r_S$-derivative in figure \ref{fig1} in order to compare with the derivative of the coordinate acceleration field with respect to $r_S$. 
    Here we take $v_S = 0,9$, $\sigma = 5$, $R=1$, $t_0 = 0$ (See Appendix \ref{app:acceleration} for the norm of $\p_{r_S} A_S(t,x^k)$).}
    \label{acceleration}
\end{figure}
We now move to an alternative solution by imposing a restriction on the dynamics instead of explicitly giving the velocity model as in the example above. 

\subsection{The case of inertial motion for Alcubierre initial conditions}
\label{illustrationofinertial}

We consider a subcase of \textbf{Corollary 2}, where we specify a motion with vanishing acceleration along geodesics (Lagrangian assumption), $A=0$, henceforth called \textit{inertial motion}. Admissible velocity models will follow from this assumption.
\textbf{Corollary 2} will then specify the admissible stress-energy sources.  

In general, the system of inertial motion is characterized by Euler's equation for $V(t,x,y,z)$:
\begin{equation}
\label{Euler}
    \frac{\partial}{\partial t}{\bm V} + ({\bm V} \cdot \bm\p){\bm V} = {\bm 0} \ ,
\quad
    \frac{\partial}{\partial t} V^i + V^j V^i_{\ ,j} = 0\ .
\end{equation}
In the Lagrange point of view, we follow the trajectories ${\bm x} = {\bm f}({\bm X},t)$ of a fluid parcel, where $\bm X$ labels trajectories of fluid parcels. This transformation is possible since spatial sections are flat. 
The solution of the above coupled nonlinear system of partial differential equations (\ref{Euler}) can be solved in the Lagrangian coordinate system:
\begin{equation}
\label{solEuler1}
    \frac{\rd}{\rd t}{V^i} = \frac{\partial}{\partial t}\Big\vert_{\bm X}V^i  = 0
\,,
\quad 
    \frac{\rd}{\rd t} := \frac{\partial}{\partial t}\Big\vert_{\bm x} + ({\bm v} \cdot \bm\p)
    = \frac{\partial}{\partial t}\Big\vert_{\bm X} 
\,.
\end{equation}
The solution of (\ref{Euler}) is a constant velocity field along trajectories: $V^i(t,X^k)= V_0^i(X^k)$. 
We integrate the previous equation and assume that at $t=t_0$, the position at that moment is $f^i(t_0,X^k)=X^i$, so we obtain the mapping from Lagrangian to Eulerian coordinates,
\begin{equation}\label{solEuler3}
    x^i = f^i(t,X^j) = X^i +V_0^i(X^j)(t-t_0)
\,,
\end{equation}
with $x^i$ the Eulerian position coordinate.
We deduce from this that $X^k = h^k (t,x^j)$, where ${\bm h} := {\bm f}^{-1}$ is required to exist. With the help of the inverse mapping ${\bm h}$ we are then able to express the admissible functions of the velocity in the Eulerian frame:
\begin{equation}\label{solEuler4}
    V^i (t,x^j) = V_0^i (X^k = h^k (x^j,t)) 
\,,
\end{equation}
with $V_0^i (X^k)$ the velocity field components at some time $t_0$, henceforth called \textit{initial time}.

\begin{figure} [H]
    \centering
    \includegraphics[width=0.85\textwidth]{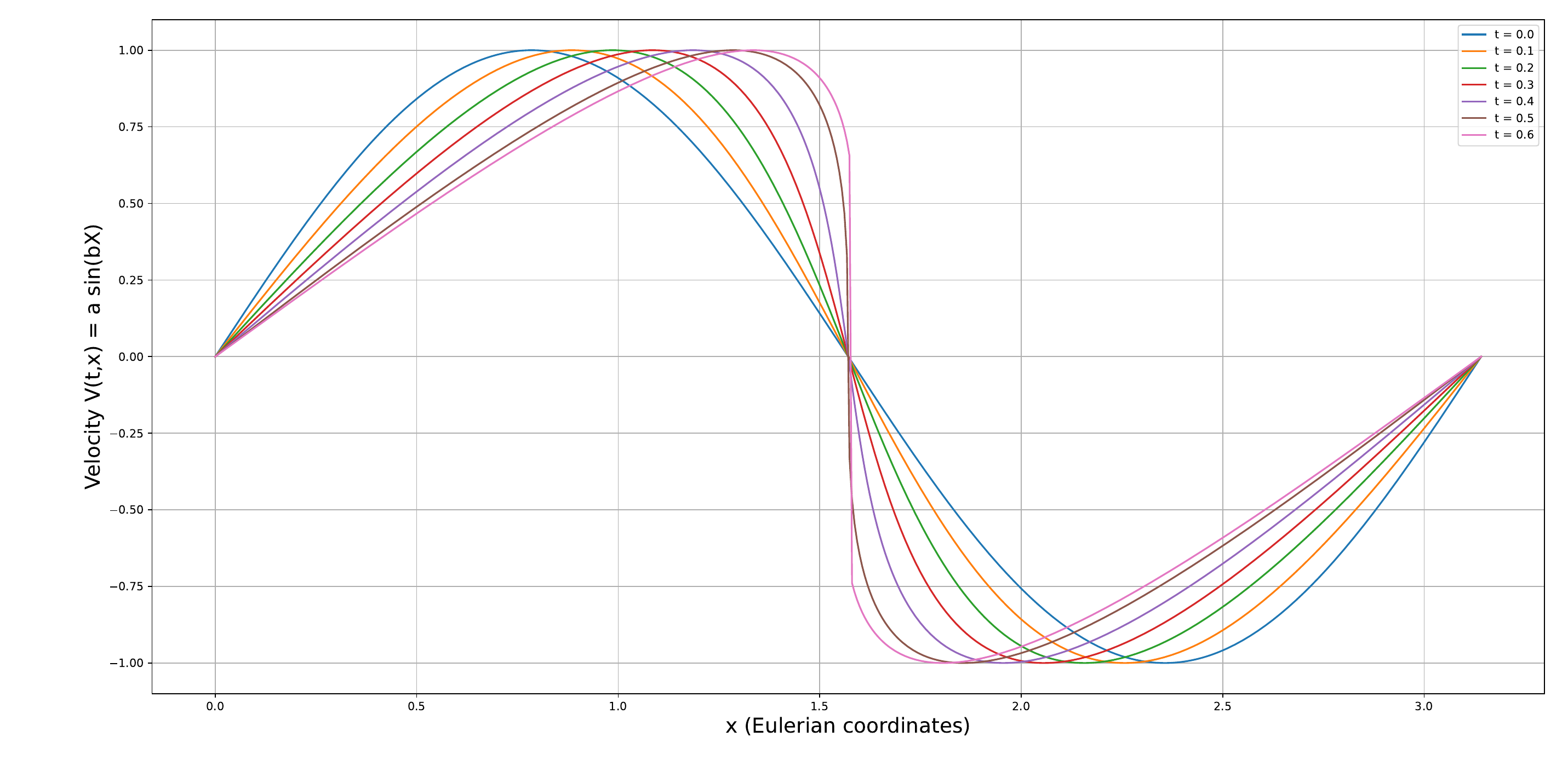}
    \caption{Example of the velocity in Eulerian space, $V(t,x)= a \sin(b h(t,x))$, over the range of times $t= [0, 0.6]$ versus $x$. Here we take $t_0=0$, $a=1$ and $b=2$.}
    \label{V_evolution}
\end{figure}

We remind here the Jacobian of the coordinate transformation,
\begin{equation}\label{jacobian}
    J(t,X^k) = \det\left( \frac{\partial f^i(t,X^k)}{\partial X^j}\right)
    = \det\left(\delta_{ij}+\frac{\partial V_0^i}{\partial X^j}(t-t_0)\right) \, .
\end{equation}
In the present case of a one-component coordinate velocity, we first give a simple example with the 
initial condition: $V_0 (X^k) = a \sin(bX) \  a=const., b=const.$,
we obtain the Eulerian positions and determine $V(t,x)$: 
\begin{equation}
    x = f(t,X) = X + a \sin(bX) (t-t_0)
\Rightarrow
    V(t,x) = a \sin(b h(t,x))
\,.
\end{equation}
For this example we solve the velocity expression in Eulerian coordinates numerically, starting from the initial condition $V_0 (X)$. For that, we must invert the map $x=f(t,X)$ to find $X=h(t,x)$, which we do for this example representing all cases that are not analytically solvable. We use numerical root-finding methods to determine $X$, and thereafter we can compute $V(t,x)$.
We choose arbitrary values for $a$ and $b$ to represent the velocity profiles in figure \ref{V_evolution}. %
\noindent
We compute the trace of $\bm\p \bm V$, the divergence of the velocity field or the expansion rate:
\begin{equation}
    {\rm I} = \p_x V = \frac{\dot J}{J} = \frac{a b \cos(bX)}{1 + a b \cos(bX) (t-t_0)}
\,,
\end{equation}
where $J$ denotes the Jacobian \eqref{jacobian}.
The expansion rate as well as other variables can be treated according to the inversion above.

\begin{figure}[H]
\centering
    \includegraphics[width=0.85\textwidth]{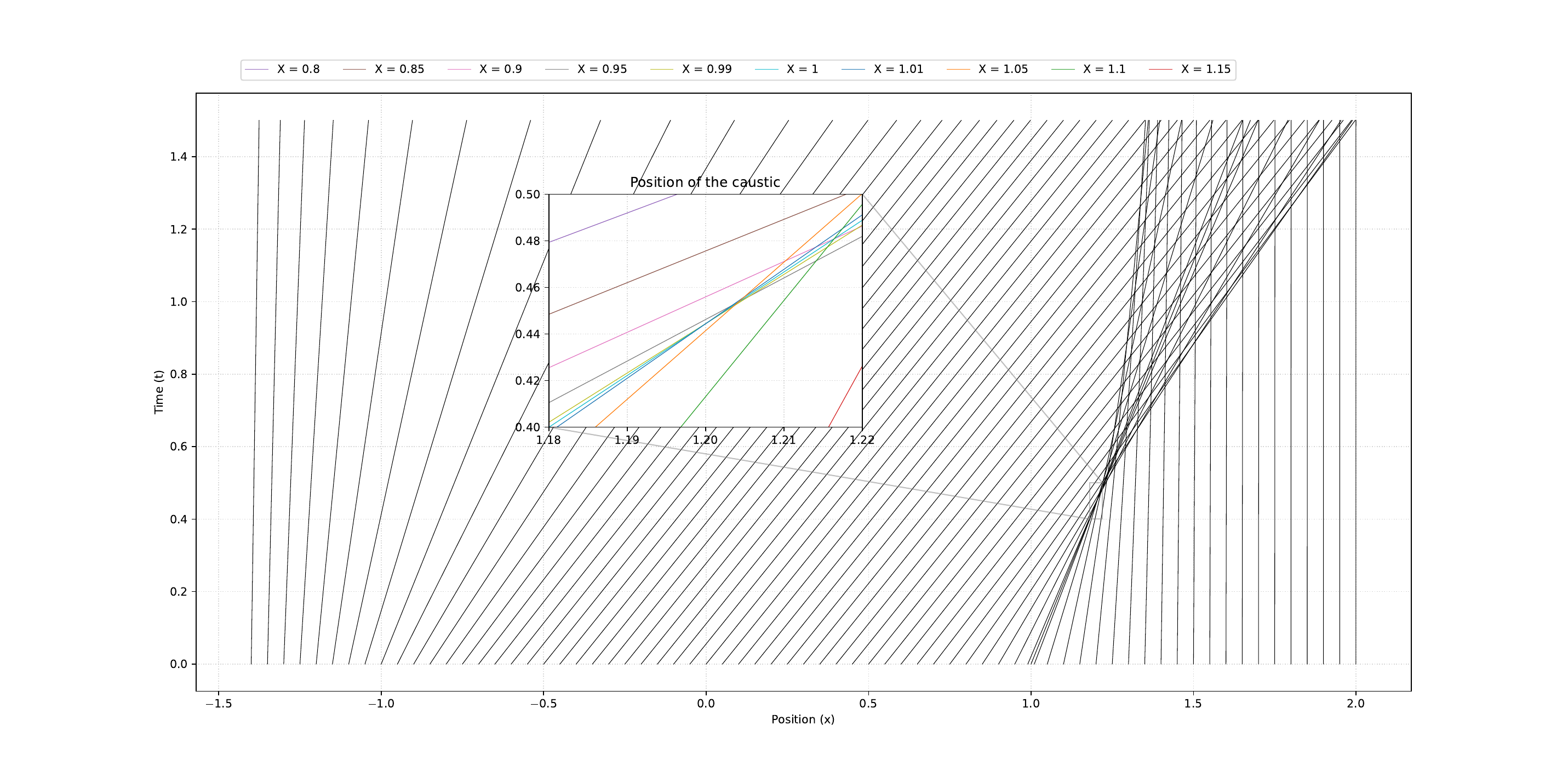}
    \caption{Family of trajectories in Eulerian space, $x = f_S = X + v_S W(r_S ({\bm X},t_0)) (t-t_0)$, with $t_0 =0$, $v_S = 0.9$, over the range of $X=[-1.4, 2.0]$ versus time. A caustic develops at a critical time when two infinitesimally close trajectories cross each other for the first time in Eulerian space.} 
\label{caustic}
\end{figure}

Our second example is still based on the dynamical assumption $A=0$, but uses the Alcubierre velocity profile as initial condition:
\begin{equation}\label{Alcubierre's Lagrange velocity}
    V_S(t,X^k) = V_S(t_0,X^k) = : V_0 (X^k)
    = v_S (t_0) W(r_S(t_0,X^k)) 
\,,
\end{equation}
where the (now Lagrangian) radial distance from the trajectory is:
\begin{equation}
    r_S(t_0, X^k) = [(X-(X_S+v_S(t_0) W(r_S(t_0,X))))^2+Y^2+Z^2)]^{1/2}
\,, 
\end{equation}
and the shape of the initial warp field window function is:
\begin{equation}\label{warpshape1}
    W(r_S) = \frac{\tanh{(\sigma(r_S + R))}-\tanh{(\sigma(r_S-R))}}{2 \tanh{(\sigma R)}}
\,,
\end{equation}
with $R$ a fixed initial (Lagrangian) radius, and $\sigma$ a constant that determines the inverse thickness of the wall of the `warp bubble'.
Here $X, Y, Z $ are the Cartesian components of $\bm X$.
$X_S$ is the fixed position of the center of the warp field in Lagrangian space, and $x_S$ is its position in the Eulerian space; we illustrate the family of trajectories in figure \ref{caustic}:
\begin{equation}
\label{trajectoryS}
    x_S(t,{X^k}) 
    = f_S (t,{X^k}) 
    = X + V_0({X^k})(t-t_0)
    = X + v_S(t_0) W(r_S(t_0,{ X^k}))(t-t_0)
\,.
\end{equation}
As expected, a caustic is formed (at $t \approx 0.445$); the velocity is multi-valued after this time. The shape of the velocity profile is changing, unlike the situation in Alcubierre's model.

\begin{figure}[H]
    \centering
        \includegraphics[
        width=\textwidth]{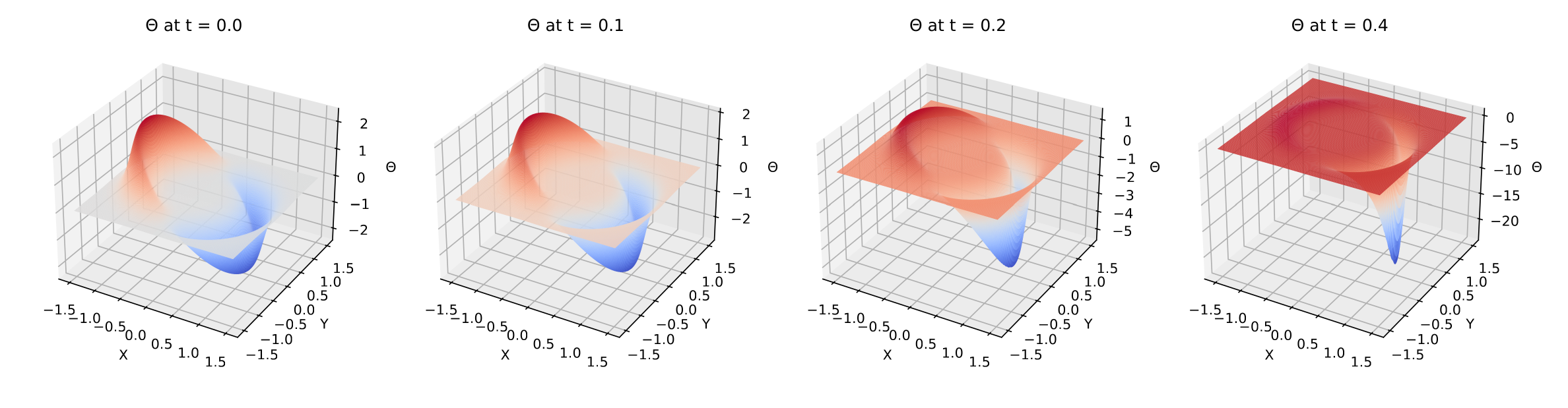}
%
        \centering
        \includegraphics[
        width=\textwidth]{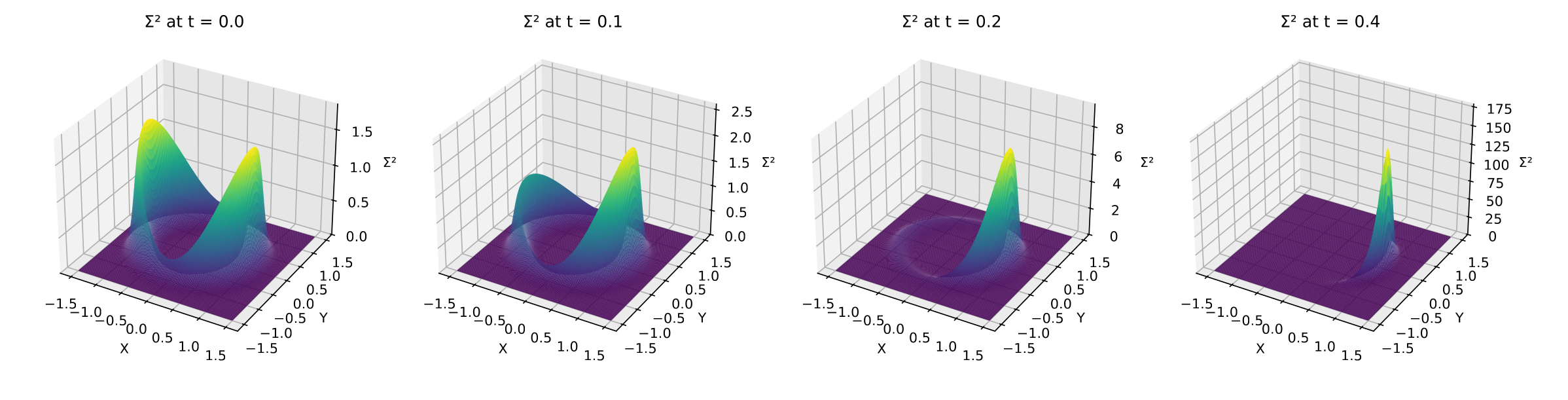}

        \centering
        \includegraphics[
        width=\textwidth]{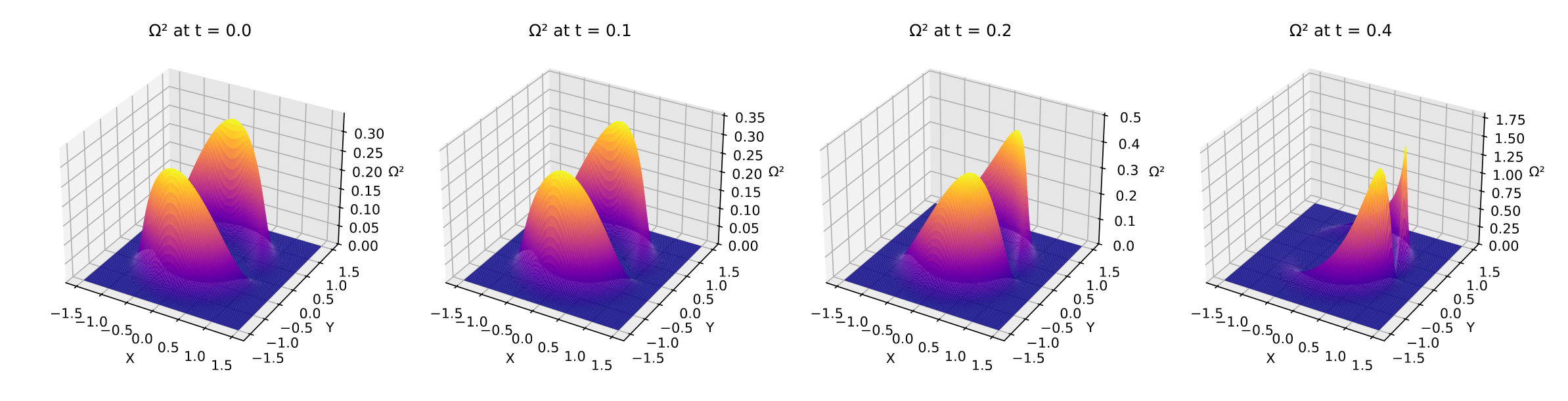}

        \centering
        \includegraphics[
        width=\textwidth]{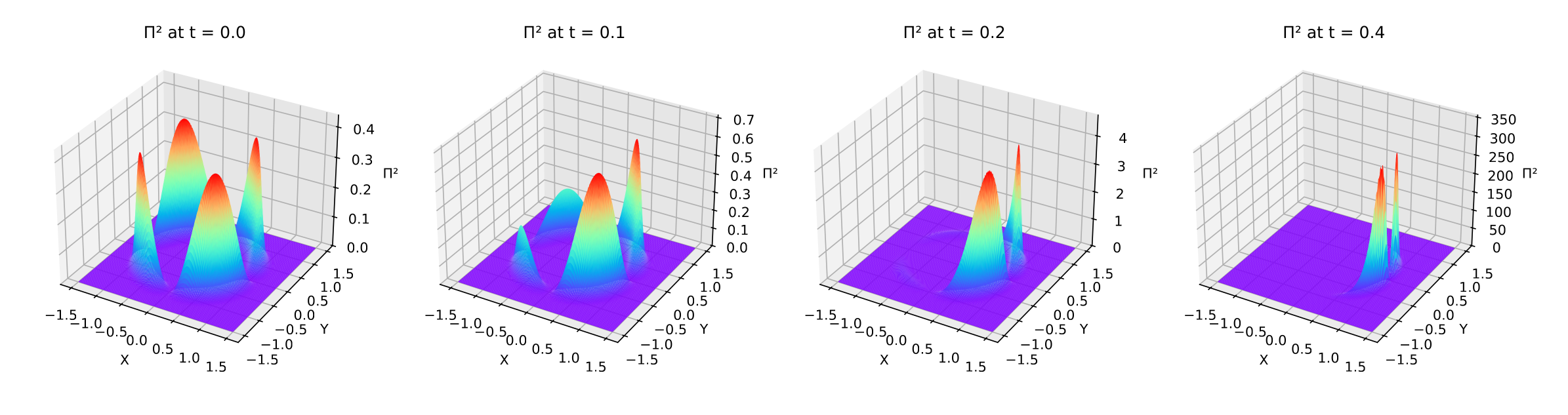}

    \caption{From top to bottom: evolution of the $\Theta$, $\Sigma^2$, $\Omega^2$ and $\Pi^2$ fields at different times $t$. The different color coding for the different fields has no significance here. At initial time, here $t=0$, left figure, Eulerian and Lagrangian coordinates coincide and we see the initial data. At $t=0.10$ and $t=0.25$ we can observe a marked deformation of the fields. The vorticity scalar and shear scalar fields move significantly with the rate of expansion field. At a time close to the computable limit $t=0.4$, right figure, the deformation of $\Theta$ at the front will become infinite. We can see the same phenomenon on all the other scalar fields (Here we set the gravitational constant $G=1$ and the others variables: $v_S = 0,9$, $\sigma = 5$, $R=1$, $t_0=0$).}
    \label{Lagrangian_evolution_fields}
\end{figure}

We express the Lagrangian evolution equations of the kinematic properties, notably with the help of the Jacobian 
 $J = 1 + (t-t_0)\Theta (t_0, {X^k})$,
and we obtain all the kinematical invariants. For the expansion rate we have:
\begin{equation}
    \label{theta_evolution}
    \Theta(t, X^k) = \frac{\Theta(t_0, X^k)}{1+(t-t_0)\Theta(t_0, X^k)} 
\,.
\end{equation}
For the shear scalar we use \textbf{Remark 3} stating that the second scalar invariant of the velocity gradient vanishes for one-component velocity fields, ${\rm II} = 0 = \frac{1}{3}\Theta^2- \Sigma^2 + \Omega^2$, so that we have:
\begin{equation}
    \Sigma^2(t, X^k)= \frac{1}{3}\Theta^2(t, X^k)+\Omega^2(t, X^k)
\,.
\end{equation}
To calculate $\Omega(t,X^k)^2$ we know the integral of the Kelvin-Helmholtz vorticity transport in terms of the vorticity vector \eqref{conservation_equation_vorticity}. Since we demand $\bm\p \times \bm A = \bm 0$, we have Cauchy's integral \cite{serrin}, \cite{BZeldovich} in three dimensions:
\begin{equation}
    \bm\Omega = \frac{(\bm \Omega (t_0) \cdot \bm \p_0)\bm f}{J}
\,,
\end{equation}
where $\bm\p_0$ denotes the nabla operator with respect to Lagrangian coordinates, and where $J$ is given by the Jacobian determinant for inertial motion. 
In the present case, vorticity is in the $(y,z)$-plane, displacement only occurs along the $x$-direction, and therefore the Lagrangian convective term reduces to the initial vorticity and Cauchy's integral reads:
\begin{equation*}
\Omega^x (t,X^k) = 0 \ , \quad
\Omega^y(t, X^k) = \frac{\Omega^y(t_0)}{ 1 + (t-t_0)\Theta (t_0, {X^k})} \ , \quad
\Omega^z(t, X^k)  \frac{\Omega^z(t_0)}{ 1 + (t-t_0)\Theta (t_0, {X^k})}\ .
\end{equation*}
From the definition $\Omega^2 = \frac{1}{2}\Omega_{ij}\Omega^{ij} = \frac{1}{2}\bm\Omega^2$ we obtain:
\begin{align}
    \label{vorticity_evolution}
    {\Omega}^2(t, X^k) 
    &= \frac{\Omega^2(t_0, X^k)}{J^2}
    = \frac{\Omega^2(t_0, X^k)}{(1+(t-t_0)\Theta(t_0,X^k))^2} 
\,.
\end{align}
We can thus explicitly express the stress components from \eqref{stresses}, restricted to the present case with null-coordinate acceleration:
\begin{equation}
    \tensor{\pi}{^x_x} (t, X^k) = - \frac{\Omega^2(t, X^k)}{4\pi G}\quad,\quad \tensor{\pi}{^x_y} (t, X^k) = \frac{-\Omega^y}{8\pi G} \Theta \quad, \quad \tensor{\pi}{^x_y} (t, X^k) = \frac{\Omega^z}{8\pi G} \Theta \,.
\end{equation}
From \textbf{Corollary 2}, the energy density and the pressure attain this form:
\begin{align}
    \epsilon (t, X^k) = -\frac{1}{8\pi G} [\Lambda +  \Omega^2(t,X^k)]
\,,\quad
    p(t,X^k) = \frac{1}{8\pi G} [\Lambda - \Omega^2(t,X^k)]
\,.
\end{align}
We compute the anisotropic stress scalar in Appendix~\ref{app:pi_calculation}. From \textbf{Corollary 2}, with the assumption of inertial motion we obtain:
\begin{equation}
    \Pi^2(t,X^k) =  \frac{1}{(8 \pi G)^2} [4\Omega^4(t, X^k) + \Omega^2(t,X^k) \Theta^2(t, X^k)]
\,.
\end{equation}
In figure \eqref{Lagrangian_evolution_fields} we show the evolution of the Lagrangian fields.
Note that this second model defines the Eulerian coordinate velocity field only implicitly through the inverse geodesics,
$V_S (t, x^k) = V_0 (t_0, X^j = h^j (t, x^k)), {\bm h} = {\bm f}^{-1}$;
(we did not map the fields back to Eulerian space here, since the field's principal evolution properties are not altered.)\\

With the dynamical assumption $A=0$, we find the equation of state relations:  
\begin{equation} 
    p - \epsilon = \frac{\Lambda}{4\pi G}\,,\quad  p + \epsilon = \pi^x_{\ x} = -\frac{\Omega^2}{4\pi G}
    \, ,
\end{equation}
so that we can infer the figures for $\epsilon$, $p$, and $\pi^x_{\ x}$ as the inversion of the previous figure \ref{Lagrangian_evolution_fields} for the vorticity scalar. This latter is generated by anisotropic stress components and the expansion scalar:
\begin{equation}
    \Omega = \sqrt{(\Omega^y)^2 + (-\Omega^z)^2} = 8\pi G \frac{\sqrt{(\pi^x_{\ y})^2 + (\pi^x_{\ z})^2}}{\p_x V}
    \,,\quad \p_x V \ne 0
\,.
\end{equation}
We finally note that for $\Lambda = 0$ the equation of state relations result in a stiff equation of state, $p=\epsilon$, and $\pi^x_{\ x} = 2p$.

\subsection{Imposing assumptions on the stress-energy sources}
\label{assumptionsources}

We here impose different assumptions on the sources within the general case for one-component coordinate velocity fields. We recall the governing equations for this discussion by isolating the sources:
\begin{equation}\label{sources1}
   8\pi G \epsilon = - \Lambda - \Omega^2 
\ ,\quad 
    8\pi G p = \Lambda - \Omega^2   - \frac{2}{3}\,\p_x A \ , \quad 8\pi G \bm q = \bm\p \times \bm \Omega \ .
\end{equation}
and for the trace-free symmetric components:
\begin{equation}\label{sources2}
 8\pi G  \pi^x_{\ x} =  \frac{2}{3}\p_x A - 2\Omega^2 \, ,\ \ 
 8\pi G  \pi^x_{\ y} =  \frac{1}{2}\p_y A  - \frac{1}{2}\Omega^z \p_x V \, ,\ \ 
 8\pi G  \pi^x_{\ z} =  \frac{1}{2}\p_z A  + \frac{1}{2}\Omega^y \p_x V \, .
\end{equation}
We immediately infer the following subcases (we include the cosmological constant):
\begin{itemize}
\item{\textbf{Dust:}} With $\epsilon = \varrho$, the dust density, and $p= 0$, $\pi^i_{\ j} = 0$, we find with the first equation of \eqref{sources2} 
a constant vorticity, $3\Omega^2 = \Lambda$, which implies a vanishing momentum flux density consistent with the dust assumption, $\bm q = \bm 0$ (which according to footnote \ref{potentialomega} would still allow for a harmonic vorticity potential); a constant dust density follows, $\varrho = -\Lambda / 6\pi G$, which for $\Lambda = 0$ reduces the spacetime
to vacuum spacetime; for $\Lambda \ne 0$, the gradient of the coordinate acceleration reduces to
$\p_x A = \Lambda$, $\p_y A =  \Omega^z \p_x V$, $\p_z A =  -\Omega^y \p_x V$, with $\Omega^y = (1/2)\p_z V = const.$ and $\Omega^z = - (1/2)\p_y V = const.$, which can be absorbed into a (cosmological) homogeneous background that replaces Minkowski spacetime as a background to the warp field. No inhomogeneous warp field can exist.

\item{\textbf{Perfect Fluid:}} We have to assume $\pi^i_{\ j} = 0$, but also $\bm q = \bm 0$. 
Using \eqref{sources2} we obtain $8\pi G \epsilon = - (\Lambda + \Omega^2)$ and 
$8\pi G p = \Lambda - 3\Omega^2$, implying the equation of state $p = 3\epsilon + \frac{\Lambda}{2\pi G}$. 
Again, $\bm q = \bm 0$ implies a harmonic vorticity potential $Z$ according to footnote \ref{potentialomega}. To deal with this case exactly, we'd have to specify fall-off and boundary conditions. A possibility is to again introduce a homogeneous (irrotational) background as in the dust case, splitting the sources accordingly, $\epsilon = :\epsilon_b + \epsilon_{\Omega}$, $p = :p_b + p_{\Omega}$.
Imposing the equation of state of a cosmological constant for the background, $p_b = - \epsilon_b = -\frac{\Lambda}{8\pi G}$, we may consider the example of a compact warp field and impose periodic boundary conditions on the harmonic vorticity potential for which the only periodic solution to the Laplace equation, $\Delta Z = 0$, is constant, we again obtain a constant vorticity. This in turn would imply the same conclusions as for the dust case above. If vorticity is set to zero, \textbf{Corollary 3} shows that the spacetime is then Minkowski spacetime, since the stress-energy tensor and the full energy-momentum tensor \eqref{EMT} must vanish.

\end{itemize}
In both cases above there is more than one function specified to close the Einstein equations, therefore implying too strong restrictions on possible solutions.
If we think of the example of Alcubierre's model to close the Einstein equations, it is impossible to support Alcubierre's velocity profile with a dust or perfect fluid source, see \textbf{Remark 6} below.
The anisotropic stresses and a non-vanishing momentum flux density are mandatory.\\

\noindent
\textbf{Remark 6} (A note on the literature)

\textit{
The full set of Einstein's equations, evaluated subsequently for sources of increasing generality, has been considered  for one-component velocity fields specified to Alcubierre's model in a series of articles,
see the references in the short summary by the authors \cite{SAR3} and in Santos Pereira's PhD thesis \cite{SARPHD}. 
In the present article, we investigated the general three-component case for the Nat\'ario class of metrics, and we discussed the restriction to one-component velocity fields as a subcase of the general 3+1 equations. 
Our short derivation of the general three-component case starting from subsection~\ref{Einstein3D} trivially covers the calculations in the one-component case in these series of articles.
While in the above papers the 4D Einstein equations have been calculated in components,\footnote{We have repeated the component calculation of the 4D Einstein equations with \textit{Sagemath} (see the data availability statement at the end), and found some differences to the components presented in the above papers.}
we here followed a direct tensorial derivation and discussed the subcase in terms of spatial components. 
Note that if the specific Alcubierre profile is determined, we have no freedom to make a further choice, which would result in a restriction on top of those dictated by the restrictions already made. In other words,  any further specific assumption on the sources in addition to Alcubierre's model may easily lead to contradictions with the result from the Einstein equations. As also Santiago \textit{et al.} \cite{Santiago} noted, we then obtain ``Minkowski spacetime in disguise", which is the case when no momentum flux and anisotropic stresses are allowed that, for the Alcubierre model, must have a specific form.}

\medskip

The study of assumptions on sources in the spirit of the Cauchy problem yields more general results when considering another metric setting that we shall discuss in the next section.

\section{\textbf{R1}-Warp: study of warp field dynamics from vector space theories}
\label{V}

In this section, we shall propose a simplified strategy to study warp field dynamics in a flow-orthogonal foliation by starting from solutions or approximations in Newtonian gravity and Newtonian cosmology. 

\subsection{Direct correspondence between Newtonian gravity and general relativity}
\label{correspondence}

Working in a flow-orthogonal foliation, many investigations in relativistic cosmology provide powerful approximation schemes and exact solutions for given initial data. Before we touch upon these works, we will below explain a way to construct approximations and solutions of general relativity for the matter model \textit{irrotational dust} from approximations and solutions of Newtonian gravity. We think that this is particularly useful to make contact to the literature on warp drive spacetimes, since preference is given to work out warp field models within flat space sections using global inertial coordinates.
We will thus keep the restriction ${\rm\bf R1}$ (flow-orthogonality) and the first part of restriction ${\rm\bf R2}$ (constant lapse function, $N=1$, i.e. geodesic slicing).  The main difference of our approach compared with the literature on warp drive models is to align the time-vector $\bm\p_t = N \p_t + {\bm N}$ with the spaceship, i.e. for $N=1$ we put $\bm N = {\bm 0}$, hence a vanishing coordinate velocity $\bm V$. The $4$-velocity in this framework then has components ${\bm u} = {\bm n} = (1,0,0,0)$ (Lagrangian description).

\subsection{Newtonian gravity and a strategy to construct general-relativistic models}

In what follows we will briefly explain the correspondence strategy. For details we refer the reader to \cite{Bcorrespondence}. 
We will employ a hydrodynamic picture and consider a self-gravitating continuum of irrotational dust, i.e. pressureless matter with rest mass density $\varrho ({\bm x},t)$ and velocity field ${\bm v}({\bm x},t)$. Both fields are represented in terms of Eulerian (inertial, nonrotating) coordinates $\bm x$ and a time parameter $t$. The acceleration field is equivalent to the gravitational field strength $\bm{g}({\bm x},t)$ due to the equivalence of inertial and gravitational mass,
\begin{equation}
\label{euler}
    \frac{\rd}{\rd t}\bm{v}=\bm{g} \quad ; \quad  \frac{\rd}{\rd t} :=\frac{\partial}{\partial t}\Big\vert_ {\bm{x}}  + \bm{v}\cdot\bm\p = \frac{\partial}{\partial t}\Big\vert_ {\bm{X}}\ ,
    \end{equation}
where we recalled the Lagrangian time derivative $\rd / \rd t$ that reduces to a partial time derivative in a Lagrangian coordinate system $\bm X$. 
In Eulerian coordinate components, we write Euler's equation \eqref{euler} and its Eulerian spatial derivative:
\begin{equation}
\label{eulerderivative}
    \frac{\partial}{\partial t} v^{i}+v^{k} v^i_{\ , k}=g^{i} \quad , \quad  \frac{\mathrm d}{{\mathrm d} t} v^i_{\ , j}+v^i_{\ , k} v^k_{\ , j} = g^i_{\ , j}\ .
    \end{equation}
The Newtonian continuum theory of gravitation is a vector theory, so that the sources of the curl and divergence of  $\bm{g}$ suffice to define the complete set of field equations (up to a harmonic vector field): 
\begin{equation}
\label{fieldequations}
 \bm\p \times \bm{g}=\bm{0}  \quad;\quad  \bm\p\cdot\bm{g}=\Lambda-4 \pi G \varrho \ ,
\end{equation}
where $G$ denotes Newton's gravitational coupling constant; for completeness we included the cosmological constant $\Lambda$ (here with dimension ${\rm time}^{-2}$). The rest mass density $\varrho$ obeys the continuity equation,
\begin{equation}
\label{continuity}
\frac{\rd}{\rd t} \varrho+\varrho \; \bm\p\cdot\bm{v}=0 \ .
\end{equation}
The Euler-Newton system comprises \eqref{euler}, \eqref{fieldequations} and \eqref{continuity}. 
The restriction to irrotational flows, $v_{[i,j]} = 0$, is necessary since the correspondence strategy directs to Einstein's equations in a flow-orthogonal foliation with vanishing covariant and coordinate vorticities. 

The correspondence strategy itself consists first in transforming the (extrinsic) representation of the self-gravitating fluid in terms of the Euler-Newton system to its Lagrangian (intrinsic) representation (Step 1). For this we introduce a one-parameter family of spatial diffeomorphisms to Lagrangian coordinates $\bm{X}$, labeling fluid parcels along their trajectories, identified with their Eulerian positions at some initial time $t_{\rm i}$ \cite{ehlersbuchert}:
\begin{equation}
\label{diffeo}
(\bm{X},t)\mapsto (\bm{x},t) = ({\bm f}({\bm{X},t}),t) \ ; \ \bm{X}:= \bm{f} (\bm{X}, t_{\rm i}) \ .
\end{equation}
$\bm{f}$ is the vector field of trajectories and ${\bf d} f^a = f^a_{\ |k}{\bf d} X^k$ the \textit{Lagrangian deformation gradient}, represented in the exact Lagrangian basis and being the only dynamical field variable in the Lagrangian representation.\footnote{Henceforth, we use the indices $a,b,c \ldots$ as counters, since Eulerian vector components and Eulerian derivatives no longer refer to an exact coordinate basis after step 2 of the correspondence (below) is executed, while $i,j,k \ldots$ remain coordinate indices referring to an exact basis.}

Transforming the gradients of the Eulerian velocity and acceleration, ${\bm v} = \dot{\bm f}({\bm X},t)$ and ${\bm g} = \ddot{\bm f}({\bm X},t)$ involves the inverse transformation of, ${\bm h}({\bm x},t): = \bm f^{-1}$, 
\begin{align}
\label{trafogradients}
v^a_{\ ,b}=v^a_{\ |k}h^k_{\ ,b} = {\dot f}^a_{\ |k}h^k_{\ ,b} \ ,\\
g^a_{\ ,b}=g^a_{\ |k}h^k_{\ ,b} = {\ddot f}^a_{\ |k}h^k_{\ ,b} \ .
\end{align}
Step 2 of the correspondence strategy then relaxes the integrability of the Lagrangian deformation gradient, introducing general one-form fields $\boldsymbol{\eta}^a$.
\begin{equation}
\label{step2}
\bd f^a \; \mapsto \; \boldsymbol{\eta}^a \  .
\end{equation}
For this we define three {\it frame fields}, ${\bm e}_b$
(the {\it Dreibein} at the worldlines of fluid parcels), being dual to Cartan's coframe 
fields $\boldsymbol{\eta}^a$. We express both in the respective local basis systems ($\bd X^k$ for forms and $\bpartial_{X^k}$ for vectors):
\begin{equation}
\label{framefields}
\boldsymbol{\eta}^a = {\eta}^a_{\ k}\,\bd X^k \ ,\quad {\bm e}_b =e_b^{\ k}\,\bpartial_{X^k}
\ , , \quad{\eta}^a_{\ k} e_b^{\ k} = \delta^a_{\ b}\ , \quad{\eta}^a_{\ k} e_a^{\ \ell} = \delta_k^{\ \ell}\ .
\end{equation}
Moving to the nonintegrable forms of the velocity and acceleration gradients, 
$v^a_{\ ,b} \mapsto \dot{\eta}^a_{\ k} e^k_{\ b}=:\theta^a_{\ b}$ and $g^a_{\ ,b} \mapsto \ddot{\eta}^a_{\ k} e^k_{\ b}=:\CF^a_{\ b}$, we finally transform these objects into our local exact (Lagrangian) basis with the help of the transformation
matrices (\ref{framefields}) and arrive at spatial tensors parametrized by the coordinate time $t$ appearing in Einstein's equations:
\begin{align}
\label{analogvelocitygradient2}
\Theta^i_{\ j}:= e_a^{\ i}\eta^b_{\ j}\ , \quad \Theta^a_{\ b}=e_a^{\ i}{\dot\eta}^a_{\ j}\ , \quad
\CF^i_{\ j}:= e_a^{\ i}\eta^b_{\ j}\,\CF^a_{\ b}=e_a^{\ i}{\ddot\eta}^a_{\ j}\ .
\end{align}
We can express the frame coefficients through the coframe coefficients using the algebraic identity,
\begin{equation}
\label{frametransformation}
e_a^{\ i} = \frac{1}{2J}\;\epsilon_{abc}\epsilon^{ik\ell}\,\eta^b_{\;k}\,\eta^c_{\;\ell}\ ,
\end{equation}
with the Levi-Civit\`a symbol $\epsilon_{abc}$, and the nonintegrable analog of the Jacobian of the spatial diffeomorphism,
\begin{equation}
\label{Jacobian_art}
J: =\det (\eta^a_{\ i}) = \frac{1}{6}
\epsilon_{abc}\epsilon^{ijk}\,\eta^a_{\ i}\eta^b_{\ j}\eta^c_{\ k}\ .
\end{equation}
As shown in \cite{Bcorrespondence}, the transformed Euler-Newton system,
\begin{align}
\label{generalizedeuler}
{\CF}^{i}_{\ j} = \dot{\Theta}^i_{\ j} + \Theta^i_{\ k}\Theta^k_{\ j} \ ,\\
\label{generalizedfieldequations}
{\CF}^k_{\ k} = \Lambda - 4\pi G \varrho \quad;\quad
{\CF}_{[ij]} = 0 \ ,
\end{align}
is equivalent to the Einstein equations with the \textit{correspondence-field relation}:
\begin{equation}
\label{key1}
{\CF}^{i}_{\ j} = - \CR^i_{\ j} + (\Lambda + 4\pi G\varrho)\delta^i_{\ j} +
 \Theta^i_{\ k}\Theta^k_{\ j} - \Theta \Theta^i_{\ j} \ .
\end{equation}
In the geometrical context of general relativity, the spatial Ricci tensor, $\CR_{ij} = \delta_{ab}\eta^a_{\ i}\eta^b_{\ k}\CR^k_{\ j}$ is introduced instead of $\CF^i_{\ j}$; the key equation \eqref{key1} is known to emerge from the Gau{\ss} embedding equation using the nonintegrable Euler equation \eqref{generalizedeuler}. 
The components of the generalized Newtonian field strength gradient form components of the spacetime Riemann tensor, 
\begin{equation}
-{\CF}^{i}_{\ j} = {^4}R^i_{\ 0 j 0} = {^4}R^\mu_{\ \nu\rho\sigma}n^\rho n^\sigma \ .
\end{equation}
Imposing the field equations \eqref{generalizedfieldequations}, the trace of \eqref{key1} becomes the energy constraint, and the antisymmetric part of \eqref{key1} vanishes due to the vanishing of the vorticity in a flow-orthogonal foliation: ${\CF}_{[ij]} = \delta_{ab}\,\eta^a_{\ [i}{\ddot\eta}^b_{\ j]} =  (\delta_{ab}\,\eta^a_{\ [i}{\dot\eta}^b_{\ j]})^{\bdot} = 0$. 
(The momentum constraints can also be derived from the commutation of second derivatives in the integrable situation, see \cite{Bcorrespondence}.)

For the metric, Step 1 transforms the Euclidean metric into the diffeomorphically flat Lagrange-Euclidean metric, 
\begin{equation}
\label{metricE}
{^3}{\bm\delta} = \delta_{ij}\bd x^i \otimes \bd x^j \mapsto \delta_{ab}\,f^a_{\;|i} f^b_{\;|j} \bd X^i \otimes \bd X^j \ .
\end{equation}
Step 2 then yields the spatial Riemannian metric, 
\begin{equation}
\label{metricR}
{^3}{\bm s}:=\delta_{ab}\;\boldsymbol{\eta}^a \otimes  \boldsymbol{\eta}^b = 
\delta_{ab} \;\eta^a_{\;i}\,\eta^b_{\;j} \,\bd X^i \otimes \bd X^j \ . 
\end{equation}
For the comparison between Newtonian and GR solutions it is convenient to work with orthogonal coframes instead of orthonormal ones. The Riemannian metric is then written as
\begin{equation}
\label{metricRorthogonal}
{^3}{\bm s}:={\hat g}_{ab}\;\boldsymbol{\eta}^a \otimes  \boldsymbol{\eta}^b = 
{\hat g}_{ab} \;\eta^a_{\;i}\,\eta^b_{\;j} \,\bd X^i \otimes \bd X^j \ , 
\end{equation}
where we have kept the same symbol for the coframes for notational ease. Gram's matrix ${\hat g}_{ab}$ replaces $\delta_{ab}$ to allow for the coframe coefficients to be initially $\eta^a_{\ i} (t_0)= \delta^a_{\ i}$, and the initial metric is thus encoded in ${\hat g}_{ab}$: $g_{ij}(\bm X,t_0)= {\hat g}_{ab}\delta^a_{\ i}\delta^b_{\ j}$, for details see Appendix A in \cite{rza3}.\\

The two-step strategy presented in this section enables us to consider well-known solutions or approximations to the Euler-Newton system and find corresponding solutions and approximations to Einstein's equations in a flow-orthogonal foliation. 
We shall now exemplify this strategy having in mind that a particular shape of the warp field may be given as initial condition. 

\subsection{Dynamics of warp fields}

We will give two examples of exact solutions of Einstein's equations that exploit the above-outlined correspondence, (i) for inertial motion, i.e. where the gravitational field is neglected so that the Euler-Newton system reduces to the system for inertial motion, leading to an interesting solution of Einstein's equations, and (ii) a particular solution with nonvanishing gravitational field strength in Newton's theory leading to a corresponding Szekeres class II solution of general relativity. 
Both classes of solutions have intrinsic curvature.

\subsubsection{Inertial motion in Newtonian theory}
\label{inertialNewton}

Setting the gravitational field strength ${\bm g} = {\bm 0}$, the Euler-Newton system \eqref{euler}, \eqref{fieldequations} and \eqref{continuity} reduces to the system
\begin{equation}
\label{solEuler1inertial}
    \frac{\rd}{\rd t}{v^i} = 0 \quad , \quad 
    \frac{\rd}{\rd t} \varrho+\varrho v^k_{\ , k}=0 \ .
\end{equation}
The first equations form a coupled system of nonlinear partial differential equations that reduces to a decoupled system of ordinary differential equations in Lagrangian coordinates, $\partial_t \vert_{\bm X} v^i = 0$, which is readily solved by a family of Galileian transformations:
\begin{equation}
\label{solEuler2}
    v^i(X^k,t)= v_0^i(X^k) \  ,  \ x^i = f^i(X^j,t) = X^i +v_0^i(X^j)(t-t_0)\, ,
\end{equation}
where we have put $f^i(X^k,t_0):=X^i$ at some initial time $t_0$, 
with $v_0^i (X^k)$ denoting the initial velocity field components of a given warp field.

We deduce from this that $X^k = h^k (x^j,t)$, where ${\bm h} := {\bm f}^{-1}$. With the help of the inverse mapping ${\bm h}$ we are then able to express the solutions for velocity and density in the Eulerian global coordinates,
\begin{equation}
\label{solEuler3v}
    v^i (x^j,t) = v_0^i (X^k = h^k (x^j,t)) \ ,
\end{equation}
\begin{equation}
\label{solEuler4x}
    \varrho (x^j,t) = \frac{\varrho_0(X^k = h^k (x^j,t))}{J(X^k = h^k(x^j,t),t)}\ ,
\end{equation}
where we used the exact integral of the continuity equation, derived from the conservation of the total rest mass $M_{D_t}$ in a spatial domain $D_t$: 
\begin{equation}
\label{Proof_of_7.1}
    0 = \frac{\rd} {\rd t}M_{D_t} = \frac{\rd}{\rd t} \int_{D_t}\varrho(x^j,t)\rd^3x\ . 
\end{equation}   
Changing to a Lagrangian domain $D_{t_0}$,  
\begin{equation}
\label{Proof_of_7.2}
    \int_{D_{t_0}} \frac{\rd}{\rd t}(\varrho(X^k,t)J(X^k,t))\rd^3X = 0 \ , 
\end{equation}
we obtain the solution
\begin{equation}
\label{Proof_of_7.3}
    \frac{\rd}{\rd t} (\varrho J) = 0 \Rightarrow \varrho J = C(X^k)\ , 
\end{equation}
with the Jacobian:
\begin{equation}
\label{Lagrange2(J)}
    J(X^k,t) = \det\Big( \frac{\partial f^i(X^k,t)}{\partial X^j}\Big) = \det\Big(\delta_{ij}+\frac{\partial v_0^i}{\partial X^j}(t-t_0)\Big) \ .
\end{equation}
At $t = t_0$, we have the initial density $\varrho (X^k,t_0) = \varrho_0(X^k)$ and we find (\ref{solEuler4x}). $\Box$

The Jacobian \eqref{Lagrange2(J)} can be expressed in terms of principal scalar invariants of the initial velocity gradient $\bm\p_0 {\bm v}_0$, where $\bm\p_0$ denotes the nabla operator with respect to Lagrangian coordinates: 
\begin{equation}
\label{Jacobianinertial}
    J(t,X^k) = 1 + {\rm I}(\bm\p_0 {\bm v}_0)(t-t_0) + {\rm II}(\bm\p_0 {\bm v}_0)(t-t_0)^2 + {\rm III}(\bm\p_0 {\bm v}_0)(t-t_0)^3 \ .
\end{equation}

Thus far, we discussed the general solution for inertial motion. Assuming now a one-component velocity field, $\bm v^i = V \delta_x^i$, we can easily show (easiest when using the representations \eqref{divII} and \eqref{divIII}) that ${\rm II}(\bm\p_0 {\bm v}_0) = {\rm III}(\bm\p_0 {\bm v}_0) = 0$, and the solution reads:
\begin{equation}
\label{Alcubierre's Lagrange density}
V ({\bm X},t)= V ({\bm X},t_0)\ , \  \varrho(\bm X,t) = \frac{\varrho(\bm X,t_0)}{J(\bm X,t)} \ ,
\end{equation}
with the reduced Jacobian,
\begin{equation}
\label{Alcubierre'sJacobian}
    J = 1 + \Theta ({\bm X},t_0)(t-t_0) \ , \ \Theta({\bm X},t_0) =  V_{|X} + V_{|Y} + V_{|Z} \ .
\end{equation}
Illustrations of this solution for Alcubierre initial data result in the same figures as those presented in Subsection~\ref{illustrationofinertial}. 

\subsubsection{A class of solutions in Newtonian theory}

Including now gravity, a class of 3D solutions without symmetry is known \cite{buchertgoetz}, characterized by a generalization of the exact one-dimensional solution \cite{zentsovachernin} in terms of the family of trajectories, 
\begin{align}
\label{buchertgoetzsolution}
\bm f &= \bm X + \bm V (\bm X)(t-t_0) + \bm G (\bm X)\frac{(t-t_0)^2}{2}\ ,\\
\bm v &= \dot{\bm f} = \bm V (\bm X) + \bm G (\bm X)(t-t_0) \ ,\quad \bm g = \ddot {\bm f} = \bm G (\bm X) \ ,
\end{align}
where the two independent initial fields are $\bm V ; = \bm v (\bm x,t_0)$ and $\bm G := \bm g (\bm x,t_0)$. In order to be a solution, these initial fields are highly restricted \cite{buchertgoetz}. In particular, the second and third principal scalar invariants of the initial velocity and field-strength gradients have to vanish. It implies that the 3D motion is \textit{locally} one-dimensional, 
i.e. there is only one eigendirection of the fluid deformation at each $\bm X$.
This class is also known for a nonvanishing cosmological constant \cite{barrowgoetz, buchert:integral}, as a subcase of the corresponding cosmological solutions on a FLRW (Friedmann-Lema\^\i tre-Robertson-Walker) background \cite{buchert1989},\footnote{The class of solutions without background admits vorticity that is, however, constraint to keep the local one-dimensionality of motion. The presence of a background removes vorticity and the motion is potential.}
\begin{align}
\bm f = \bm X &+\frac{1}{2\Lambda}\left(\bm G (\bm X) + \sqrt{\Lambda}\bm V (\bm X)\right)[e^{\sqrt{\Lambda}(t-t_0)}-1 ] \nonumber \\&+ \frac{1}{2\Lambda}\left(\bm G (\bm X)- \sqrt{\Lambda}\bm V(\bm X)\right)[e^{-\sqrt{\Lambda}(t-t_0)}-1] \ .
\end{align}
We can make the link to the already discussed illustrations for inertial motion as follows. We shall adopt the so-called \textit{slaving condition}, i.e. we remove the independence of the initial fields $\bm V$ and $\bm G$ by aligning the initial velocity in the direction of high densities, i.e. $\bm V = \bm G t_0$. (This is a common practice in cosmology to simplify initial data, justified by the asymptotic feature of linearized perturbations on the background to align both fields.) This in turn implies that we can re-parametrize the time to 
match the inertial solution, e.g. for the solution curves \eqref{buchertgoetzsolution}:
\begin{equation}
\label{T}
T: = (t-t_0) + \frac{(t-t_0)^2}{2t_0}\; , \; \dot T = \frac{t}{t_0} \; , \; \ddot T = \frac{1}{t_0} \, .
\end{equation}

\subsubsection{Correspondent solutions in GR}
\label{inertialGR}

We now construct the corresponding solutions to the Einstein equations. 
Considering first the solutions \eqref{buchertgoetzsolution}. 
The correspondent solution for the spatial (orthogonal) coframe coefficients in Eq.~\eqref{metricRorthogonal}, using the slaving condition and the time variable \eqref{T}, reads:
\begin{align}
\label{GRsolution1}
\eta^a_{\ i} &= \delta^a_{\ i} + V^a_{\ i} T \ ,\\
\label{GRsolutionderivatives}
{\dot\eta}^a_{\ i} &= V^a_{\ i} \dot T = G^a_{\ i} t \ , \ {\ddot\eta}^a_{\ i} = V^a_{\ i} \ddot T = G^a_{\ i}\ ,
\end{align}
where $V^a_{\ i}$ denote the coefficients of the (nonintegrable) matrix corresponding to the initial velocity gradient (and, via slaving, to the initial field strength gradient (in nonintegrable form: $G^a_{\ i} t_0$). 
The expansion and field tensor coefficients read:
\begin{align}
\Theta_{ij} &= {\hat g}_{ab} \dot\eta^a_{\ i} \eta^b_{\ j}= {\hat g}_{ab} V^a_{\ i}(\bm X)V^b_{\ j}(\bm X) ({\dot T} T) \ ,\\
\CF_{ij} &= {\hat g}_{ab} \ddot\eta^a_{\ i} \eta^b_{\ j}= {\hat g}_{ab} V^a_{\ i}(\bm X)V^b_{\ j}(\bm X) ({\ddot T} T) \ .
\end{align}
For the mixed tensors we follow the transformation rules \eqref{analogvelocitygradient2} and \eqref{frametransformation}, e.g. for the transformation of the velocity gradient,
\begin{equation}
v^a_{\ ,b}=v^a_{\ |k}h^k_{\ ,b} = {\dot f}^a_{\ |k}h^k_{\ ,b} \ ,
\end{equation}
yielding the corresponding expansion tensor coefficients,
$\Theta^i_{\ j}= e_a^{\ i}\eta^b_{\ j}\,\Theta^a_{\ b}=e_a^{\ i}{\dot\eta}^a_{\ j}$, 
with $J$ given in \eqref{Jacobian_art}. (Similarly for $\CF^i_{\ j}$). For our solution \eqref{GRsolution1} we obtain:
\begin{equation}
\Theta^i_{\ j} = \frac{1}{2J}\;\epsilon_{abc}\epsilon^{ik\ell}\,\eta^b_{\;k}\,\eta^c_{\;\ell}{\dot\eta}^a_{\ j} 
\ , \quad 
\CF^i_{\ j} = \frac{1}{2J}\;\epsilon_{abc}\epsilon^{ik\ell}\,\eta^b_{\;k}\,\eta^c_{\;\ell}{\ddot\eta}^a_{\ j} \ ,
\end{equation}
with \eqref{GRsolution1} and \eqref{GRsolutionderivatives} inserted, and $J= \det (\delta^a_{\ i} + V^a_{\ i}T)$. The above fields solve 
\eqref{generalizedeuler}, \eqref{generalizedfieldequations}, \eqref{key1}, and determine the Ricci curvature coefficients:
\begin{align}
&\dot{\Theta}^i_{\ j} + \Theta^i_{\ k}\Theta^k_{\ j} = \CF^i_{\ j} \ ,\ \CF^k_{\ k}=\Lambda - 4\pi G \varrho \ , \ \CF_{[ij]} = 0 \ , \nonumber \\
&\CR^i_{\ j} = - \CF^i_{\ j} + (\Lambda + 4\pi G\varrho)\delta^i_{\ j} +
 \Theta^i_{\ k}\Theta^k_{\ j} - \Theta \Theta^i_{\ j} \ .\nonumber
\end{align}

\subsubsection{Contact with relativistic cosmology}
\label{cosmology}

What has been investigated above can be seen as part of a larger framework that investigates solutions and approximations of general relativity for perturbations at a FLRW background cosmology.
The Newtonian perturbative ansatz in Lagrangian perturbation theory, 
\begin{equation}
\label{Nansatz}
\bm f = a(t) \bm F (\bm X,t)\quad {\rm with} \quad \bm F = \bm X + \bm P (\bm X,t)\ ,
\end{equation}
generalizes to the relativistic ansatz for spatial coframe coefficients,
\begin{equation}
\label{GRansatz}
\eta^a_{\ i} = a(t) F^a_{\ i} (\bm X,t)\quad {\rm with} \quad F^a_{\ i} = \delta^a_{\ i} + P^a_{\ i} (\bm X,t)\ .
\end{equation}
The Einstein equations for irrotational dust in a flow-orthogonal setting can be entirely written in terms of spatial coframes, parametrized by the coordinate time $t$, and the resulting system of equations can be linearized in the deformation $P^a_{\ i}$; for details see the series of papers starting with \cite{rza1}; the most general linearized system is investigated in \cite{rza4}.

Contact to the motion of a warp field is made if this latter is considered as a Lagrangian perturbation at a cosmological background rather than a field on Minkowski space. Solutions and approximations of relativistic cosmology in the Lagrangian framework can be interpreted at a vanishing background, i.e. the scale factor is static and the Hubble function $H=\dot a / a \approx 0$. The ansatz \eqref{GRansatz} then reduces to the solutions discussed in the previous subsection.

It is a special feature of the linearized Lagrange-Einstein system in a flow-orthogonal foliation \cite{rza1,rza4} that it can be written in the form \eqref{GRansatz} and thus the linear approximations coincide with the 3D classes of solutions discussed in \cite{buchert1989} with their subclasses investigated in the previous subsection, provided the restrictions on initial data are respected. The class of solutions corresponding in general relativity to the class discussed in \cite{buchert1989} is known as Szekeres class II solutions \cite{rza6,beyond}. 

In cosmology it is, however, common practice to set generic intial data, thus using the ansatz \eqref{GRansatz} as an approximation. The successful performance of this approximation was already obvious in the Newtonian perturbation ansatz \eqref{Nansatz}, if the linearized approximation is compared with full Newtonian N-body simulations \cite{melottshandarin89,melottpellmanshandarin}, \cite{buchertmelottweiss,melottbuchertweiss}.
The linearized Newtonian approximation is known as the Zel'dovich approximation, which can be transformed to the inertial system by rescaling the time and space coordinates, as we also did in the previous section using the slaving assumption; for early investigations and reviews, see 
\cite{ShandarinZeldovich} and \cite{BZeldovich}; for a recent review on the Newtonian and relativistic perturbation schemes, including a historical account, together with the relations to the Szekeres class of solutions, the reader may consult the review \cite{Universe}.

\section{Summary and Outlook}
\label{VI}

In this paper we have investigated the governing equations for so-called \textbf{R}-motion in general relativity, having identified in \cite{BBL}  this class of motions as being relevant for the study of warp drive spacetimes in the literature.
We kept the coordinate velocity field general and discussed two particular solutions for one-component velocity fields:
1. A given velocity model using the warp drive spacetime of Alcubierre, and 2. Specifying the velocity model to a constant velocity along the geodesics. We have first  discussed the Alcubierre model illustrating its kinematical but also dynamical properties in terms of the coordinate acceleration. For comparison we exploited the solution 2 using Alcubierre's model as initial condition. We discussed that the warp field is then unstable in time, i.e. caustics will generically develop. We then proposed a simplified strategy to study warp fields dynamically, exploiting a correspondence between the vector theory of Newtonian gravity and general relativity. This allows us to study warp fields with intrinsic curvature in the context of exact solutions and approximation schemes for physical matter models making contact to well-known results in relativistic cosmology. 

We shall now touch upon some possibilities of investigation, still within the flow-orthogonal framework. We recall that this should be considered as an intermediate step towards the study of warp fields in a tilted slicing, when we can expect to also understand possible warp mechanisms that link covariant kinematical and dynamical properties to the dynamical creation of warp fields. 

\subsection{Stability of warp fields}

In this paper we learned about the instability of the Alcubierre warp field, taken as initial data for another class of solutions for \textbf{R}-motion.  It is clear that Alcubierre's field, even if taken as initial condition only, would represent a fully developed warp field, not being generated from reasonable starting conditions. Within the suggested new framework with vanishing shift,  we can model the future and past evolution, hence assuming low-amplitude warp fields initially.

In general, a criterion for the stability of a warp field may guide us to design the physics needed for stability. As an example, we may adopt the criterion of volume preservation.
Obviously, the volume is not preserved if it is not forced to be so as in Alcubierre's model. In all other cases, we are led to consider stabilizing the warp field through pressure-supported matter models.
For perfect fluids within a flow-orthogonal foliation we have to adopt a lapse function whose gradient is related to the gradient of the isotropic pressure \cite{rza5}. An equation of state has to be assumed and adjusted such that the warp field remains stable, e.g. preserves its volume. It is in this context where physical warp fields can be constructed that obey energy conditions and that are subject to a fully dynamical and morphological change of the warp field. Stability becomes a matter of understanding the physical evolution as an interplay between energy, pressure, kinematic fields and curvature, as opposed to forcing conditions on a particular stationary velocity profile as in Alcubierre's model.

\subsection{Gravitational wave emission from warp fields}

Within a flow-orthogonal setting, we have proposed to study the dynamics as well as the density and curvature distribution of simple warp fields by appealing to well-studied solutions and approximation schemes in relativistic cosmology. We are able to also work out the Weyl curvature in terms of the spatially projected gravitoelectric and gravitomagnetic parts of the Weyl tensor.  Without realizing this idea in this paper, we refer to results of \cite{rza4} within the linearized Lagrangian approximation. With these tools it is straightforward to calculate the gravitational wave emission from warp fields in this restricted situation, see also \cite{warpcollapse}.

\subsection{\textbf{T}-motion}

To go beyond the limitations of flow-orthogonality that implies vanishing covariant vorticity and acceleration fields,  we have to consider a new representation of motion within a covariant framework. In \cite{BBL} we have proposed tilted warp drives, called \textbf{T}-Warp. In this representation $\bm u$ is not aligned with the normal $\bm n$ of the foliation. Moreover, choosing a Lagrangian coordinate system,  the $4$-velocity of the spaceship and the time vector coincide, ${\bm u}_S = 
\bm\p_t$,  i.e. we have $\bm u_S = (1, \bm 0)$ and $\underline{\bm u}_S = (-1, \gamma(|\bm v_S|)\underline{\bm v}_S)$.
We can compare these two warp representations (see figure 1 in \cite{BBL}).

The main challenge of understanding \textbf{T}-motion lies in the relations between an Eulerian observer in the prescribed foliation and the covariantly moving spaceship, in which case a number of interesting new features arise. For example, while an Eulerian observer would see the spaceship moving at a proper velocity $\gamma(\bm v_S)\underline{\bm v}_S$, which  may exceed the speed of light, the spaceship itself is at rest within the restframe Lagrangian description of its motion. 
The development of horizons, as seen by the Eulerian observer, is expected if the warp field is supported by intrinsic curvature, which implies that the spaceship leaves the initial tangent space of the Eulerian observer. 
Furthermore, the presence of covariant acceleration and vorticity and their mutual coupling enables the study of warp mechanisms, i.e. a control of e.g. vorticity by the spaceship would entail acceleration. 

Within a tilted setting, the distinction between geometrical (passive) variables of the foliation such as the Eulerian energy $E$ vs. the fluid (active) variables such as the restframe energy $\epsilon$ becomes important. The presentation of the equations in this paper applies to the active description of motion of the spaceship. Both descriptions coincide only in the flow-orthogonal setting.

The rich consequences of such a more general description is certainly enticing and will open up new physical perspectives on warp drives.
Future efforts will be dedicated to this tilted setting.

\bigskip\bigskip

\authorcontributions{TB: conceptualization, methodology, derivation of general kinematics and dynamics of Einstein's equations for the general cases, writing, supervision, funding acquisition; AF: derivation of component expressions for one-component solutions, writing, software, visualization. 
All authors have read and agreed to the published version of the manuscript.}

\funding{This work forms a spin-out of a project that has received funding from the European Research Council (ERC) under the European Union's Horizon 2020 research and innovation programme (grant agreement ERC advanced grant 
740021--ARTHUS, PI: TB).}

\dataavailability{All the \textit{Python}, \textit{Mathematica} and \textit{Sagemath} codes for this study can be consulted if required at the
github link (Author: AF): \url{https://github.com/AntonyFrackowiak/Warp_drive.git}.
The component expressions for Einstein's equations are listed within internship reports by Antony Frackowiak \cite{AFM1, AFM2}, which can be also obtained from the github link.}


\acknowledgments{We would like to thank Hamed Barzegar for useful discussions.}

\conflictsofinterest{
The authors declare no conflict of interest.}
\clearpage
\appendixstart
\appendix

\section{Kinematical variables for one-component coordinate velocities and for the Alcubierre model}
\label{app:alcubierrekinematics}

We give the analytical expressions for the shear and vorticity tensors as well as for the rates of expansion, shear and vorticity, first for general one-component coordinate velocities and then for the specific choice of the Alcubierre model (indicated by the index Alc).
We assume a motion in $x$ direction only, and we have one component of the velocity $\bm V({\bm x},t) = V({\bm x},t)\delta_x^i$ (for the Alcubierre solution we denote $V=V_S$):
\begin{align}
\label{sigma_Alcubierre}
   (\Sigma_{ij}) & = \begin{pmatrix}
   \frac{2}{3}\frac{\p V}{\p x} &
   \frac{1}{2}\frac{\p V}{\p y} &
   \frac{1}{2}\frac{\p V}{\p z} &
   \\[10pt]
   \frac{1}{2}\frac{\p V}{\p y} &
   -\frac{1}{3}\frac{\p V}{\p x} &
   0 &
   \\[10pt]
   \frac{1}{2}\frac{\p V}{\p z} &
    0 &
   -\frac{1}{3}\frac{\p V}{\p x} &
   \end{pmatrix} \ , \ 
   (\Sigma_{ij})^{\rm Alc} =
   v_S \frac{\p W(r_S)}{\p r_S} \begin{pmatrix}
   \frac{2}{3}\frac{\p r_S}{\p x} &
   \frac{1}{2}\frac{\p r_S}{\p y} &
   \frac{1}{2}\frac{\p r_S}{\p z} &
   \\[10pt]
   \frac{1}{2}\frac{\p r_S}{\p y} &
   -\frac{1}{3}\frac{\p r_S}{\p x} &
   0 &
   \\[10pt]
   \frac{1}{2}\frac{\p r_S}{\p z} &
    0 &
   -\frac{1}{3}\frac{\p r_S}{\p x} &
   \end{pmatrix}
   \nn
   \\[10pt]
    & = \frac{v_S}{r_s} \frac{\p W(r_S)}{\p r_S} \begin{pmatrix}
   \frac{2}{3} (x-x_s) &
   \frac{1}{2} y &
   \frac{1}{2} z &
   \\[10pt]
   \frac{1}{2} y &
   -\frac{1}{3} (x-x_s) &
   0 &
   \\[10pt]
   \frac{1}{2} z &
    0 &
   -\frac{1}{3} (x-x_s) &
   \end{pmatrix}.
   \\[10pt]
\label{omega_Alcubierre}
    (\Omega_{ij}) & = \begin{pmatrix}
   0 &
   \frac{1}{2}\frac{\p V}{\p y} &
   \frac{1}{2}\frac{\p V}{\p z} &
   \\[10pt]
   -\frac{1}{2}\frac{\p V}{\p y} &
   0 &
   0 &
   \\[10pt]
   -\frac{1}{2}\frac{\p V}{\p z} &
    0 &
    0 &
   \end{pmatrix} \ , \  \begin{matrix} {\bm\Omega}^T = \frac{1}{2}\Big(0 ,\quad \p_z V ,\quad - \p_y V\Big)  &\\[10pt] 
   ({\bm \p\times \bm \Omega})^T = \frac{1}{2}(-\p_y^2 V -\p_z^2 V,\  \p_x(\p_y V), \  \p_x(\p_z V)) \end{matrix}.
   \\[10pt]
    (\Omega_{ij})^{\rm Alc} &= v_S \frac{\p W(r_S)}{\p r_S}\begin{pmatrix}
   0 &
   \frac{1}{2}\frac{\p r_S}{\p y} &
   \frac{1}{2}\frac{\p r_S}{\p z} &
   \\[10pt]
   -\frac{1}{2}\frac{\p r_S}{\p y} &
   0 &
   0 &
   \\[10pt]
   -\frac{1}{2}\frac{\p r_S}{\p z} &
    0 &
    0 &
   \end{pmatrix}
   = \frac{v_S}{r_S}\frac{\p W(r_S)}{\p r_S}\begin{pmatrix}
   0 &
   \frac{1}{2} y &
   \frac{1}{2} z &
   \\[10pt]
   -\frac{1}{2} y &
   0 &
   0 &
   \\[10pt]
   -\frac{1}{2} z &
    0 &
    0 &
   \end{pmatrix}.
   \end{align}
   \begin{align}
\label{theta_Alcubierre}
    \Theta &= \p_x V \ , \quad \Theta^{\rm Alc} = v_S \frac{\p W(r_S)}{\p r_S}\frac{\p r_S}{\p x} 
    =  v_S \frac{\p W(r_S)}{\p r_S}\frac{x-x_s}{r_S}\ ,
    \\[10pt]
\label{sigma_scalar_Alcubierre}
    (\Sigma^2) &= \frac{1}{3}(\p_x V)^2 + \frac{1}{4}\Big[ (\p_y V)^2 + (\p_z V)^2 \Big] =\frac{1}{3}\Theta^2 + \Omega^2 \ ,\nn
    \\[10pt]
    (\Sigma^2)^{\rm Alc} &=  
    v_S^2\Big(\frac{\p W(r_S)}{\p r_S}\Big)^2\left[\frac{1}{3}\Big(\frac{\p r_S}{\p x}\Big)^2+\frac{1}{4}\Big(\frac{\p r_S}{\p y}\Big)^2+\frac{1}{4}\Big(\frac{\p r_S}{\p z}\Big)^2\right]\
    \nn
    \\[10pt]
    &= \frac{v_S^2}{r_S^2}\Big(\frac{\p W(r_S)}{\p r_S}\Big)^2\Big(\frac{1}{3}(x-x_S)^2+\frac{1}{4} y^2 +\frac{1}{4} z^2 \Big).
    \\[10pt]
\label{omega_scalar_Alcubierre}
    (\Omega^2) &= \frac{1}{4} \Big[ (\p_y V)^2 + (\p_z V)^2 \Big]  \ , \nn
    \\[10pt]
    (\Omega^2)^{\rm Alc} &=
     v_S^2\Big(\frac{\p W(r_S)}{\p r_S}\Big)^2
    \left[\frac{1}{4}\Big(\frac{\p r_S}{\p y}\Big)^2+\frac{1}{4}\Big(\frac{\p r_S}{\p z}\Big)^2\right]
    =
    \frac{v_S^2}{r_S^2}\Big(\frac{\p W(r_S)}{\p r_S}\Big)^2
    \Big(\frac{1}{4} y^2 +\frac{1}{4} z^2 \Big)\ .\
\end{align}

\section{Proof of Lemma 1}
\label{ProofofLemma1}

By definition we have for the coordinate acceleration $A^i = \frac{\mathrm{d}}{\mathrm{d} t} V^i$.
Taking the derivative with respect to Eulerian coordinates results in $A^i_{\ , j}=\frac{\mathrm{d}}{\mathrm{d} t} V^i_{\ ,j} + V^i_{\ ,k}V^k_{\ ,j}$. We compare this with 
the covariant field tensor, defined as ${\mathcal F}^i_{\ j} := \frac{\mathrm{d}}{\mathrm{d} t} \Theta^i_{\ j} + \Theta^i_{\ k}\Theta^k_{\ j}$, and formally obtain the relation:
\begin{equation}
\label{Ftensor}
A^i_{\ , j} = {\mathcal F}^i_{\ j} + \frac{\mathrm{d}}{\mathrm{d} t}\Omega^i_{\ j} + \Omega^i_{\ k}\Omega^k_{\ j} +
\Theta^i_{\ k}\Omega^k_{\ j} + \Omega^i_{\ k} \Theta^k_{\ j}\ ,
\end{equation}
with\footnote{Notice that $\Omega^i_{\ k}\Omega^k_{\ i} = - \Omega_{ik}\Omega^{ik} = - 2\Omega^2$ according to \eqref{vorticity_def}.}
\begin{equation}
\label{fieldequationsA}
A_{(i,j)}=\CF_{ij}+ \tensor{\Omega}{_i_k}\Omega{^k}{_j} \ \ , \ \ 
A_{[i,j]}= \ddt{}{\Omega_{ij}} + 2\Theta_{k [i} \tensor{\Omega}{^k_{j]}} \ \ , \ \ 
A^k_{\ , k}= \CF^k_{\ k} - 2\Omega^2 \ ,
\end{equation}
and the trace-free symmetric part,
\begin{equation}
\label{trace-freeA}
A_{(i,j)} - \frac{1}{3} A^k_{\ ,k}\delta_{ij}=\CF_{ij} - \frac{1}{3} \CF^k_{\ k} \delta_{ij}
+ \tensor{\Omega}{_i_k}\Omega{^k}{_j} + \frac{2}{3}\Omega^2 \delta_{ij} \ .
\end{equation}
We aim at calculating the field tensor $\CF^i_{\ j}$ from the Einstein evolution equation \eqref{Einsteinevolution2},\footnote{Here we extend a correspondence between Newtonian gravitation and general relativity including a shift vector field, \textit{c.f.} Section~\ref{IV}.} setting $N=1$ and $N^i = - V^i$, \textbf{R2}, \eqref{Einsteinevolution2} becomes (note that the terms proportional to $V^i$ also entail vorticity):  
\begin{align}
\partial_t \Theta^i_{\ j} = - {\mathcal R}^i_{\ j} - \Theta 
\Theta^i_{\ j} + [\Lambda - 4\pi G (p-\epsilon)]\delta^i_{\ j} + 8\pi G \pi^i_{\ j}  \nonumber\\
- V^k \Theta^i_{\ j||k} - \Theta^i_{\ k}V^k_{\ || j} + V^i_{\ ||k}\Theta^k_{\ j} \ .
\end{align}
Imposing the further restriction $\CR^i_{\ j} = 0$, \textbf{R3}, the covariant spatial derivatives become partial derivatives (denoted by a comma); we absorb the term 
$V^k \Theta^i_{\ j,k}$ into the total time-derivative to obtain:
\begin{equation}
\label{evolutionTheta}
    \ddt{}{\tensor{\Theta}{^i_j}} =
    - \Theta \tensor{\Theta}{^i_j} + \tensor{\Omega}{^i_k}\tensor{\Theta}{^k_j} 
    - \tensor{\Theta}{^i_k} \tensor{\Omega}{^k_j}  + 
    [\Lambda - 4\pi G (p-\epsilon)]\delta^i_{\ j} + 8\pi G \pi^i_{\ j} \ .
\end{equation}
We identify the field tensor ${\mathcal F}^i_{\ j}$ in this equation and obtain the relation:
\begin{equation}
\label{fieldtensorF}
    \tensor{\CF}{^i_j} =
    - \Theta \tensor{\Theta}{^i_j} + \tensor{\Theta}{^i_k} \tensor{\Theta}{^k_j}
    + \tensor{\Omega}{^i_k}\tensor{\Theta}{^k_j} - \tensor{\Theta}{^i_k} \tensor{\Omega}{^k_j} 
    + [\Lambda - 4\pi G (p-\epsilon)]\delta^i_{\ j} + 8\pi G \pi^i_{\ j} \ .
\end{equation}
resulting in the (covariant) field equations for $\CF^i_{\ j}$ (using \eqref{relationtoOmega2}, the energy constraint \eqref{energyconstraint}), and the trace-free property of $\pi^i_{\ j}$:
\begin{equation}
\label{fieldeqF}
\tensor{\CF}{^k_k} = \Lambda - 4\pi G (\epsilon + 3p)\quad ; \quad
{\mathcal F}_{[ij]} = 0 \ ,
\end{equation}
where the second is granted by definition of the field $\tensor{\CF}{^i_j}$. 

We combine the covariant field equation for the trace \eqref{fieldeqF} with the trace equation for the coordinate acceleration \eqref{fieldequationsA} to obtain the result \eqref{divA} of Lemma 1. $\Box$

Equation \eqref{rotA} is the transport equation for the coordinate vorticity, which is the identity already derived in \eqref{partsofgradA} using the kinematical decomposition $\Theta_{ij} = (1/3)\Theta \delta_{ij} + \sigma_{ij}$. $\Box$

The trace-free symmetric part follows by using \eqref{fieldtensorF}, \eqref{fieldeqF}, \eqref{relationtoOmega2}, and the energy constraint \eqref{energyconstraint} for $\CR =0$:
\begin{equation}
\label{trace-freeF}
 \tensor{\CF}{^i_j} - \frac{1}{3} \CF^k_{\ k} \delta^i_{\; j} =
    \tensor{\Theta}{^i_k} \tensor{\Theta}{^k_j} - \Theta \tensor{\Theta}{^i_j} - \frac{1}{3}
    \left( \Theta^{ij}\Theta_{ij} - \Theta^2 \right)\delta^i_{\; j} 
    + \tensor{\Omega}{^i_k}\tensor{\Theta}{^k_j} - \tensor{\Theta}{^i_k} \tensor{\Omega}{^k_j} 
    + 8\pi G \pi^i_{\ j} \ .
\end{equation}
Inserting the above result into \eqref{trace-freeA} we obtain the result \eqref{tidalA} of the Lemma. $\Box$\\

\section{Elements for and Proof of Corollary 2 in components}
\label{elementsCorollary2}

Since in the literature, Einstein's equations are often considered in components, 
this appendix may give useful formulas by providing the components of the stress-energy tensor and the Einstein equations. The full 4D Einstein system has been calculated in components using the Sagemath tool and the reader is directed to the data-availability statement at the end of this paper. 
    
We start with the metric \eqref{metricTheorem1}, where the coordinate velocity has the following form: $V(t,x,y,z)$.
After calculating the elements of the stress-energy tensor decomposition for a fluid \eqref{EMT}, we obtain from Einstein's equations the following purely spatial expressions\footnote{The spatial nature follows from the orthogonality relations for the momentum density vector and the stress tensor, $n_{\mu} q^\mu = 0$, $n^\mu p_{\mu\nu} = 0$.}:
\begin{equation}
    \epsilon = -\frac{1}{32 \pi G}\left[(\p_y V)^2 + (\p_z V)^2 + 4 \Lambda \right] 
\,,
\end{equation}
\begin{equation}
\begin{split}
    q^i = \frac{1}{16 \pi G} \begin{pmatrix}
    & - \p{_y^2} V -  \p{_z^2}  V\,, 
    &  \p_x(\p_y V)\,, 
    & \p_x (\p_z V)
    \end{pmatrix} \,.
\end{split}
\end{equation}
Notice that the spatial momentum flux vector $q^i$ is equal to the curl of the vorticity vector, which is the content of the momentum constraints.

From the components of $\tensor{p}{^i_j}$ we can read off the components of the anisotropic stress tensor; its trace and its non-diagonal elements are written as follows:
\begin{equation}
    \tensor{p}{^i_j} = p \delta{^i_j} + \tensor{\pi}{^i_j} 
\quad
\Rightarrow
\quad
    \tensor{p}{^k_k} = 3 p\,,
\quad
    \tensor{\pi}{^k_k} = 0\,,
\end{equation}
\begin{align*}
    p + \tensor{\pi}{^x_x} &= \frac{1}{8 \pi G} (-3\Omega^2 + \Lambda)
\quad (i)\,,
\\[4pt]
    p + \tensor{\pi}{^y_y} &=  \frac{1}{8 \pi G} \left[ -\p_x A + \Lambda + \Omega^2 - 2 (\Omega^y)^2\right]
\quad (ii)\,,
\\[4pt]
    p + \tensor{\pi}{^z_z}& = \frac{1}{8 \pi G} \left[- \p_x A + \Lambda + \Omega^2 - 2 (-\Omega^z)^2\right]
\quad (iii)\,,
\\[4pt]
    \tensor{\pi}{^x_y} &= \tensor{\pi}{^y_x} = \frac{1}{8 \pi G} \left( \frac{1}{2}\p_y A - \Omega^z \p_x V \right)\,,
\\[4pt]
    \tensor{\pi}{^x_z}& = \tensor{\pi}{^z_x} =  \frac{1}{8 \pi G} \left( \frac{1}{2}\p_z A + \Omega^y \p_x V   \right)\,,
\\[4pt]
    \tensor{\pi}{^y_z} &= \tensor{\pi}{^z_y} = -\frac{1}{4\pi G}\Omega^y \Omega^z\,.
\end{align*}
From the trace of $\tensor{p}{^i_j}$, we have with $(i)$, $(ii)$ and $(iii)$:
\begin{align*}\label{p_expression}
    3p = \frac{1}{8 \pi G} (-3\Omega^2 + 3\Lambda - 2\p_x A)
\quad&\Rightarrow\quad
    p = \frac{1}{8 \pi G} \left(-\Omega^2 + \Lambda - \frac{2}{3}\p_x A \right)\,,
\\[4pt]
    4\pi G (3p+\epsilon) &=  (-2\Omega^2 + \Lambda - \p_x A)
\quad(iv)\,.
\end{align*}
From $(iv)$ and the expression for $\epsilon$ we have:
\begin{equation}
\label{v_case2}
    \bm\p \cdot \bm A = \p_x A = \Lambda - 4\pi G (\epsilon + 3 p) - 2 \Omega^2
\quad(v)\,.
\end{equation}
The trace-free symmetric part of $A_{ij}$ defines the trace-free part $\pi_{ij}$ in terms of the sources of curl and divergence of $\bm A$ (recall that the curl of $\bm A$ is an identity via the Kelvin-Helmholtz transport equation for vorticity and not an additional constraint equation, since Einstein's equations are symmetric.) 

For the trace-free symmetric part we have in general (\textit{c.f.} \textbf{Theorem 1}):
\begin{equation}
    A_{(i,j)} - \frac{1}{3} \tensor{A}{^k_{,k}} \delta_{ij} = V_{i,k} \tensor{V}{^k_{,j}} -  \tensor{V}{^k_{,k}} V_{i,j} + \frac{2}{3} \II (\bm\partial \bm V) \delta_{ij} - 2 V_{i,k} \tensor{\Omega}{^k_j} + \tensor{V}{^k_{,k}} \Omega_{ij} + 2 \Omega_{ik} \tensor{\Omega}{^k_j} + 8 \pi G \pi_{ij} \,.
\end{equation}
Specified to $\bm V(t,x,y,z)= (V(t,x,y,z), 0, 0)$, the second principal scalar invariant vanishes, $\rm{II}(\bm\partial \bm V)=0 $, and we also have: $V_{i,k} \tensor{V}{^k_{,j}} -  \tensor{V}{^k_{,k}} V_{i,j} = V_{,x} V_{,j} -  V_{,x}V_{,j} = 0$.
For $\bm A(t,x,y,z)= (A(t,x,y,z), 0, 0)$ we are left with
(recall the components of $\bm\Omega$ and $\Omega_{ij}$ in Appendix~\ref{app:alcubierrekinematics}):
\begin{equation}
\begin{aligned}
\frac{2}{3}A_{,x} &= - 2 (V_{,y} \tensor{\Omega}{^y_x} + V_{,z}\tensor{\Omega}{^z_x}) +   2 (\Omega_{xy} \tensor{\Omega}{^y_x} + \Omega_{xz} \tensor{\Omega}{^z_x} )+ 8 \pi G \pi_{xx} 
\\[4pt]
 &    = (V_{,y})^2 + (V_{,z})^2 - \frac{1}{2} \left[  (V_{,y}^2 + V_{,z}^2) \right] + 8 \pi G \pi_{xx} 
  = 2\Omega^2 +  8 \pi G \pi_{xx}\,;
\\[4pt]
\frac{1}{2}A_{,y} =& -2 (V_{,x}  \tensor{\Omega}{^x_y}) + V_{,x} \Omega_{xy} + 2 \Omega_{xx} \tensor{\Omega}{^x_y} + 8 \pi G \pi_{xy} = -\frac{1}{2} V_{,x} V_{,y} +  8 \pi G \pi_{xy} \,;
\\[4pt]
\frac{1}{2}A_{,z} =&  -2 (V_{,x}  \tensor{\Omega}{^x_z}) + V_{,x} \Omega_{xz} + 2 \Omega_{xx} \tensor{\Omega}{^x_z} + 8 \pi G \pi_{xz} = -\frac{1}{2} V_{,x} V_{,z} +  8 \pi G \pi_{xz} \,.
\end{aligned}
\end{equation}
This shows that the divergence and the curl of $\bm A$ are determined through the trace-free symmetric components, but also that $\bm A$ is determined (up to a harmonic vector field) by its divergence and rotation, as follows:
\begin{equation}
\label{Corollary2a}
   \bm\p \cdot \bm A = \p_x A =  3 \Omega^2 + 12\pi G \tensor{\pi}{^x_x} \,,
\end{equation}
\vspace{-17pt}
\begin{equation}
\label{Corollary2b}
\bm\p \times \bm A =  \begin{pmatrix}
    & 0 
\\[4pt]
    & \p_z A
\\[4pt]
    &- \p_y A
\end{pmatrix}
= 2\begin{pmatrix}
    0&
\\[4pt]
    \p_t \Omega^y + V \p_x \Omega^y + \Omega^y \p_x V &
\\[4pt]
    \p_t \Omega^z + V \p_x \Omega^z + \Omega^z \p_x V &
    \end{pmatrix}
= \begin{pmatrix}
    & 0  
\\[4pt]
    & 16 \pi G \tensor{\pi}{^x_z} - \p_x V \p_z V
\\[4pt]
    & -16\pi G \tensor{\pi}{^x_y} +  \p_x V \p_y V 
\end{pmatrix}.
\end{equation}
\begin{equation}
A_{(i,j)} - \frac{1}{3} \tensor{A}{^k_{,k}} \delta_{ij} 
\\[4pt]
    = \begin{pmatrix}
    \frac{2}{3} \partial_x A&
    \frac{1}{2}\partial_y A&
    \frac{1}{2}\partial_z A&
\\[4pt]
    \frac{1}{2}\partial_y A&
    -\frac{1}{3}\partial_x A&
    0&
\\[4pt]
    \frac{1}{2}\partial_z A&
    0&
    -\frac{1}{3}\partial_x A&
    \end{pmatrix}\, ,
\end{equation}    
which overall proves \textbf{Corollary 2}. $\Box$\\

Turning now to the conservation laws, the energy conservation equation \eqref{energy_conservation_law} reads:
\begin{align*}
    &\dot\epsilon + \Theta (\epsilon + p) + \tensor{\pi}{^i_j}\tensor{\sigma}{^i_j} = 0
\\[4pt]
&\Leftrightarrow \ 
    \dot\epsilon + \Theta (\epsilon + p) = (-\frac{2}{3}\tensor{\pi}{^x_x}+\frac{1}{3}\tensor{\pi}{^y_y}+\frac{1}{3}\tensor{\pi}{^z_z})\p_x V -\frac{1}{2}(\tensor{\pi}{^x_y}+\tensor{\pi}{^y_x}) \p_y V -\frac{1}{2}(\tensor{\pi}{^x_z}+\tensor{\pi}{^z_x}) \p_z V
\\[4pt]
&\Leftrightarrow \ 
    \dot\epsilon + \p_x V (\epsilon + p +\frac{2}{3}\tensor{\pi}{^x_x}-\frac{1}{3}\tensor{\pi}{^y_y}-\frac{1}{3}\tensor{\pi}{^z_z}) + \tensor{\pi}{^x_y} \p_y V +\tensor{\pi}{^x_z} \p_z V = 0
\quad(vi)\,,
\end{align*}
because $\tensor{q}{^\mu_{;\mu}} = 0$. The total time derivative of the energy density reads:
\begin{align*}
    \dot\epsilon 
    &= \frac{\rd}{\rd t}\epsilon = \frac{\p}{\p t}\epsilon + V \p_x \epsilon
    = -\frac{1}{8\pi G} \left(\frac{\p}{\p t}\Omega^2 + V \p_x \Omega^2 \right) = -\frac{1}{8\pi G} \frac{\rm d}{{\rm d}t} \left(\Omega^2\right)\,.
\end{align*}
Using the vorticity transport equation \eqref{conservation_equation_vorticity}, respectively \eqref{Corollary2b}, we obtain for the evolution of the rotational energy:
\begin{align*}
\frac{\rm d}{{\rm d}t} \left(\frac{1}{2}\Omega^2\right) = \p_y A \,\p_y V + \p_z A \,\p_z V - \Omega^2 \p_x V \,,\\
\text{which yields:}\quad\dot\epsilon = -\frac{1}{4\pi G} [ \p_y A\, \p_y V + \p_z A\, \p_z V  - \p_x V \Omega^2]
\quad(vii)\,.
\end{align*}
We calculate separately the elements of $(vi)$ with the components expression of $p+\tensor{\pi}{^i_i}$: $(i)$, $(ii)$ and $(iii)$. We start with $(vii)$. Later we join the terms together:
\begin{align*}
(\alpha)\quad \dot\epsilon &= - \tensor{\pi}{^x_y} \p_y V -\tensor{\pi}{^x_z} \p_z V
    +\frac{1}{8\pi G} \p_x V (4\Omega^2)\,,
\\
    (\beta)\quad \p_x V (\epsilon +p) &= \p_x V \left[-\frac{1}{8\pi G}(\Lambda + \Omega^2)+p\right] = \p_x V \left[\frac{1}{8\pi G} \left(-2\Omega^2-\frac{2}{3} \p_x A\right)\right]
\\
    \text{using in the second step} \quad p &= \frac{1}{8\pi G} \left(\Lambda -\Omega^2 - \frac{2}{3} \p_x A\right) \,.
\end{align*}
\begin{align*}
(\gamma)\quad \tensor{\pi}{^i_j}\tensor{\sigma}{^i_j}&=\p_x V \left(\frac{2}{3}\tensor{\pi}{^x_x}-\frac{1}{3}\tensor{\pi}{^y_y}-\frac{1}{3}\tensor{\pi}{^z_z}\right)
    + \p_y V \tensor{\pi}{^x_y} + \p_z V \tensor{\pi}{^x_z} \\
    &= \p_x V \tensor{\pi}{^x_x} + \p_y V \tensor{\pi}{^x_y} + \p_z V \tensor{\pi}{^x_z}\,,
\end{align*}
using $\tensor{\pi}{^x_y} = \tensor{\pi}{^y_x}$ and $\tensor{\pi}{^x_z}= \tensor{\pi}{^z_x}$, 
and the trace-free condition for $\tensor{\pi}{^i_j}$. Overall we have:
\begin{align*}
    (\alpha)+(\beta)+(\gamma)=0 \quad &\Leftrightarrow \quad \frac{1}{8\pi G}\p_x V \left(2\Omega^2 - \frac{2}{3} \p_x A + 8\pi G\tensor{\pi}{^x_x}\right) = 0 \\
    &\Leftrightarrow \quad \tensor{\pi}{^x_x} = \frac{1}{8\pi G}\left(\frac{2}{3}\p_x A - 2 \Omega^2 \right)\,.
\end{align*}
We note that this relation is equivalent to $(i)$ if we introduce the $p$ expression from $(iv)$, so that the energy conservation equation does not provide any additional restriction.\\

The momentum conservation law \eqref{momentum_conservation_law}, for vanishing covariant acceleration (following from  the assumed constant lapse function, $a^\mu = h^{\mu\nu} [\ln (N)]_{||\nu} = 0$), 
and vanishing covariant vorticity in the flow-orthogonal foliation, reads:
\begin{align*}
    0 = h^{\alpha\mu} p_{;\alpha} + \tensor{h}{^\mu_\alpha} \dot q^{\alpha} + \frac{4}{3} \Theta q^{\mu} + q^{\alpha} \tensor{\sigma}{_\alpha^\mu} + \tensor{h}{^\mu_\alpha} \tensor{\pi}{^{\alpha\beta}_{;\beta}}
\quad(x)
\ .
\end{align*}
Below, we give the component expressions for each term (note $h^0_{\;0} = 0$):
\begin{align*}
    \tensor{h}{^{\alpha\mu}} p_{;\alpha} &= (\p_x p, \quad \p_y p, \quad \p_z p)\,,
\\
    \tensor{h}{^\mu_\alpha} \dot q^{\alpha} &= \tensor{h}{^\mu_\alpha} u^\beta  q{^\alpha_{\ ;\beta}}
    = \tensor{h}{^\mu_\alpha} u^\beta \left( \partial_\beta q^\alpha + {}^{(4)}\Gamma^\alpha_{\beta\gamma} q^\gamma \right)
    = \tensor{h}{^\mu_\alpha}  \left( u^\beta \partial_\beta q^\alpha + ({}^{(4)}\Gamma^\alpha_{0 \gamma}+ {}^{(4)}\Gamma^\alpha_{k \gamma} V^k) q^\gamma \right)
\\
    &=  \partial_t q^i+V^k \partial_k q^i -  V^i V^k \Theta_{jk}q^j - \tensor{\Omega}{^i_j} q^j+ V^i \Theta_{jk}q^j V^k  =  \frac{\rd q^i}{\rd t} - \tensor{\Omega}{^i_j}  q^j  
\\   
   &= \frac{1}{8\pi G}\left(({\bm\partial} \times {\bm\Omega})^x (1 + \frac{1}{2} \partial_y V + \frac{1}{2} \partial_z V) \ , \   ({\bm\partial} \times {\bm\Omega})^y (1 - \frac{1}{2} \partial_y V )\ , \ ({\bm\partial} \times {\bm\Omega})^z + \frac{1}{2}\partial_z V ({\bm\partial} \times {\bm\Omega})^x\right) 
\\
&= \frac{1}{8\pi G}\left( q^x(1 + \frac{1}{2} \partial_y V + \frac{1}{2} \partial_z V) \ , \  q^y (1 - \frac{1}{2} \partial_y V )\ , \ 
q^z + \frac{1}{2}\partial_z V q^x\right) \, ,
\end{align*}
\begin{align*}
    \frac{4}{3} \Theta q^{\mu} &= \frac{1}{6\pi G} \p_x V \left[({\bm\p} \times {\bm\Omega})^x, \quad  ({\bm\p} \times {\bm\Omega})^y, \quad ({\bm\p} \times {\bm\Omega})^z \right]
    = \frac{4}{3}  (q^x ,\  q^y ,\ q^z)\ \p_x V\,,
\\
    q^{\alpha} \tensor{\sigma}{_\alpha^\mu} 
   &= \left( q^x \frac{2}{3}\p_x V + q^y \frac{1}{2}\p_y V + q^z \frac{1}{2}\p_z V \ \ 
    ,\ \ -q^y \frac{1}{3} \p_x V + q^x \frac{1}{2} \p_y V \ \
    ,\ \  -q^z \frac{1}{3} \p_x V + q^x \frac{1}{2} \p_z V  \right)
\,,
\\
    \tensor{h}{^\mu_\alpha} \tensor{\pi}{^{\alpha\beta}_{;\beta}} &= 
    \left( \tensor{\pi}{^{xx}_{,x}}+  \tensor{\pi}{^{xy}_{,y}} + \tensor{\pi}{^{xz}_{,z}}
    ,\quad 
    \tensor{\pi}{^{yx}_{,x}} + \tensor{\pi}{^{yy}_{,y}} + \tensor{\pi}{^{yz}_{,z}}
    ,\quad
    \tensor{\pi}{^{zx}_{,x}} + \tensor{\pi}{^{zy}_{,y}} +  \tensor{\pi}{^{zz}_{,z}}\right)\,.
\end{align*}
As more compactly outlined in the proof to \textbf{Corollary 2}, we can rewrite Equation (x) solely in terms of spatial components (replacing $\dot q^i = \frac{\rd}{\rd t} q^i - \tensor{\Omega}{^i_j}q^j$):
\begin{equation*}
     -\bm\p \cdot \bm p = \frac{\rd}{\rd t} q^i + \frac{4}{3} \Theta q^i + \tensor{\sigma}{^i_j} q^j - \tensor{\Omega}{^i_j} q^j  \,.
\end{equation*}
Finally, $(x)$ becomes:
\begin{align*}
\Leftrightarrow
    -&\begin{pmatrix}
    &\p_x p + \tensor{\pi}{^{xx}_{,x}}+  \tensor{\pi}{^{xy}_{,y}} + \tensor{\pi}{^{xz}_{,z}}
\\[4pt]
    &\p_y p +\tensor{\pi}{^{yx}_{,x}} + \tensor{\pi}{^{yy}_{,y}} + \tensor{\pi}{^{yz}_{,z}}
\\[4pt]
    &\p_z p +\tensor{\pi}{^{zx}_{,x}} + \tensor{\pi}{^{zy}_{,y}} +  \tensor{\pi}{^{zz}_{,z}}
\end{pmatrix}
    =\begin{pmatrix}
    & \frac{\rd}{\rd t} q^x + 2 q^x \p_x V   
\\[4pt]
    &\frac{\rd}{\rd t} q^y + q^y \p_x V  + q^x \p_y V 
\\[4pt]
    &\frac{\rd}{\rd t} q^z + q^z \p_x V + q^x \p_z V 
    \end{pmatrix}\quad(xi)\,,
\end{align*}
or, $ -\bm\p \cdot \bm p = \frac{\rd}{\rd t}{\bm q} + (\p_x V) {\bm q} + q^x \bm\p V$.

\noindent
We express the above using $\bm A$ and calculate the curl of the vorticity transport equation:
\begin{align*}
    \frac{1}{2}\bm\p \times (\bm\p \times \bm A) &= \bm\p \times \left(\frac{\rd}{\rd t} \bm\Omega + \bm\Omega(\bm\p\cdot \bm V)-(\bm\Omega\cdot\bm\p) \bm V \right)\,,
\\
\Leftrightarrow
    \frac{1}{8\pi G}
    \begin{pmatrix}
    & -\p^2_y A - \p^2_z A 
\\[4pt]
    & \p_x (\p_y A)
\\[4pt]
    & \p_x (\p_z A)
    \end{pmatrix}
    &=
    \begin{pmatrix}
    & \frac{\rd}{\rd t} q^x - \frac{5}{2} q^y \p_y V - \frac{5}{2} q^z \p_z V + q^x \p_x V 
\\[4pt]
    & \frac{\rd}{\rd t} q^y + 2 q^y \p_x V + \frac{1}{2} q^y \p_y V + \frac{1}{16\pi G}\p_y V \p_x^2 V 
\\[4pt]
    & \frac{\rd}{\rd t} q^z + 2 q^z \p_x V + \frac{1}{2} q^z \p_z V + \frac{1}{16\pi G} \p_z V \p_x^2 V 
    \end{pmatrix}.
\end{align*}
Thus, $(xi)$ simplifies and reads: 
\begin{equation*}
     -\bm\p \cdot \bm p 
    =\begin{pmatrix}
    & \frac{-\p_y^2 A - \p_z^2 A}{8\pi G} +\frac{5}{2} q^y \p_y V + \frac{5}{2} q^z \p_z V + q^x \p_x V
\\[4pt]
    & \frac{\p_x(\p_y A)}{8\pi G} - q^y \p_x V -  \frac{1}{2}q^y \p_z V + q^x \p_y V -\frac{1}{16\pi G} \p_y V \p_x^2 V
\\[4pt]
    & \frac{\p_x(\p_z A)}{8\pi G} - q^z \p_x V -  \frac{1}{2}q^z \p_z V + q^x \p_z V - \frac{1}{16\pi G}\p_z V \p_x^2 V
    \end{pmatrix}
\quad(xii)\,.
\end{equation*}\\

\section{Coordinate acceleration for the Alcubierre solution}
\label{app:acceleration}

\begin{figure}[H]
    \centering
  \includegraphics[trim={0cm 0cm 0cm 0cm},clip, width=0.46\textwidth, height=0.49\textwidth]{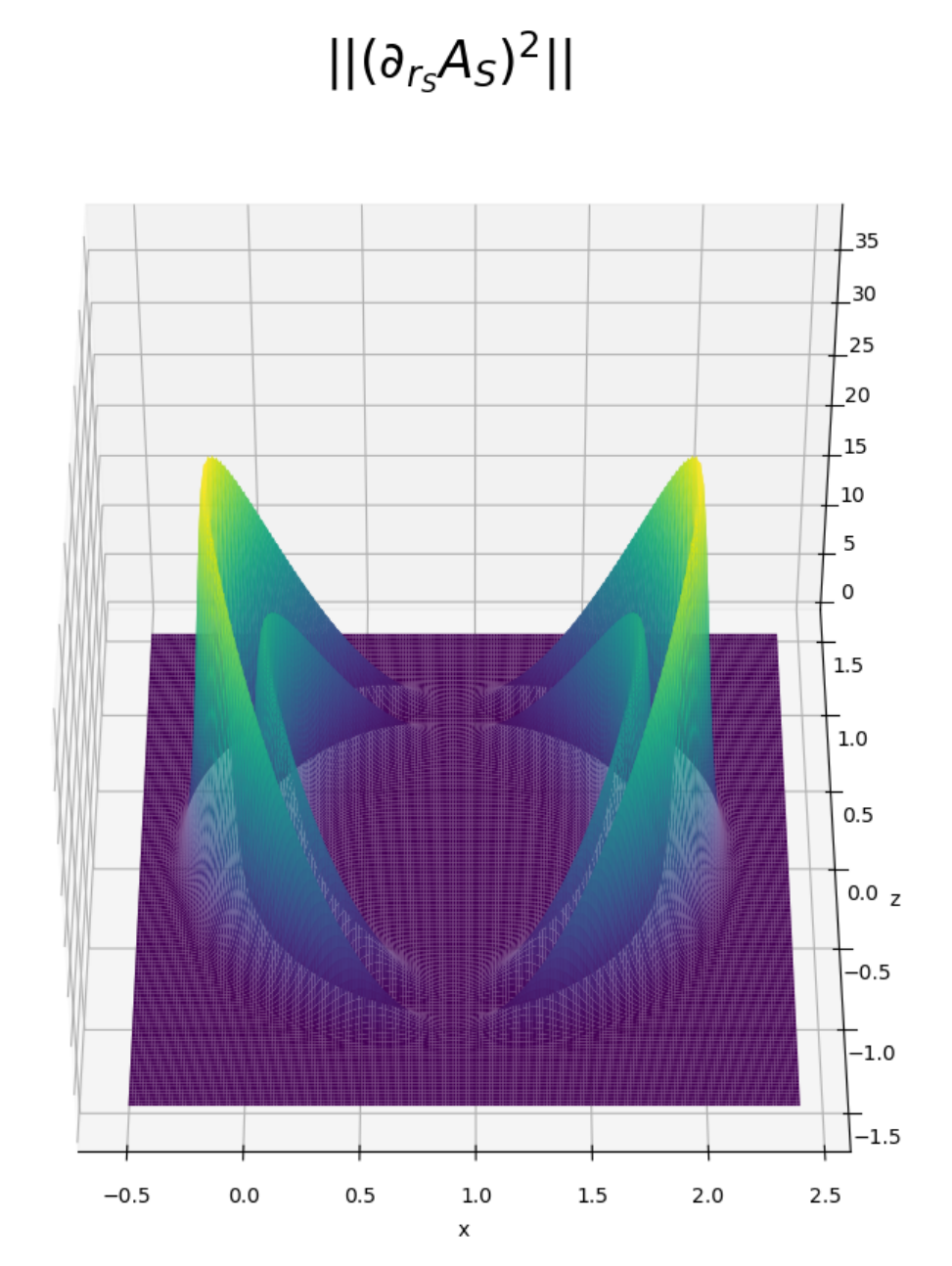}
     \hspace{10pt}\centering
  \includegraphics[trim={0cm 0cm 0cm 0cm},clip, width=0.46\textwidth, height=0.49\textwidth]{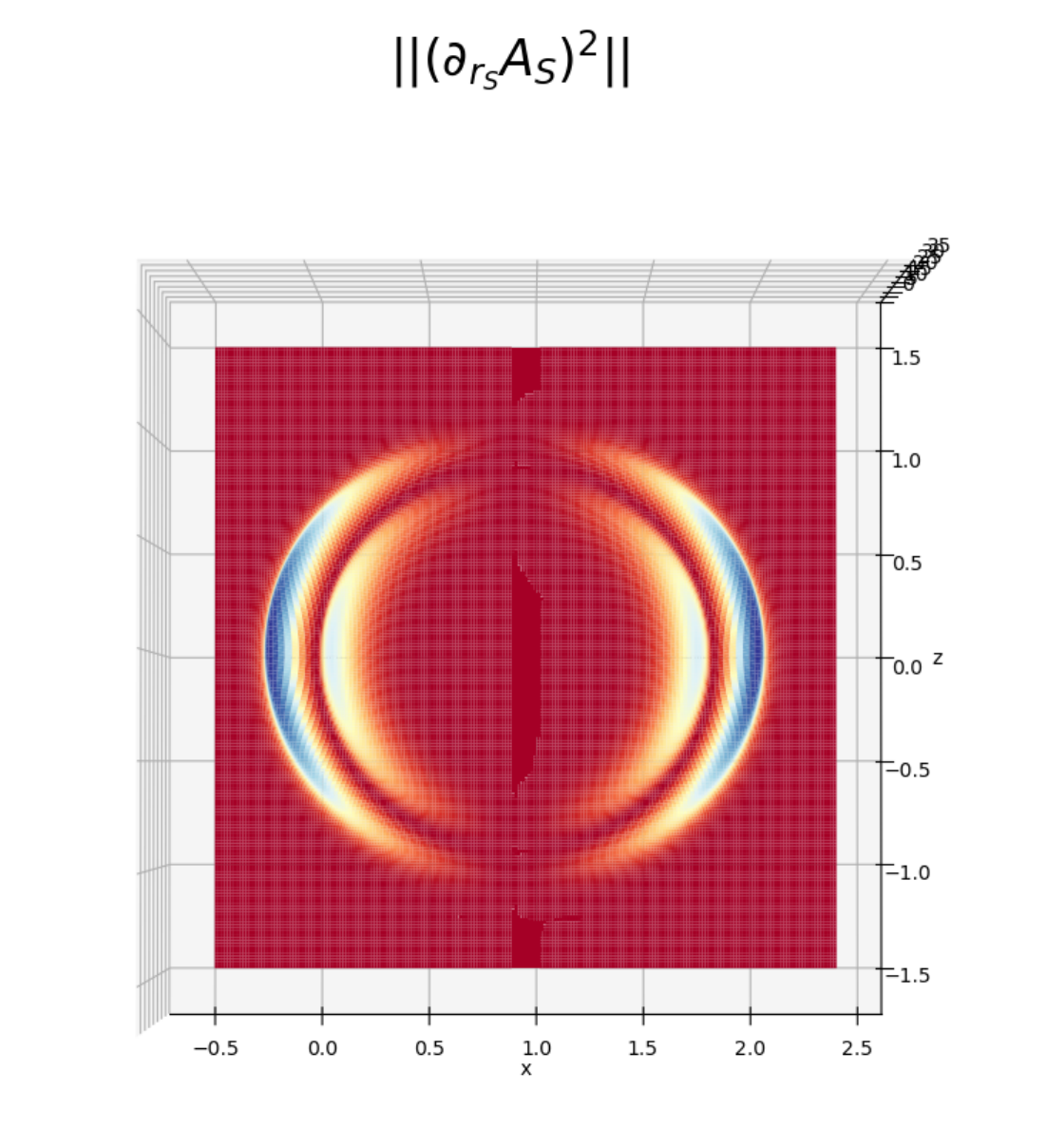}
  \caption{The norm of the $r_S$-derivative of the coordinate acceleration field. Two views are proposed, one facing us (direction of movement is along x from left to right), and another from above.}
    \label{gradA}
\end{figure}

We first calculate the coordinate acceleration directly in the Eulerian frame \footnote{We use the following derivation laws: $(\tanh{u})' = u' \sech^2 (u)\,,\quad (\sech^2 (u))' = -2 u' \sech^2 (u)\tanh{(u)}$.}: 
\begin{align*}
    A_S(t,x^k) &=\frac{\rd V_S (t,x^k)}{\rd t} = \frac{\partial v_s(t)}{\partial t} W(r_S) + v_S(t) \frac{\rd W(r_S)}{\rd t}
    =: a_S(t) W(r_S) + v_S(t) \frac{\partial W}{\partial r_S}\frac{\rd rs}{\rd t} \,,
\\
&= a_S(t) W(r_S) + v_S^2(t) g(r_S) \frac{(x-x_S (t))}{r_S} [W(r_S)-1]\,,
\\
\text{where}\quad
& g(r_S) = \frac{\p W}{\p r_S} = \frac{\sigma}{2 \tanh{(\sigma R)}} [\sech^2{(\sigma(r_S + R))} -  \sech^2{(\sigma(r_S - R))}]\,,
\end{align*}
with $a_S$ the acceleration of the warp bubble and we deduce for the gradient of $A_S$:
\begin{equation}\label{gradient acceleration Alcubierre}
\begin{aligned}
     \p_x A_S(t,x^k) &= a_S(t) \frac{(x-x_S)}{r_S} g(r_S) + \Theta^2 \\ 
     &+ v_S^2(t) \left[h(r_S)\frac{(x-x_S)^2}{r_S^2} (W(r_S)-1) + g(r_S)\left( \frac{r_S^2 -(x-x_S)^2}{r_S^3}(W(r_S)-1)  \right) \right],
\\[4pt]
 \p_y A_S(t,x^k) &=  a_S(t) \frac{y}{r_S} g(r_S) + \frac{\Theta^2 y}{x-x_S}\\
    &+ v_S^2(t) \left[h(r_S)\frac{(x-x_S)y}{r_S^2} (W(r_S)-1) +g(r_S)\left(- \frac{(x-x_S)y}{r_S^3}(W(r_S)-1) \right) \right],
\\[4pt]
    \p_z A_S(t,x^k) &=  a_S(t) \frac{z}{r_S} g(r_S)  + \frac{\Theta^2 z}{(x-x_S)}\\
    &+ v_S^2(t) \left[h(r_S)\frac{(x-x_S)z}{r_S^2} (W(r_S)-1) + g(r_S)\left(- \frac{(x-x_S)z}{r_S^3}(W(r_S)-1)  \right)\right],
\\[4pt]
\textrm{where}\quad h(r_S) &= \frac{\p^2 W}{\p  r_S^2}=\frac{\sigma^2}{\tanh{(\sigma R)}} [\sech^2{(\sigma(r_S +R))} \tanh{(\sigma(r_S + R))}\\
&- \sech^2{(\sigma(r_S -R))} \tanh{(\sigma(r_S -R))}]\ .
\end{aligned}
\end{equation}
To understand how the acceleration field changes when we move the warp, we also calculate the following derivative and illustrate it in figure~\ref{gradA}:
\begin{equation}
    \p_{r_S} A_S(t,x^k) = \frac{x-x_S}{r_S} \p_x A_S(t,x^k) + \frac{y}{r_S} \p_y A_S(t,x^k) + \frac{z}{r_S}\p_z A_S(t,x^k)\,.
\end{equation}
Finally, we briefly show that the total coordinate acceleration averaged over the warp field implies that the warp field itself does not have a net contribution to acceleration for vanishing external acceleration $a_S (t)$.
We have:
\begin{equation}
    A_S = \frac{d V_S}{d t} = \partial_t V_S + V_S \partial_x V_S = a_S(t) W(r_S) + v_S(t) \frac{\partial W}{\partial r_S} \frac{d r_S}{d t}\,,
\end{equation}
and define 
\begin{equation}
    \bm{q} = \left\{
    \begin{array}{ll}
        q_x = x-x_S(t) \\
        q_y =y \\
        q_z =z \\
    \end{array}
\right.
    \Rightarrow r_S = \sqrt{q_x^2 + q_y^2 + q_z^2}\ , \ \text{so}\quad 
    V_S(\bm{q},t) = v_S(t) W(r_S) \ .
\end{equation}
Taking spherical coordinates, $\rd^3 q = r_S^2 \sin{\theta}  \,\rd\theta \, \rd\phi\, \rd r_S$, the volume average yields:
\begin{equation}
 \langle A \rangle_{\mathbb{S}} =   \frac{1}{\textrm{Vol}(\mathbb{S})} \int_{\mathbb{S}} A \rd^3 q  
 =  \frac{a_S(t)}{\textrm{Vol}(\mathbb{S})} \int_{\mathbb{S}} W(r_S) r_S^2 \, \rd^3 q
 = \frac{3 a_S(t)}{R^3} \int_0^R W(r_S) r_S^2 \, \rd r_S \ .
 \end{equation}
 This means that for $a_S (t) = 0$ the average of the coordinate acceleration vanishes.\\

\section{Stress anisotropic scalar for inertial motion}\label{app:pi_calculation}

We take the $\tensor{p}{^i_j}$ and $p$ expressions from Appendix \ref{elementsCorollary2} and deduce the anisotropic stress components for the second example solution (\textit{idem}
$\p_i A =0$). With this we can calculate the stress anisotropic scalar $\Pi^2$:
\begin{align*}
    \tensor{\pi}{^x_x} &= \frac{1}{8 \pi G} (-3\Omega^2 + \Lambda) - p
\quad\Rightarrow\quad 
    \tensor{\pi}{^x_x} = \frac{1}{8 \pi G} (-2\Omega^2)\,,
\\[4pt]
    \tensor{\pi}{^y_y} &=  \frac{1}{8 \pi G} \left[\Lambda + \Omega^2 - 2 (\Omega^z)^2\right] - p
\quad\Rightarrow\quad 
    \tensor{\pi}{^y_y} = \frac{1}{8\pi G} \left[2\Omega^2 - 2 (\Omega^z)^2\right]\,,
\\[4pt]
    \tensor{\pi}{^z_z}& = \frac{1}{8 \pi G} \left[ \Lambda + \Omega^2 - 2 (-\Omega^y)^2\right] - p
\quad\Rightarrow\quad 
    \tensor{\pi}{^z_z} = \frac{1}{8\pi G} \left[2\Omega^2 - 2 (\Omega^y)^2\right]\,.
\end{align*}
We have $\pi^{\mu\nu} = h^{\mu\alpha}h^{\nu\beta} \pi_{\alpha\beta}$ and we see immediately that the $0$-components are null. Furthermore, $\tensor{\pi}{^\mu_\nu} = h^{\mu\alpha}\pi_{\alpha\nu}$. We only have spatial components and $\pi^{ij} = \tensor{\pi}{^i_j} = \pi_{ij}$,  so that we can simplify the scalar expression as follows:
\begin{equation}
\Pi^2 = \frac{1}{2} \pi^{ij}\pi_{ij} = \frac{1}{2} (\pi_{xx}^2 + \pi_{yy}^2 + \pi_{zz}^2 + 2 \pi_{xy}^2 + 2\pi_{xz}^2  + 2 \pi_{yz}^2)\,.
\end{equation}
Now we replace the pressure, $p = \frac{1}{8\pi G} (-\Omega^2 + \Lambda) $, and simplify:
\begin{align*}
    \pi_{xx}^2  &= \frac{1}{(8\pi G)^2} (4\Omega^4)\,,
\\[4pt]
    \pi_{yy}^2 &=  \frac{1}{(8\pi G)^2} \left[4\Omega^4 - 8 \Omega^2 (\Omega^z)^2+ 4 (\Omega^z)^4\right]\,,
\\[4pt]   
    \pi_{zz}^2 &= \frac{1}{(8\pi G)^2} \left[4\Omega^4 -8\Omega^2 (\Omega^y)^2  + 4 (\Omega^y)^4\right]\,,
\\
\Rightarrow\quad 
    &\pi_{xx}^2+ \pi_{yy}^2 + \pi_{zz}^2 =  \frac{1}{(8\pi G)^2} \left[8 \Omega^2 -  8 (\Omega^z)^2(\Omega^y)^2\right]\,,
\end{align*}
\begin{equation*}
    \pi_{xy}^2 = \frac{1}{(8 \pi G)^2} ( \Omega^z)^2 (\p_x V)^2
 , \ 
     \pi_{xz}^2 = \frac{1}{(8 \pi G)^2} ( \Omega^y)^2 (\p_x V)^2
 , \ 
    \pi_{yz}^2 = \frac{2}{(8\pi G)^2} (\Omega^y)^2 (\Omega^z)^2
  ,
\end{equation*}
\begin{align*}
\Rightarrow\quad
    2\pi_{xy}^2 + 2 \pi_{xz}^2 + 2\pi_{yz}^2 = \frac{1}{(8 \pi G)^2} \left[ 2\Omega^2 \Theta^2 + 8  (\Omega^z)^2(\Omega^y)^2 \right]
\,,
\\[4pt]   
\Rightarrow\quad
    \pi_{xx}^2+ \pi_{yy}^2 + \pi_{zz}^2 + 2\pi_{xy}^2 + 2 \pi_{xz}^2 + 2\pi_{yz}^2 = 
    \frac{1}{(8 \pi G)^2} \left(8\Omega^4 + 2\Omega^2 \Theta^2 \right)
\,.
\end{align*}
We finally obtain for the stress anisotropic scalar:
\begin{equation}
    \Pi^2 = \frac{1}{(8 \pi G)^2} (4\Omega^4 + \Omega^2 \Theta^2 )
\,.
\end{equation}

\renewcommand{\thesection}{References}
\newcommand\eprintarXiv[1]{\href{http://arXiv.org/abs/#1}{arXiv:#1}}
\section*{References}

\end{document}